\documentclass{aa}
\usepackage[varg]{txfonts}
\usepackage{natbib}
\usepackage{amsmath, amssymb}
\usepackage{graphicx}
\usepackage[breaklinks=true]{hyperref}
\newcommand{\pdt}[1]{{{\partial #1}\over {\partial t}}}

\newcommand{\fup}{F_{\uparrow,b}}
\newcommand{\fdown}{F_{\downarrow,b}}

     \def\aap{A\&A}
     \def\apjl{ApJL}
     \def\apjs{ApJS}

\hypersetup{
	pdftoolbar=false,
	pdfmenubar=true,
	pdfstartview={FitP},
	breaklinks=true,
	frenchlinks=false,
        colorlinks=true,
        linkcolor=red,
        citecolor=blue,
        filecolor=cyan,
        urlcolor=magenta,
        pdftitle={Dynamic mineral clouds on HD 189733b. I. 3D RHD with kinetic, non-equilibrium cloud formation},
	pdfsubject={hydrodynamics, radiative transfer, planets and satellites: atmospheres, planets and satellites: individual: HD 189733b, methods: numerical},
	pdfauthor={Graham Lee, Ian Dobbs-Dixon, Christiane Helling, Kristof Bognar, Peter Woitke}
        }

\begin{document}

\title{Dynamic mineral clouds on HD 189733b}
\subtitle{I. 3D RHD with kinetic, non-equilibrium cloud formation}
\author{G. Lee\inst{1}\thanks{E-mail:
gl239@st-andrews.ac.uk} \and I. Dobbs-Dixon\inst{2} \and Ch. Helling\inst{1} \and K. Bognar\inst{3} \and P. Woitke\inst{1}}
\institute{SUPA, School of Physics and Astronomy, University of St Andrews, North Haugh, St Andrews, Fife KY16 9SS, UK \and
NYU Abu Dhabi, PO Box 129188, Abu Dhabi, UAE \and
Department of Physics, University of Toronto, Toronto, ON M5S 1A7, Canada}

\date{Received: 30 March 2016 / Accepted: }

\abstract{
Observations of exoplanet atmospheres have revealed the presence of cloud particles in their atmospheres.
3D modelling of cloud formation in atmospheres of extrasolar planets coupled to the atmospheric dynamics has long been a challenge.
}
{
We investigate the thermo-hydrodynamic properties of cloud formation processes in the atmospheres of hot Jupiter exoplanets.
}
{
We simulate the dynamic atmosphere of HD 189733b with a 3D model that couples 3D radiative-hydrodynamics with a kinetic, microphysical mineral cloud formation module designed for RHD/GCM exoplanet atmosphere simulations.
Our simulation includes the feedback effects of cloud advection and settling, gas phase element advection and depletion/replenishment and the radiative effects of cloud opacity.
We model the cloud particles as a mix of mineral materials which change in size and composition as they travel through atmospheric thermo-chemical environments.
All local cloud properties such as number density, grain size and material composition are time-dependently calculated.
Gas phase element depletion as a result of cloud formation is included in the model.
In-situ \textit{effective medium theory} and Mie theory is applied to calculate the wavelength dependent opacity of the cloud component.
}
{
We present a 3D cloud structure of a chemically complex, gaseous atmosphere of the hot Jupiter HD 189733b.
Mean cloud particle sizes are typically sub-micron (0.01-0.5 $\mu$m) at pressures less than 1 bar with hotter equatorial regions containing the smallest grains. 
Denser cloud structures occur near terminator regions and deeper ($\sim$ 1 bar) atmospheric layers. 
Silicate materials such as MgSiO$_{3}$[s] are found to be abundant at mid-high latitudes, while TiO$_{2}$[s] and SiO$_{2}$[s] dominate the equatorial regions.
Elements involved in the cloud formation can be depleted by several orders of magnitude.
}
{
The interplay between radiative-hydrodynamics and cloud kinetics leads to an inhomogeneous, wavelength dependent opacity cloud structure with properties differing in longitude, latitude and depth.
This suggests that transit spectroscopy would sample a variety of cloud particles properties (sizes, composition, densities).
}


\maketitle

\section{Introduction}
\label{sec:Intro}
The detection of atoms and molecules such as Na \citep{Charbonneau2002}, K \citep{Sing2011a}, H$_{2}$O \citep{Swain2009}, CO \citep{Snellen2010} and CH$_{4}$ \citep{Swain2008} in the atmospheres of hot Jupiter exoplanets have provided a first glimpse into the chemical composition of these planets.
One of the major results of observational efforts is the detection of clouds/hazes in exoplanet atmospheres.
Two key signatures of atmospheric cloud in transmission spectra are an almost featureless optical spectrum, interpreted as Rayleigh scattering by cloud particles, and a relatively flat infrared spectrum with weakened molecular signatures.
Hubble and Spitzer observations of exoplanets such as HD 189733b \citep{Pont2013}, WASP-31b \citep{Sing2015a}, WASP-6b \citep{Nikolov2015} and GJ 1214b \citep{Kreidberg2014} show evidence for clouds in their atmosphere.
\citet{Deming2013} present detailed HST observations of HD 209458b, and show that the muted spectral H$_{2}$O and Na features can be explained by the additional grey opacity of clouds.
\citet{Sing2016}'s study of 10 hot Jupiters, incl. HD 189733b, in their Hubble and Spitzer observing program, provides compelling evidence that the muted water lines in the infrared can be attributed to high altitude mineral clouds.
They show that the signatures of clouds are present across a diverse range of planet equilibrium temperatures (T$_{\rm eq}$ = 960$\,\ldots\,$2510 K), radii (R$_{\rm pl}$ = 0.96$\,\ldots\,$1.89 R$_{\rm J}$) and mass (M$_{\rm pl}$ = 0.21$\,\ldots\,$1.40 M$_{\rm J}$).
\citet{Lecavelier2008} and \citet{Wakeford2015} have shown that a sub-micron cloud composition of silicate minerals such as MgSiO$_{3}$[s] can fit the observed Rayleigh slope of HD 189733b.
Star spots \citep{McCullough2014} have been discussed as possible explanation for the HD 189733b spectral features.
However, this does not rule out the presence of atmospheric cloud.
\citet{Evans2013}'s HST analysis of secondary transit observations for HD 189733b show a blueward slope in the geometric albedo at optical wavelengths, suggesting the presence of wavelength dependent backscattering cloud particles. 

In recent years, global circulation models (GCM) and radiative-hydrodynamic (RHD) models have been developed in order to understand the global dynamics of extrasolar atmospheres.
To date, most groups have focused on modelling the hot Jupiters, HD 189733b \citep[e.g.][]{Showman2009,Dobbs-Dixon2013}, HD 209458b \citep[e.g.][]{Showman2009, Dobbs-Dixon2010,Rauscher2010,Heng2011,Mayne2014} and the warm sub-Neptune GJ 1214b \citep[e.g.][]{Kataria2014, Charnay2015a}.
Codes differ in their implementation and hydrodynamic assumptions (see summary in \citealt{Mayne2014}) and radiative-transfer schemes \citep{Amundsen2014} as well as the complexity with which gas and cloud chemistry is implemented.
However, the overall global thermal structures and jet patterns remain similar for HD 209458b simulations \citep{Heng2015}.
Hot Jupiter simulations of HD 189733b \citep[e.g.][]{Showman2009,Dobbs-Dixon2013} have consistently reproduced offsets of maximum temperature to the East of the sub-stellar point, in line with \citet{Knutson2009} Spitzer thermal maps of atmospheres.
\citet{Kataria2016} modelled the hydrodynamics, thermal structure of 9 hot Jupiters from the \citet{Sing2016} observing program.
They post-process their simulations and examine the differences in the gas phase chemical equilibrium abundances between each planet.
Other simulations have investigated the effect of eccentricity \citep{Lewis2010, Lewis2014}, orbital distance \citep{Showman2015}, rotation \citep{Rauscher2014, Showman2015} and irradiation \citep{Perna2012, Komacek2016} on the hydrodynamics of the atmosphere.

Thus far, investigations of atmospheric cloud properties of hot Jupiters using GCM/RHD simulations have not included a model for describing the formation of clouds.
\citet{Parmentier2013} simulated the mixing of constant $\mu$m sized (a = 0.1$\,\ldots\,$10 $\mu$m) cloud tracer particles for HD 209458b.
They showed that sub-micron sized grains are likely to not settle out of the upper atmosphere.
\citet{Charnay2015b} simulated the warm sub-Neptune GJ 1214b.
They applied a constant particle size approach for KCl[s] and ZnS[s] using the phase equilibrium condensate chemistry from \citet{Morley2012}.
\citet{Oreshenko2016} used a phase equilibrium scheme for MgSiO$_{3}$[s], Mg$_{2}$SiO$_{4}$[s], Fe[s], TiO[s] and Al$_{2}$O$_{3}$[s].
They post-process their GCM to model the optical and infrared wavelength phase curves of Kelper-7b \citep{Demory2013}.
\citet{Parmentier2016} simulated a suite of hot Jupiter atmospheric models at equilibrium temperatures ranging from T$_{\rm eq}$ = 1300$\,\ldots\,$2200 K.
They applied a phase equilibrium approach to model single, homogeneous species Fe[s], Al$_{2}$O$_{3}$[s], MgSiO$_{3}$[s], Cr[s], MnS[s] and Na$_{2}$S[s] clouds for prescribed, constant grain sizes.
They then post-process their suite of models to investigate optical phase curve offsets in the Kepler bandpass.
A common feature of the above cloud modelling approaches is the prescription of a constant grain size and homogeneous mineral composition of cloud particles.

We present the first 3D hot Jupiter RHD model with a coupled kinetic cloud formation, opacity and transport of cloud particles.
We combine the HD 189733b atmosphere simulation of \citet{Dobbs-Dixon2013} with a time-dependent, kinetic non-equilibrium cloud formation model based on \citet{Woitke2003, Woitke2004, Helling2006, Helling2008}.
Section \ref{sec:Model} provides a summary of the RHD model and the cloud formation model.
Section \ref{sec:approach} outlines our cloud formation approach, numerical approach and a discussion on the present state of convergence.
In Sect. \ref{sec:Results} we present the 3D temperature and horizontal velocity field and local cloud properties such as cloud number density, local cloud particle sizes, material composition and element depletion.
Section \ref{sec:CloudOpacity} presents wavelength dependent cloud, gas and total opacities and examines the stellar heating rates resulting from the cloud opacity.
Section \ref{sec:Discussion} contains the discussion and Sect. \ref{sec:Conclusion} outlines our summary and conclusions.

\section{Model description}
\label{sec:Model}

The 3D modelling of dynamic clouds for hot Jupiter atmospheres requires a coupled, time-dependent hydrodynamic, radiative-transfer and cloud formation model scheme. 
We time dependently evolve the 3D Navier-Stokes equations and two-stream radiative transfer scheme as in \citet{Dobbs-Dixon2013} coupled with a time dependent, 3D, kinetic phase-non-equilibrium cloud formation and gas/dust phase chemistry module based on \citet{Woitke2003, Woitke2004, Helling2006, Helling2008}.
Gravitational settling of the cloud particles and element conservation under the influence of cloud formation is taken into account.

\subsection{3D radiative-hydrodynamic model}
The radiative-hydrodynamic (RHD) model applied in this study combines the fully compressible 3D Navier-Stokes equations to a two-stream, frequency dependent radiative transfer scheme.
Equations \eqref{eq:continuity}, \eqref{eq:momentum} and \eqref{eq:thermalenergy} represent the continuity, momentum and energy conservation of a local fluid element respectively,

\begin{equation}
\pdt{\rho_{\rm gas}} + \nabla\cdot\left(\rho_{\rm gas}{\bf u_{\rm gas}}\right) = 0 ,
\label{eq:continuity}
\end{equation}

where $\rho_{\rm gas}$ [g cm$^{-3}$] is the hydrodynamic gas density and ${\bf u_{\rm gas}}$ [cm s$^{-1}$] the hydrodynamic gas velocity.

\begin{multline}
\pdt{{\bf u_{\rm gas}}} + \left({\bf u_{\rm gas}}\cdot\nabla\right) {\bf u_{\rm gas}}= -
\frac{1}{\rho_{\rm gas}}\nabla{P_{\rm gas}} + {\bf g} \\ -2{\bf \Omega\times u_{\rm gas}} - 
     {\bf\Omega \times} \left({\bf\Omega\times r}\right)+\nu\nabla^2{\bf
       u_{\rm gas}} +\frac{\nu}{3}\nabla\left(\nabla\cdot{\bf u_{\rm gas}}\right) ,
\label{eq:momentum}
\end{multline}

where $P_{\rm gas}$ [dyn cm$^{-2}$] is the local gas pressure, $\vec{g}$=g{\bf e}$_{r}$ [cm s$^{-2}$] the gravitational acceleration in the radial direction, ${\bf \Omega}$ [rad s$^{-1}$] the rotational frequency of the planet, ${\bf r}$(r, $\phi$, $\theta$) [cm] the spherical coordinate radial distance vector and $\nu$ = $\eta / \rho_{\rm gas}$ [cm$^{2}$ s$^{-1}$] the constant kinematic viscosity.

\begin{equation}
\left[ \pdt{e_{\rm gas}} + ({\bf u_{\rm gas}}\cdot\nabla) e_{\rm gas} \right] = - P_{\rm gas} \,
\nabla \cdot {\bf u_{\rm gas}} - \frac{1}{r^2}\frac{d}{dr}\left(r^2 F_r\right) +
S_{\star} + D_\nu ,
\label{eq:thermalenergy}
\end{equation}

where $e_{\rm gas}$ [erg cm$^{-3}$] is the internal energy density of the fluid element, F$_{r}$ [erg cm$^{-2}$ s$^{-1}$] the radiative flux in the radial direction, $S_{\star}$ [erg cm$^{-3}$ s$^{-1}$] the incident stellar energy density per second and $D_{\nu}$ [erg cm$^{-3}$ s$^{-1}$] the local energy density per second from gaseous viscous heating.
For the radiation field, the heating due to stellar irradiation $S_{\star}$ [erg cm$^{-3}$ s$^{-1}$] is given by

\begin{equation}
S_{\star}= \left(\frac{R_{\star}}{a}\right)^2
\displaystyle\sum\limits_{b=1}^{nb}\frac{d\tau_{b,\star}}{dr}
e^{-\tau_{b,\star}/\mu_{\star}}\int_{\nu_{b,1}}^{\nu_{b,2}}\pi
B_{\nu}\left(T_{\star},\nu\right)d\nu,
\label{eq:stellarheat}
\end{equation}

where $R_{\star}$ [cm] and $a$ [cm] are the stellar radius and semi-major axis, respectively, $r$ [cm] the radial coordinate, $\mu_{\star}$ = $\cos \theta$ the cosine of the angle between the normal and the incident stellar photons, accounting for the slant path from a vertically defined optical depth and $\tau_{b, \star}$ the optical depth of stellar photons for wavelength band $b$ given by 

\begin{equation}
 \tau_{b, \star}(r) = -1.66 \int^{r}_{R_{\rm p}}\rho_{\rm gas}\kappa_{b, \star}\textrm{d}l.
\end{equation}

Where 1.66 is the diffusivity factor, an approximation that accounts for an exponential integral that arises when taking the first moment of the intensity to calculate the flux \citep{Elsasser1942} and $\kappa_{b}$  [cm$^{2}$ g$^{-1}$] the opacity for wavelength band $b$ (Eq. \ref{eq:kappa}).
In the two-stream approximation, the net radial flux, $F_{r}\left(\tau_b\right)$ [erg cm$^{-2}$ s$^{-1}$], for wavelength band $b$ in the upward or downward direction is 

\begin{equation}
F_{r}\left(\tau_b\right) = \displaystyle\sum\limits_{b}
\left(\fup\left(\tau_b\right)-\fdown\left(\tau_b\right)\right).
\label{eq:ftotal}
\end{equation}

The propagation of radiation intensity in the downward $\fdown$ and upward $\fup$ direction at each cell (two-steam approximation) is given by

\begin{equation}
\fdown \left(\tau_b\right) =
\int_0^{\tau_b}S_b(T_{\rm gas}) e^{-\left(\tau_b-\tau_b^{\prime}\right)}
d\tau_b^{\prime},
\label{eq:fdownsolution}
\end{equation}

and

\begin{equation}
\fup\left(\tau_b\right) =
S_{b}\left(T_{bot}\right)e^{-\left(\tau_{b,bot}-\tau_b\right)} + \\
\int_{\tau_{b,bot}}^{\tau_b}S_b(T_{\rm gas})e^{-\left(\tau_b^{\prime}-\tau_{b}\right)}
d\tau_b^{\prime},
\label{eq:fupsolution}
\end{equation}

respectively, where $S_{b}$ [erg cm$^{-2}$ s$^{-1}$] is the source function given by the wavelength integrated Planck function at the local temperature (T = T$_{\rm gas}$),

\begin{equation}
 S_{b}(T_{\rm gas}) = \int_{\nu_{b,1}}^{\nu_{b,2}}\pi
B_{\nu}(T_{\rm gas},\nu)d\nu .
\end{equation}

$S_{b}(T_{bot})$ in Eq. \eqref{eq:fupsolution} is the contribution of energy emitted from the interior of the planet, below which the planet is assumed to emit as a blackbody.
This interior contribution to the upward flux is fixed to match the observed radius \citep{Dobbs-Dixon2013}.
Stellar heating is fully accounted for in Eq. \eqref{eq:stellarheat}.

More details on the RHD model implementation of the 3D Navier-Stokes equations and the two-stream approximation for radiative transfer can be found in \citet{Dobbs-Dixon2010} and \citet{Dobbs-Dixon2013}, respectively.

\subsection{Cloud formation model}
\label{sec:Moments}

On Earth, water clouds are formed when particles made of sand, ash or ocean spray salt are lofted up into atmospheric layers where water vapour is supersaturated.
These cloud condensation nuclei or aerosols are the start of the formation of water droplets due to the surface area they provide.
Only a small supersaturation ratio (S $\sim$ 1.001; \citealt{Korolev2003}) is required to start water condensation on these surfaces in the Earth's atmosphere.
Since no solid crust exists on hot Jupiters, cloud formation has to start from the gas phase by the nucleation of seed particles.
Possible seed formation species have been reviewed by \citet{Helling2013} which found TiO$_{2}$ to be an efficient nucleating gas species for Brown Dwarf and hot Jupiter atmospheres.
TiO$_{2}$[s] is a high-temperature condensate which forms seed particles from a chemical pathway of successive (TiO$_2$)$_{N}$ stable clusters through homogeneous nucleation. 
Based on the results of \citet{Jeong2003} and \citet{Lee2015a}, we are in the position to apply a modified classical nucleation theory where we use the thermodynamical data for the individual cluster sizes (TiO$_2$)$_{N}$ to derive the seed formation rate dependent on the local gas temperature and gas density, and hence calculate the number of locally formed seed particles.
After the first seed particles are nucleated from the gas phase, chemical surface reactions grow the grain bulk.
This surface can grow or evaporate while the particles are transported across the globe, depending on the local thermodynamic and chemical conditions.
The composition of the grain bulk changes when various materials become thermally stable or unstable.
This process results in a solid mantle of mixed mineral composition where the seed particle contributes a negligible amount of volume compared to the total volume of the whole cloud particle.
The local growth and evaporation chemical surface processes occur at second-to-minute timescales \citep{Helling2001, Helling2004} meaning the properties (size, composition) of the cloud particles adapt quickly to the local gaseous thermodynamic properties. 

To describe the cloud formation process we use the set of dust moment equations derived by \citet{Woitke2003,Woitke2004,Helling2006,Helling2008}.
The dust moments $L_{j}(\vec{r})$ [cm$^{j}$ g$^{-1}$] ($j$ = 0, 1, 2, 3) are the local integrated particle size distribution, weighted by a power of the grain volume $V^{j/3}$, defined as
\begin{equation}
 \rho_{\rm gas} L_{j}(\vec{r}, t) = \int_{V_{l}}^{\infty} f(V, \vec{r}, t)V^{j/3} \textrm{d}V, 
\end{equation}

where $f(V,\vec{r})$ [cm$^{-6}$] is the distribution of particles in volume space and $V_{l}$ [cm$^{3}$] the volume of a seed particle.
The conservation equation of dust moments is given by \citep{Woitke2003} 

\begin{equation}
\label{eq:dustmom}
\pdt{\left(\rho_{\rm gas} L_{j}\right)} + \nabla \cdot \left(\rho_{\rm gas} L_j {\bf u_{\rm d}} \right) = V_{l}^{j/3}J_{*} + \frac{j}{3}\chi^{\rm net}\rho_{\rm gas} L_{j - 1} ,
\end{equation}

which has the same conservation transport equation structure as the Navier-Stokes equations.
The local time evolution of cloud properties are represented by the source/sink terms (r.h.s.) describing the seed formation and the grain growth/evaporation chemical processes.
The second l.h.s. term denotes the advective flux of the conserved quantity ($\rho_{\rm gas} L_{j}$) through space, given by the dust phase hydrodynamic fluid velocity $\textbf{u}_{\rm d}(\vec{r})$ [cm s$^{-1}$]. 
The relative velocity of gas and dust is given by the drift velocity ${\bf v}_{\rm dr}(\vec{r})$ [cm s$^{-1}$] defined as \citep{Woitke2003} 

\begin{equation}
\label{eq:vdust}
{\bf v}_{\rm dr}(\vec{r}) = {\bf u}_{\rm d}(\vec{r}) - {\bf u}_{\rm gas}(\vec{r}) , 
\end{equation}

where $\textbf{u}_{\rm gas}(\vec{r})$ [cm s$^{-1}$] is the gas hydrodynamic velocity.
Following the analysis in \citet{Woitke2003} the mean equilibrium drift velocity $\langle$$\overset{\circ}{{\bf v}}_{\rm dr}$$\rangle(\vec{r})$ [cm s$^{-1}$] in the large Knudsen number regime (Kn $\gg$ 1) is given by

\begin{equation}
\label{eq:vdr}
\langle\overset{\circ}{{\bf v}}_{\rm dr}\rangle(\vec{r}) = - \frac{\sqrt{\pi}}{2}\frac{g\rho_{\rm d} \langle {\rm a}\rangle}{\rho_{\rm gas}c_{\rm T}} {\bf e}_{\rm r} , 
\end{equation}

where $\rho_{\rm d}$ [g cm$^{-3}$] is the bulk (material) dust density, $c_{\rm T}$ [cm s$^{-1}$] the speed of sound, $g$\textbf{e}$_{\rm r}$ [cm s$^{-2}$] the gravitational acceleration in the radial direction and $\langle$a$\rangle$$(\vec{r})$  [cm] the local mean grain size.

The r.h.s. terms in Eq. \ref{eq:dustmom} denote the local chemical processes that alter the local dust moments; nucleation: $J_{*}(\vec{r}) \geq$ 0 cm$^{-3}$ s$^{-1}$, growth: $\chi^{\rm net}(\vec{r}) >$ 0 cm s$^{-1}$ and evaporation: $\chi^{\rm net}(\vec{r}) <$ 0 cm s$^{-1}$.
The nucleation rate $J_{*}(\vec{r})$ [cm$^{-3}$ s$^{-1}$] for homomolecular homogeneous nucleation of seed particles, applying modified classical nucleation theory \citep{Jeong2003, Woitke2003, Helling2013, Lee2015a} is given by

\begin{multline}
 J_{*}(t, \vec{r}) = \frac{f(1,t)}{\tau_{\rm gr}(1,N_{*},t)}Z(N_{*}) \\ \exp\left((N_{*} - 1) \ln S(T) - \frac{\Delta G(N_{*})}{RT}\right),
\end{multline}

where $f(1,t)$ [cm$^{-3}$] is the number density of the seed forming gas species, $\tau_{\rm gr}$ [s] the growth timescale of the critical cluster size $N_{*}$, $Z(N_{*}$) the Zeldovich factor (the contribution from Brownian motion to the nucleation rate),  $\Delta G(N_{*})$ [kJ mol$^{-1}$] the change in Gibbs free energy from the formation of cluster size N$_{*}$ and $S(T)$ the supersaturation ratio defined as

\begin{equation}
\label{eq:sratio}
S(T) = \frac{n_{\rm s}kT}{p^{\rm vap}_{\rm s}(T)},
\end{equation}
where $n_{\rm s}$ is the gas phase number density of species $s$ and $p^{\rm vap}_{\rm s}$(T) the vapour pressure of species $s$, which is a function of local gas temperature.

The net growth/evaporation velocity $\chi^{\rm net}(\vec{r})$  [cm s$^{-1}$] of a grain due to chemical surface reactions \citep{Gail1986,Helling2006} is

\begin{equation}
\label{eq:chinet}
 \chi^{\rm net}(\vec{r}) = \sqrt[3]{36\pi}\sum_{s}\sum_{r = 1}^{R} \frac{\Delta V_{r}^{s}n_{r}^{\rm key}v_{r}^{\rm rel}\alpha_{r}}{\nu_{r}^{\rm key}}\left(1 - \frac{1}{S_{r}}\frac{1}{b^{s}_{\rm surf}}\right),
\end{equation}

where $r$ is the index for the chemical surface reaction, $\Delta V_{r}^{s}$ the volume increment of the solid $s$ by reaction $r$, $n_{r}^{\rm key}$ the particle density of the key reactant in the gas phase, $v_{r}^{\rm rel}$ the relative thermal velocity ($v_{r}^{\rm rel}$ = $\sqrt{kT / 2\pi m_{r}}$) of the gas species taking part in reaction $r$, $\alpha_{r}$ the sticking coefficient of reaction $r$ and $\nu_{r}^{\rm key}$ the stoichiometric factor of the key reactant in reaction $r$. 
$S_{r}$ is the reaction supersaturation ratio \citep{Helling2006} and 1/$b^{s}_{\rm surf}$ = V$_{\rm s}$/V$_{\rm tot}$ the volume of solid $s$, V$_{s}$, to the total grain volume containing all species V$_{\rm tot}$ = $\sum_{s}$ V$_{s}$.
No prior assumptions about the particular grain size distribution or grain sizes are required to compute the number density of cloud properties.
Should detailed analysis using a distribution be required, it can be reconstructed by assuming a distribution shape from the dust moments \textit{a posteriori} \citep{Helling2008,Stark2015}.

The local dust number density $n_{\rm d}(\vec{r})$ [cm$^{-3}$] and mean grain radius $\langle$a$\rangle(\vec{r})$ [cm] are calculated from the dust moments by

\begin{equation}
\label{eq:nd}
 n_{\rm d}(\vec{r}) = \rho_{\rm gas}(\vec{r}) L_{0}(\vec{r}),
\end{equation}

and

\begin{equation}
\label{eq:meana}
 \langle{\rm a}\rangle(\vec{r}) = \sqrt[3]{\frac{3}{4\pi}}\frac{L_{1}(\vec{r})}{L_{0}(\vec{r})},
\end{equation}

respectively.

The composition of material forming on the grain mantle changes as a result of local chemical and thermodynamic conditions and the thermal stability of each material in those conditions.
The volume of a specific solid mineral $s$ depends on the growth/evaporation rate of that material.
The volume of each material $s$ can be described by a separate moment conservation equation for the third dust moment, $L_{3,s}(\vec{r})$  [cm$^{3}$ g$^{-1}$], \citep{Helling2008} 

\begin{equation}
\label{eq:L3s}
\pdt{\left(\rho_{\rm gas} L_{3,s}\right)} + \nabla \cdot \left(\rho_{\rm gas} L_{3,s} {\bf u_{\rm d}} \right) = V_{l, s}J_{*} + \chi^{\rm net}_{s}\rho_{\rm gas} L_{2}, 
\end{equation}

where the growth velocity of the solid $s$, $\chi^{\rm net}_{s}(\vec{r})$ , is given by

\begin{equation}
\label{eq:chis}
 \chi^{\rm net}_{s}(\vec{r}) = \sqrt[3]{36\pi}\sum_{r = 1}^{R} \frac{\Delta V_{r}^{s}n_{r}^{\rm key}v_{r}^{\rm rel}\alpha_{r}}{\nu_{r}^{\rm key}}\left(1 - \frac{1}{S_{r}}\frac{1}{b^{s}_{\rm surf}}\right). 
\end{equation}

The local volume fraction V$_{s}$/V$_{\rm tot}$ of each species $s$ is calculated from the $L_{3, s}(\vec{r})$  dust moment using the identity \citep{Woitke2004, Helling2008}

\begin{equation}
\label{eq:midentity}
\sum_{s}L_{3, s}  = L_{3}, \sum_{s}V_{\rm s}  = V_{\rm tot}.
\end{equation}

\subsection{Element abundance and gas phase chemistry}
The local gas-phase composition is an input for the cloud formation process as it determines (along with temperature) the cloud formation nucleation and growth/evaporation rates.
The elements involved in the cloud formation process are altered by depletion or replenishment depending on the dominating cloud formation processes (nucleation, growth, evaporation).
The depletion/enrichment of gas phase elements $\varepsilon_{i}(\vec{r})$ (abundance ratio of element $i$ to Hydrogen; $\varepsilon_{i}(\vec{r})$ = $n_{i}$/$n_{\langle H\rangle}$) by cloud particle growth/evaporation \citep{Helling2006} is given by

\begin{multline}
\label{eq:eldep}
\pdt{\left(n_{\langle H\rangle}\varepsilon_{i}\right)} + \nabla \cdot \left(n_{\langle H\rangle}\varepsilon_{i}{\bf u_{\rm gas}} \right) = -\nu_{x,0}N_{l}J(V_{l}) \\
- \sqrt[3]{36\pi}\rho L_{2}\sum^{R}_{r=1}\frac{\nu_{x,s}n^{\rm key}_{r}v_{r}^{\rm rel}\alpha_{r}}{\nu^{\rm key}_{r}}\left(1 - \frac{1}{S_{r}}\frac{1}{b^{s}_{\rm surf}}\right) ,
\end{multline}

where $i$ is the index of the element that contributes to the cloud formation process and $n_{\langle H\rangle}$ the local total number density of hydrogen.
The first r.h.s. terms describes the consumption of elements from the nucleation process. 
The second r.h.s. term denotes the source (evaporation) and sink (growth) of elements as a result of cloud particle chemical surface reactions.
The second l.h.s. term describes the advection of $\varepsilon_{i}(\vec{r})$ through space at the local gas velocity \textbf{u}$_{\rm gas}(\vec{r})$.

The composition of the local gas phase is calculated assuming chemical equilibrium based on the element abundances derived from Eq. \eqref{eq:eldep}. 
Using these local element abundances, we use a computationally efficient, hierarchical chemical abundance calculation introduced in \citet{Helling2001}.

\subsection{Cloud and gas opacity}
Cloud particles are a large source of opacity, absorbing and scattering photons at optical and infrared wavelengths.
We apply spherical particle Mie theory \citep{Mie1908} in combination with effective medium theory to calculate the mixed material cloud opacity.
Extinction efficiencies, Q$_{\rm ext}(\lambda, a)$, for the local mean grain sizes $\langle$a$\rangle$ are calculated based on the \citet{Bohren1983} \textit{BHMie} routines. 
We use two approximations for the size parameter (x = 2$\pi\langle$a$\rangle$ / $\lambda$) limits of Mie theory. 
For large size parameters x $\geq$ 1000 we use the large particle, hard sphere scattering approximation where the absorption efficiency asymptotically tends towards zero, $Q_{\rm abs}$ = 0, and all extinction is assumed to be from scattering.
In the large particle limit,  extinction efficiency $Q_{\rm ext}$ is then

\begin{equation}
Q_{\rm ext} =  2.
\end{equation}

For small size parameters x $<$ 10$^{-6}$ we use the small metallic sphere particle limit approximation, as, for example, outlined by \citet{Gail2014}, where

\begin{equation}
Q_{\rm abs} = 4x\operatorname{Im}(\alpha) + \frac{2}{15}x^{3}\operatorname{Im}(\epsilon), 
\end{equation}

and 

\begin{equation}
Q_{\rm sca} = \frac{8}{3}x^{4}\alpha\alpha^{*},
\end{equation}

where $\epsilon$ is the complex dielectric function and $\alpha$ = ($\epsilon$ - 1)($\epsilon$ + 2).
The second term in $Q_{\rm abs}$ contains the effect due to induction of eddy currents on the grain surface by the electromagnetic field of the photons.
The $Q_{\rm sca}$ equation calculates the contribution to the total extinction from Rayleigh scattering. 
This approximation has been shown to produce similar results to Mie theory for very small size parameters \citep{Gail2014}.
For all other size parameters the full Mie calculation is carried out.
The mass extinction coefficient $\kappa_{\lambda,cloud}$ [cm$^{2}$ g$^{-1}$] at wavelength $\lambda$ is then given by

\begin{equation}
\kappa_{\lambda,cloud} = Q_{\rm ext}(a)\pi a^{2} n_{\rm d}/ \rho_{\rm gas}, 
\end{equation}

where a = $\langle$a$\rangle$ is the mean grain size from Eq. \eqref{eq:meana}.
We calculate the cloud opacity at wavelengths corresponding to the 31 wavelength opacity bin edges used in \citet{Showman2009}. 
The cloud opacity is then averaged across the wavelength range of the bins to calculate the cloud opacity for that band.
For gas opacities, we use the temperature, density tabulated frequency dependent results from \citet{Sharp2007} Planck averaged over the wavelength bins of \citet{Showman2009}.
The total band opacity from the gas and cloud $\kappa_{b, total}$ [cm$^{2}$ g$^{-1}$] is then given by

\begin{equation}
\label{eq:kappa}
\kappa_{b, total} = \kappa_{b, gas} + \kappa_{b, cloud}.
\end{equation}

This local total opacity is treated as a purely absorptive extinction in the radiative transfer scheme. 

\subsubsection{Effective medium theory}

Each mineral material present in the cloud particle will contribute their specific optical $(n, k)$ constants, weighted by their local volume fraction V$_{\rm s}$/V$_{\rm tot}$ (Eq. \ref{eq:midentity}) of the material in each grain.
Effective optical constants for the material mixtures are calculated using \textit{effective medium theory}.
We primarily use the numerical Bruggeman method \citep{Bruggeman1935} given by

\begin{equation}
 \sum_{s}\left(\frac{\rm V_{\rm s}}{\rm V_{\rm tot}}\right) \frac{\epsilon_{s} - \epsilon_{\rm av}}{\epsilon_{s} + 2\epsilon_{\rm av}} = 0, 
\end{equation}

where V$_{\rm s}$/V$_{\rm tot}$ is the volume fraction of solid species $s$, $\epsilon_{s}$ the dielectric function of solid species $s$ and $\epsilon_{\rm av}$ the effective, average dielectric function over the total cloud particle volume. 
A Newton-Raphson minimisation scheme is then applied to solve for $\epsilon_{\rm av}$.
In cases of rare non-convergence at the furthest infrared bands ($\lambda$ = 20.00-46.00, 46-324.68 $\mu$m) where the $(n, k)$ experimental values for materials are most uncertain; we apply the analytic Landau-Lifshitz-Looyenga [LLL] method of \citet{Looyenga1965} given by

\begin{equation}
\sqrt[3]{\epsilon_{\rm av}^{2}} = \sum_{\rm s}\left(\frac{\rm V_{\rm s}}{\rm V_{\rm tot}}\right)\sqrt[3]{\epsilon_{s}^{2}}.
\end{equation}

Values for the optical constants are the same as those used in \citet{Lee2015b}.
Most solid materials published experimental optical constants do not cover the wavelength range ($\lambda$ = 0.261$\,\ldots\,$324.68 $\mu$m) required for the radiative-transfer scheme.
To overcome this, we extrapolate the available data for each material to shorter and longer wavelengths.
For wavelengths shorter than published data we assume that $(n, k)$ remains constant.
At longer wavelengths, for the non-conducting material considered in this study, we assume that $n$ remains constant while $k$ is reduced linearly from the last data point to the longer wavelengths. 
This can make the material effective optical constants for the infrared wavelength bins, $\lambda$ = 20.00-46.00 $\mu$m and $\lambda$ = 46.00-324.68 $\mu$m uncertain depending on the volume fraction and available data of each species.

\section{Approach}
\label{sec:approach}

In this section we outline our cloud formation approach, numerical approach, initial conditions and convergence properties for our RHD and cloud formation model.
The addition of our cloud formation model the RHD model adds additional costs to the simulation times.
For example, the $\sim$60 simulated days presented here took $\sim$20 Earth days using 64 cores\footnote{Using the NYU Abu Dhabi HPC cluster, BuTinah}.

\subsection{Cloud formation and element abundance}

We consider the homogeneous nucleation of TiO$_{2}$ seed particles based on (TiO$_{2}$)$_{N}$ cluster data from \citet{Jeong2003, Lee2015a}.
We consider 5 simultaneous surface materials, TiO$_{2}$[s], SiO[s], SiO$_{2}$[s], Mg$_{2}$SiO$_{4}$[s], MgSiO$_{3}$[s] with 22 of the corresponding surface chemical reactions found in \citet{Helling2008}.
The local cloud properties are locally time-dependently computed for each computational domain while the flux of the moments through 3D space can be calculated using a normal advection scheme.
This is significant improvement over our previous 1D methods which rely on mixing timescale arguments to calculate clouds properties and do not consider transport in horizontal directions.
We solve Eq. \eqref{eq:eldep} for 4 element abundances: Ti, O, Si and Mg and assume a constant solar element abundance for all other elements.
We assume horizontal and meridional frictional coupling of the dust and gas phase (\textbf{u}$_{\rm d, h, m}$ = \textbf{u}$_{\rm gas, h, m}$). 
Vertical decoupling between the dust and gas phase is applied given by Eq. \eqref{eq:vdust} (\textbf{u}$_{\rm d, vert}$ = \textbf{u}$_{\rm gas, vert}$ + \textbf{v}$_{\rm dr}$).

\subsection{Numerical approach}
\label{sec:numapp}

\begin{figure*}
\begin{center}
\includegraphics[width=0.49\textwidth]{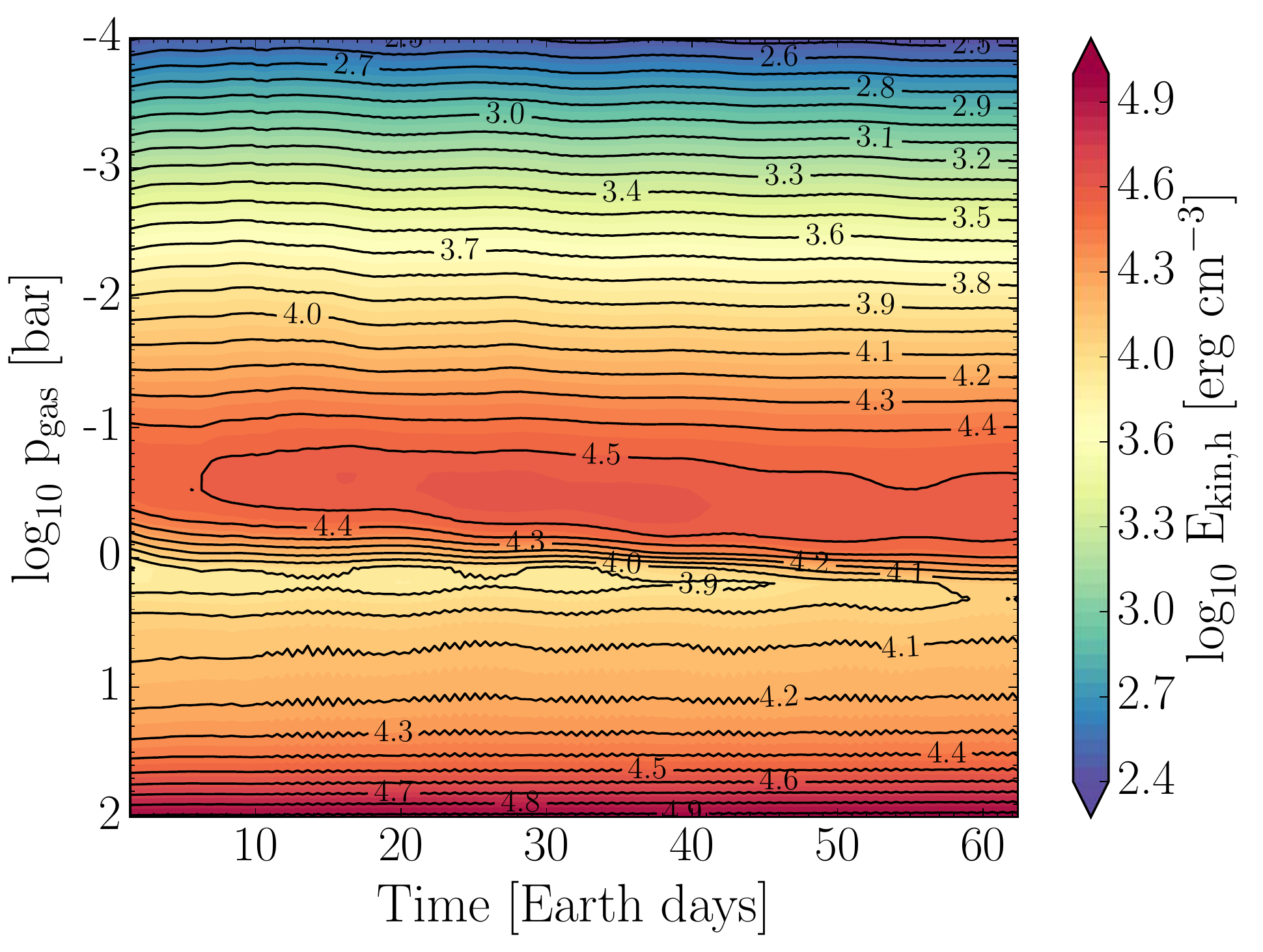}
\includegraphics[width=0.49\textwidth]{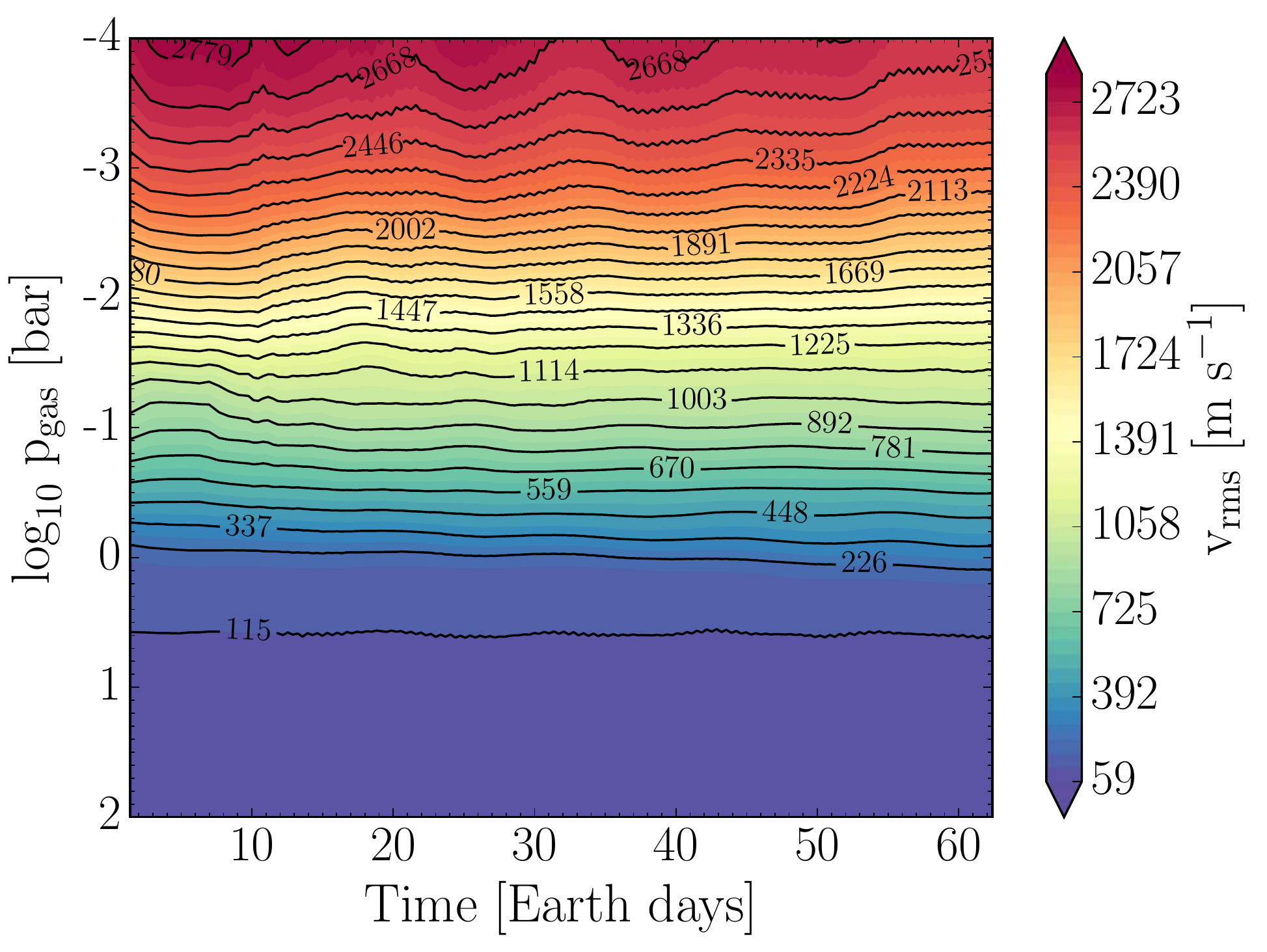}
\includegraphics[width=0.49\textwidth]{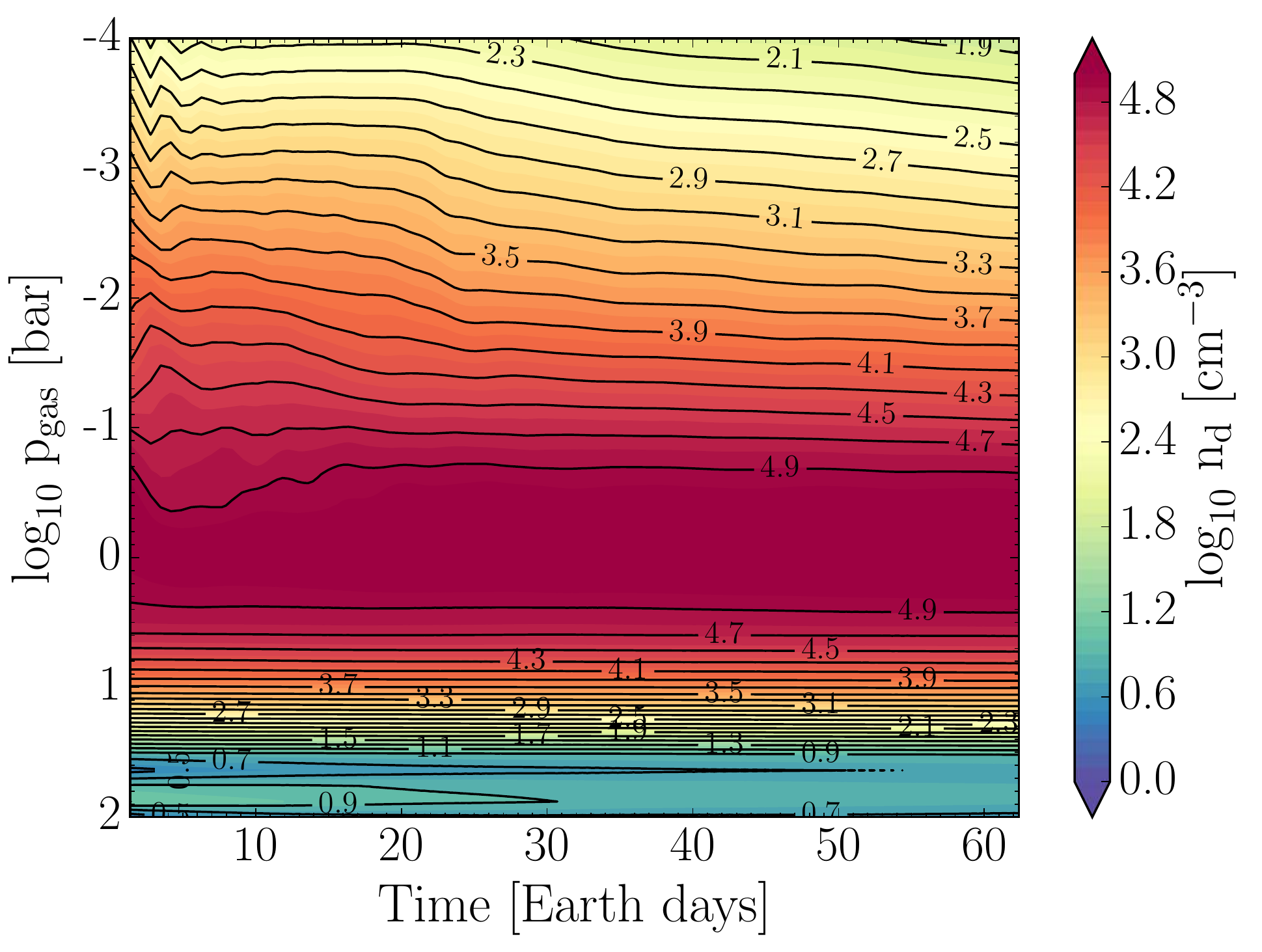}
\includegraphics[width=0.49\textwidth]{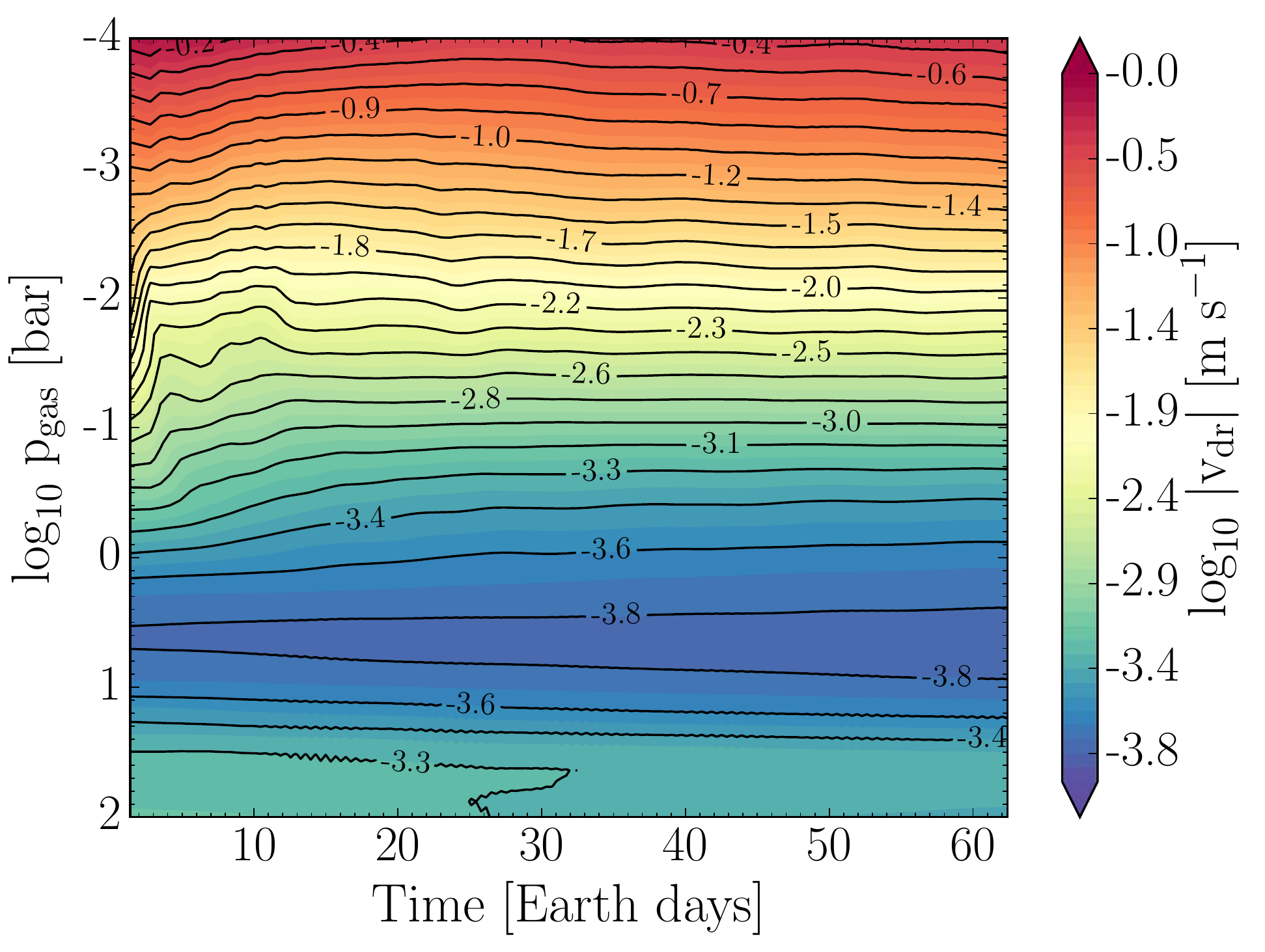}
\caption{Mean global hydrodynamic and cloud properties as function of local gas pressure during $\sim$60 Earth days of simulation.
\textbf{Top:} Kinetic energy density $\log_{10}$ E$_{\rm h, kin}$ [erg cm$^{-3}$] (left) and global mean r.m.s. velocity \textbf{v}$_{\rm rms}$ [m s$^{-1}$] (right).
\textbf{Bottom:} Global mean cloud particle number density $\log_{10}$ $n_{\rm d}$ [cm$^{-3}$] (Eq. \ref{eq:nd},  left) and drift velocity $\log_{10}$ $|$\textbf{v}$_{\rm dr}|$ [m s$^{-1}$] (Eq. \ref{eq:vdr}, right).
Parallel contour lines indicate that the global properties of the chosen value in the atmosphere, at that pressure iso-bar, are not changing significantly with time.
Note: there is no spin-up period since we use of a well converged continuation simulation of \citet{Dobbs-Dixon2013} as initial conditions.
 }
\label{fig:convergence}
\end{center}
\end{figure*}

We evolve the Navier-Stokes equations (Eq. \ref{eq:continuity}, \ref{eq:momentum} and \ref{eq:thermalenergy}) over a spherical grid with a resolution of (N$_{\rm r}$, N$_{\phi}$, N$_{\theta}$) = (100, 160, 64), where r is the radial direction, $\phi$ the longitude and $\theta$ the latitude.
The upper radial boundary is allowed to vary between $\sim$ 10$^{-5}$ and 10$^{-4}$ bar depending on dynamical properties and the lower boundary is fixed at $\sim$500 bar.
Vertical velocity dampening (sponge zone) is implemented near the upper boundaries of the simulation, common to GCM models \citep[e.g.][]{Mayne2014}.
We account for flow over polar regions by the method found in \citet{Dobbs-Dixon2012}.

One integration of the kinetic cloud formation chemistry can take $\sim$20-30 times the computational time of a single hydrodynamic timestep.
Therefore, the cloud formation chemistry (r.h.s. Eq. \ref{eq:dustmom}) is integrated every 10 hydrodynamic timesteps to update the local cloud particle properties.
The drift velocity \textrm{v}$_{\rm dr}(\vec{r})$  and cloud opacity $\kappa_{b, cloud}$ are updated after the cloud formation chemistry.
At every hydrodynamic timestep, cloud moments; L$_{j}(\vec{r})$, material volume composition; L$_{3, s}(\vec{r})$ and gaseous element abundances; $\varepsilon_{i}(\vec{r})$ are advected around the globe by calculation of the advective term in Eq. \eqref{eq:dustmom}, \eqref{eq:L3s} and \eqref{eq:eldep}.

Additionally, we take a number of physically based assumptions to reduce the number of cloud chemistry iterations required and to ensure physical solutions to the cloud properties.
During an evaporation event, the maximum integration step size is reduced by half to capture the evaporation process more consistently.
We limit evaporation of grains to the seed particle size ($\sim$0.001 $\mu$m); should the integrator attempt to evaporate the moment solutions below seed particle sizes, the end solution is assumed to be at the local seed particle values.
At this point, should the seed particles be thermally unstable ($\chi^{\rm net}_{\rm TiO_{2}[s]}(\vec{r})$ $<$ 0 cm s$^{-1}$) then all seed particles are assumed to be evaporated (i.e. $L_{j}(\vec{r})$ = 0) and the Ti and O elements returned to the gas phase by Eq. \eqref{eq:insta-evap}.
This condition is only met at the hottest, deepest parts of the atmosphere (T$_{\rm gas} >$ 2300 K) in our simulations.
Thermal stability may prevail to higher gas temperatures if other high-temperature condensates are included.

We limit the calculation of the dust chemistry to regions where the number density of grains is $n_{\rm d}(\vec{r})$ $>$ 10$^{-10}$ cm$^{-3}$ unless the local nucleation rate is $J_{*}(\vec{r})$ $>$ 10$^{-10}$ cm$^{-3}$ s$^{-1}$. 
This condition ensures that only regions that are efficiently nucleating or already contain a significant number of cloud particles are integrated. 
This criteria ensures that deeper atmospheric regions (p$_{\rm gas}$ $>$ 10 bar) which have a negligible nucleation but large growth rate, where a small number of cloud particles grow rapidly large ($>$1 cm), do not occur.
Without this criteria, these regions become computationally challenging as the drift velocity (Eq. \ref{eq:vdr}) becomes large ($\sim$ speed of sound) for these particles.
The hydrodynamic Courant-Friedrich-Levi [CFL] timestep condition then limits the hydrodynamic timestep to unfeasibly low values.
The cloud opacity and drift velocity in regions which contain very little cloud particles ($n_{\rm d}(\vec{r})$  $<$ 10$^{-10}$ cm$^{-3}$) are assumed to be zero.
For regions where $n_{\rm d}(\vec{r})$ $>$ 10$^{-10}$ cm$^{-3}$, a lower bound cloud opacity of $\kappa_{\rm cloud}$ = 10$^{-7}$ cm$^{2}$ g$^{-1}$ across all wavelength bands is implemented to aid numerical stability.
This is required since the results of Mie theory in certain seed particle regions can approach floating point limits.
This corresponds to a lower bound of TiO$_{2}$[s] seed particle opacity ($\kappa_{cloud}$($\lambda$, $a_{seed}$) $\ge$ 10$^{-7}$ cm$^{2}$ g$^{-1}$) at optical wavelengths. 

Furthermore, cells with very small growth/evaporation rates of $|\chi_{\rm net}(\vec{r})|$ $<$ 10$^{-20}$ cm s$^{-1}$ are assumed to remain constant with respect to the dust moments. 
Only cells which have local conditions that are significant departures from the equilibrium solution ($\chi_{\rm net}(\vec{r})$  = 0) are integrated in time.
We found integrating cells with $|\chi_{\rm net}(\vec{r})|$ $<$ 10$^{-20}$ cm s$^{-1}$ was computationally prohibitive and did not produce significantly different results.

The need to only update the r.h.s. of the dust moment equation every 10th hydrodynamic timestep may lead to a fast transport of cloud particles into high temperature regions where they will evaporate. 
This volatile material would evaporate rapidly at some material dependent ``evaporation window'' as it passed into these unstable regions.
The dust moment and element conservation equations become numerically stiff for such an intense evaporation, hence, the dust moments would approach phase equilibrium (i.e. when evaporation stops) in very small timesteps.
Small timesteps for the intense evaporation regions are necessary in order to solve the element conservation equation (Eq. \ref{eq:eldep}) to the best possible precision. 
To overcome this numerical challenge, we introduce a scheme where, should the integration timestep be too low at the beginning of the cloud chemistry integration, a fraction of the volatile materials are instantly evaporated back into the gas phase.
This process is repeated until integration of cloud properties can be computed in a reasonable time.
Growing or (near-)stable ($|\chi_{\rm s}(\vec{r})|$ $>$ 10$^{-30}$ cm s$^{-1}$) material are not altered and assumed to remain constant.
The return of elements to the gas phase from each evaporation species $s$ due to an instantaneous evaporation event is given by \citep{Woitke2006}

\begin{equation}
\label{eq:insta-evap}
\varepsilon_{i}(\vec{r})  = \varepsilon^{\rm b}_{i}(\vec{r}) + \frac{\nu_{s} 1.427 \textrm{amu}}{V_{0, s}} \Delta L_{3, s}(\vec{r}), 
\end{equation}

where $\varepsilon^{\rm b}_{i}(\vec{r})$ is the element abundance before the instant evaporation step, $\nu_{i, s}$ the stoichiometric coefficient of element $i$ in species $s$ and
1.427 amu the conversion factor between gaseous mass density $\rho_{\rm gas}$ and Hydrogen nuclei density $n_{\langle H\rangle}$ and $\Delta$$L_{3, s}(\vec{r})$ the difference in grain volume of species $s$ before and after instant evaporation.

This scheme has the added benefit of evenly spreading computational load, since each evaporating cell has a more equal work load.
Additionally, since the surface chemical growth of the particles occurs in second-minute timescales \citep{Helling2001, Helling2004}; if too much of a material is instantly evaporated off, the material can quickly grow back to an equilibrium solution before the end of one hydrodynamical time-step.

\subsection{Initial conditions}
\label{sec:initial_conditions}

For initial conditions, we use a well converged RHD model (total simulated time: $\sim$420 Earth days) of \citet{Dobbs-Dixon2013} which included a parameterised cloud opacity (Eq. \ref{eq:paraopac}).
This parameterised opacity is switched off in our simulations.
The initial 3D (T$_{\rm gas}$, p$_{\rm gas}$, $\textbf{v}_{\rm gas}$) structures do not vary significantly from the published results.
As in \citet{Dobbs-Dixon2013} we use a tidally locked HD 189733b set-up with the rotation rate equal to the orbital period (2.22 days).
We assume an initial solar abundance of elements throughout the globe given by \citet{Asplund2009}.

To set up the 3D RHD simulation, the dust properties are integrated each hydrodynamic timestep for the first 100 iterations and the effects of cloud opacity are neglected until $\sim$5.5 Earth days into the simulation.
During this time, larger sized grains with $\langle$a$\rangle$ $>$ 1 $\mu$m will have gravitationally settled from the upper atmosphere to their pressure supported regions ($\sim$1 bar).
After these first steps, the opacity of the cloud particles at all positions is accounted for in the radiative transfer scheme.

\subsection{Convergence tests}
\label{sec:convergence}

We investigate the present state of the model convergence by examining the the horizontal kinetic energy density E$_{\rm kin, h}$ = $\rho_{\rm gas}$ \textbf{u}$_{\rm h}^{2}$ / 2 [erg cm$^{-3}$] and root-mean-square (r.m.s.) horizontal + meridional velocity \textbf{v}$_{\rm rms}$ = $\sqrt{(\textbf{u}_{\rm h}^{2} + \textbf{u}_{\rm m}^{2}) / 2}$ [m s$^{-1}$] zonal and meridional averaged at pressure iso-bars.
These two quantities show how the global hydrodynamic velocity structure of the atmosphere is changing with atmospheric pressure and with time to check the state of the simulation with respect to a possible statistical steady state.
We introduce two properties for the cloud structure, the cloud number density $n_{\rm d}$ [cm$^{-3}$] (Eq. \ref{eq:nd}) and the equilibrium vertical drift velocity \textbf{v}$_{\rm dr}$ [m s$^{-1}$] (Eq. \ref{eq:vdr}), which are zonally and meridionally averaged at each iso-bar.
Together, these two quantities take into account the time evolution of the global density structure of cloud material as well as the grain size and composition due to their respective dependences.

Overall, the horizontal and meridionial mean gas kinetic energy density,  r.m.s. velocity and cloud property contours remain reasonably constant throughout our study integration period.
This suggests that the horizontal and meridional gas and cloud structures are not significantly varying in time during the $\sim$60 Earth days simulated here, and so further changes to the cloud structure are likely to come from the longer timescale vertical motions.
An integration time of $\sim$ 60 days does not capture the longer vertical settling timescales ($>$ 1000 days; \citealt{Parmentier2013}) of small particles ($<$ 0.1 $\mu$m) in the upper atmosphere, nor the larger ($\sim$ 1 $\mu$m) particles at the clouds base at $\sim$ 1 bar (Sect. \ref{sec:Discussion}).

\section{Results}
\label{sec:Results}

\begin{figure*}
\begin{center}
\includegraphics[width=0.49\textwidth]{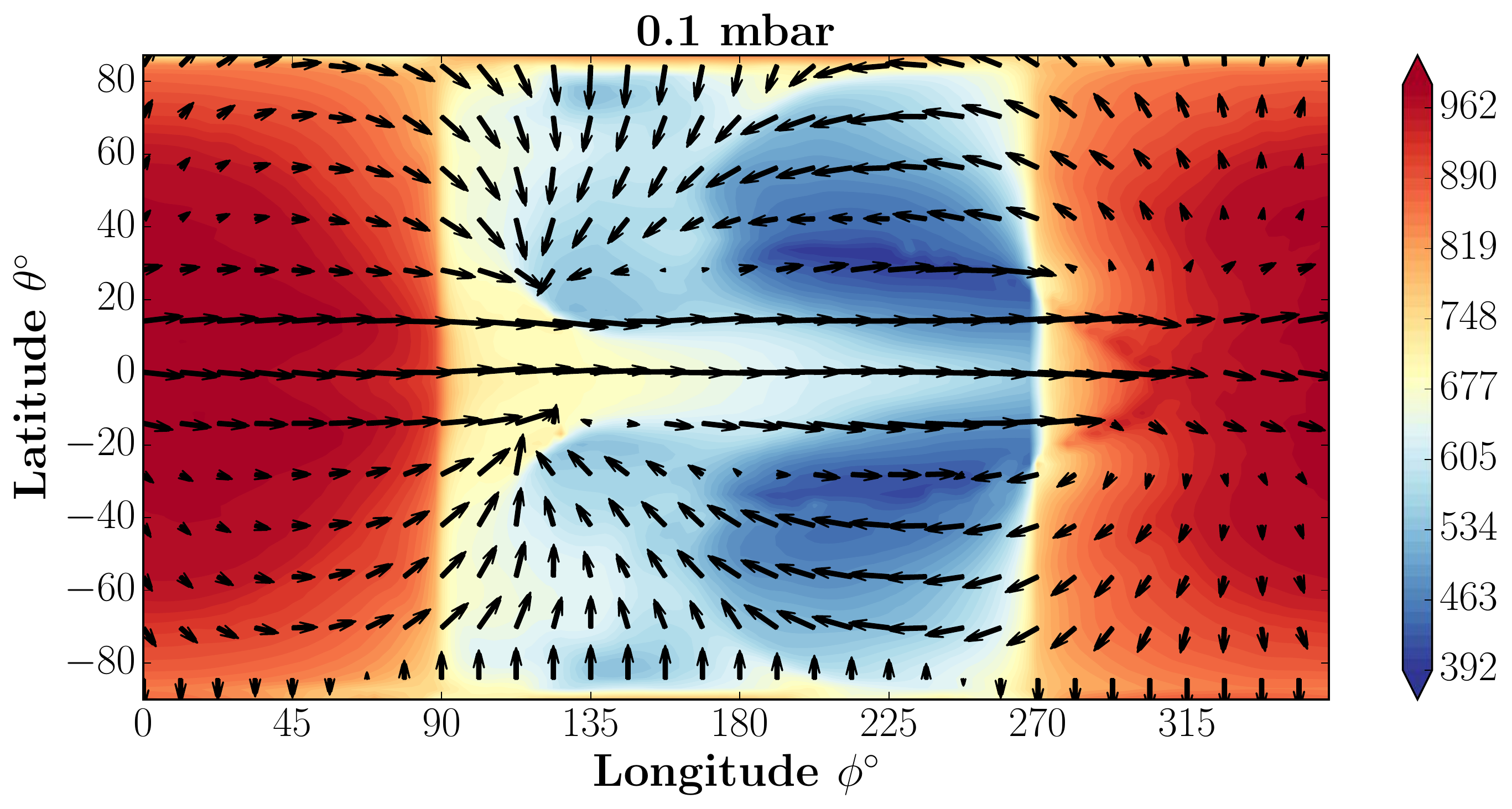}
\includegraphics[width=0.49\textwidth]{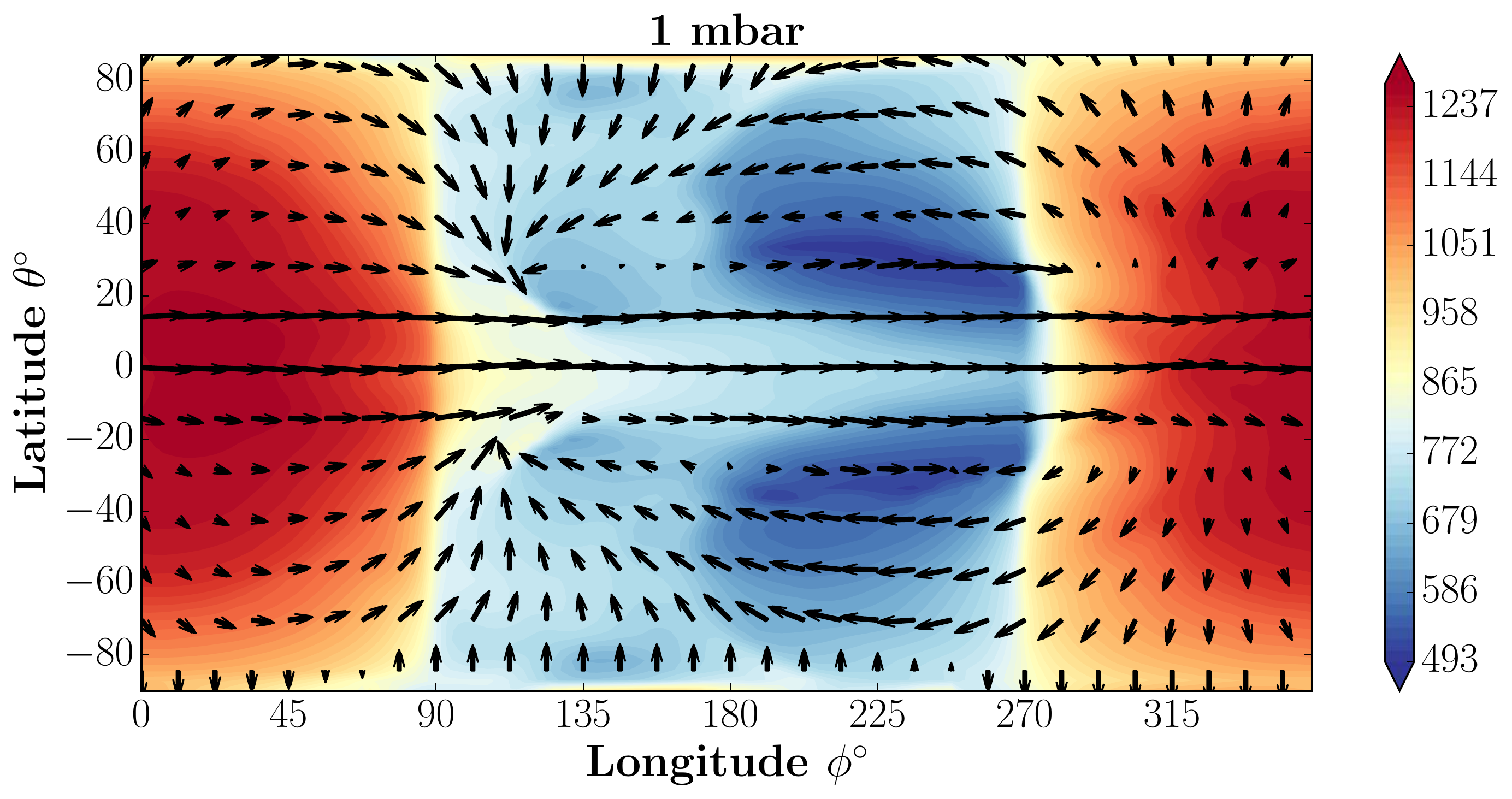}
\includegraphics[width=0.49\textwidth]{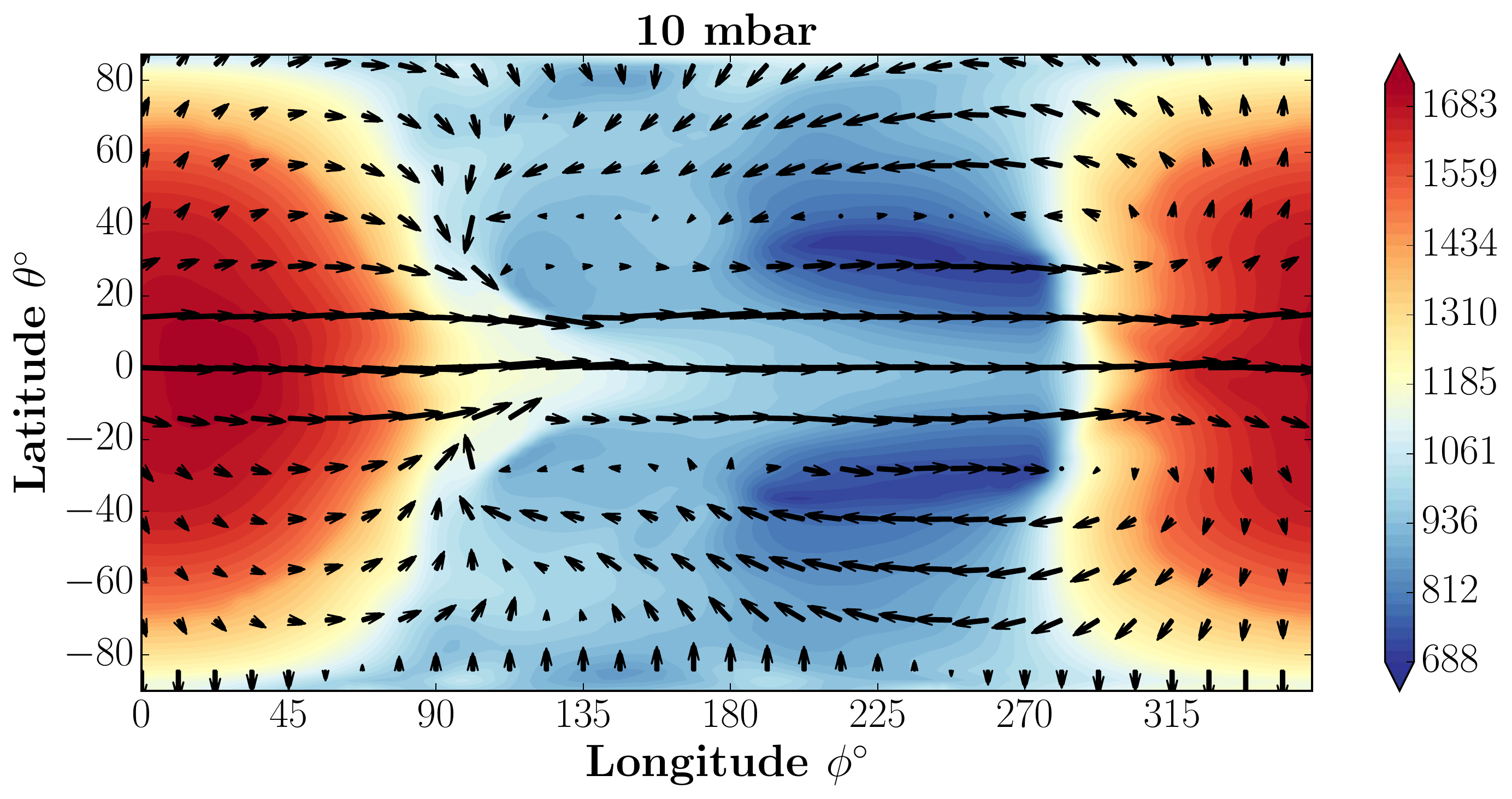}
\includegraphics[width=0.49\textwidth]{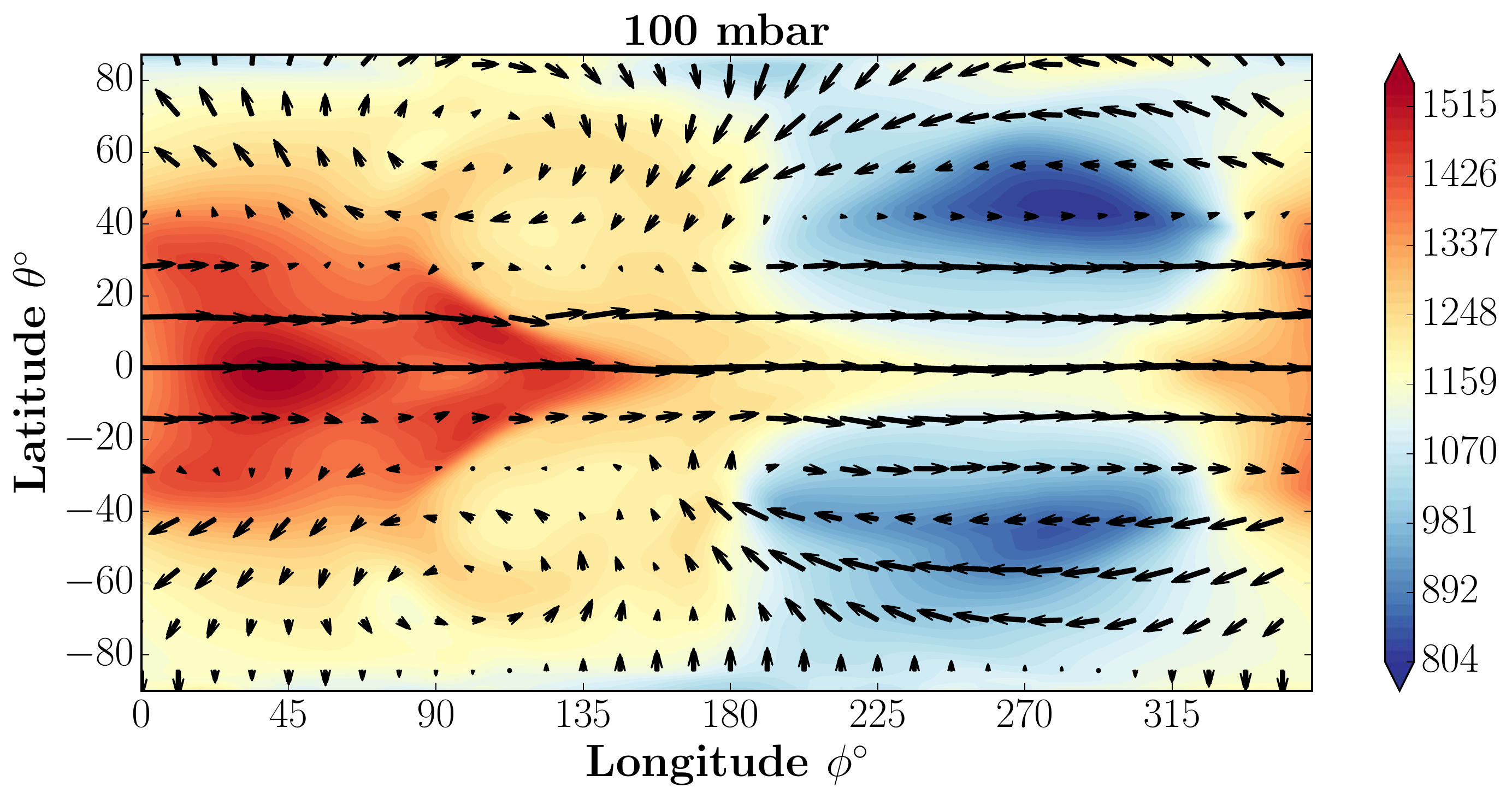}
\includegraphics[width=0.49\textwidth]{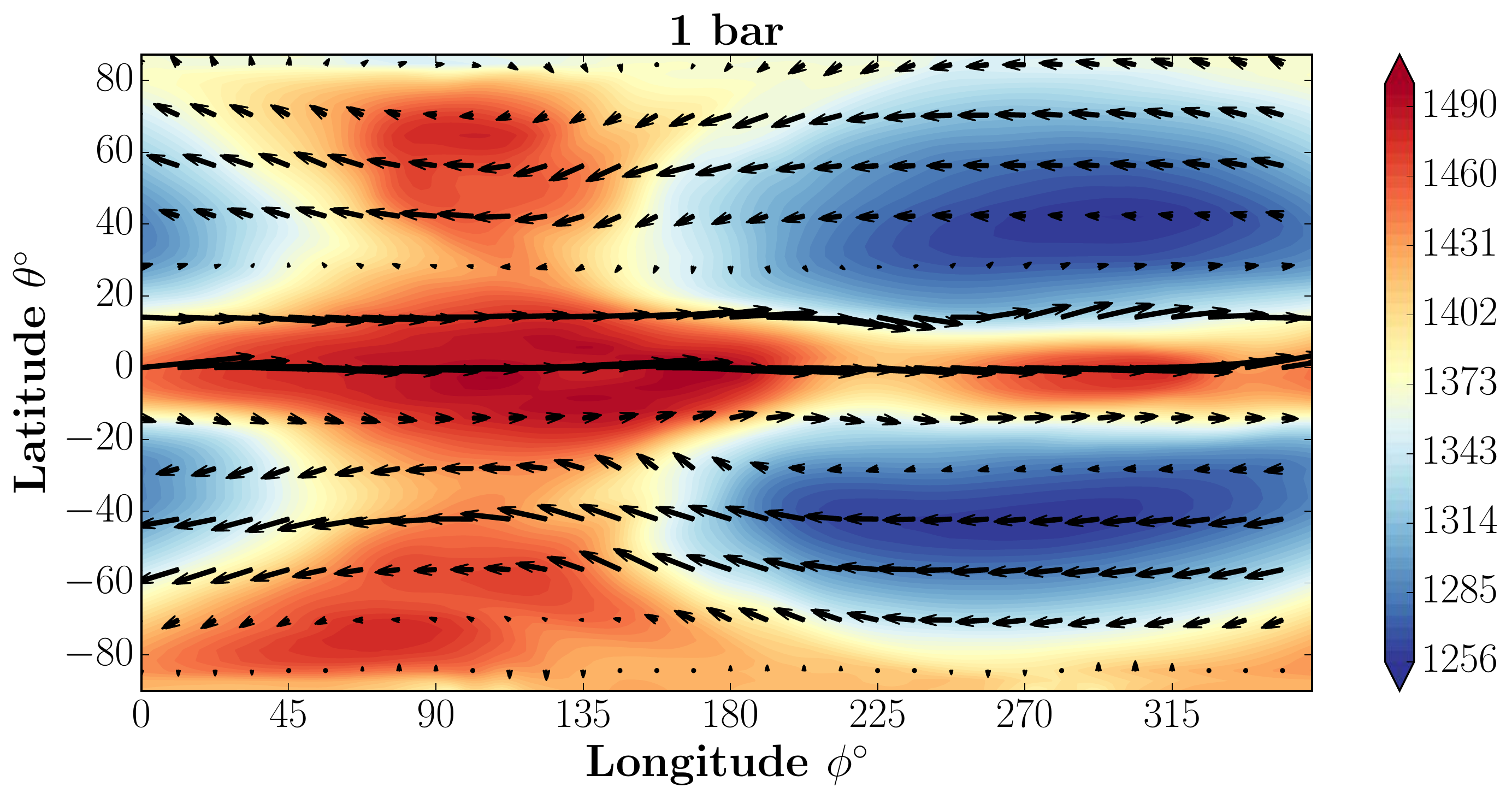}
\includegraphics[width=0.49\textwidth]{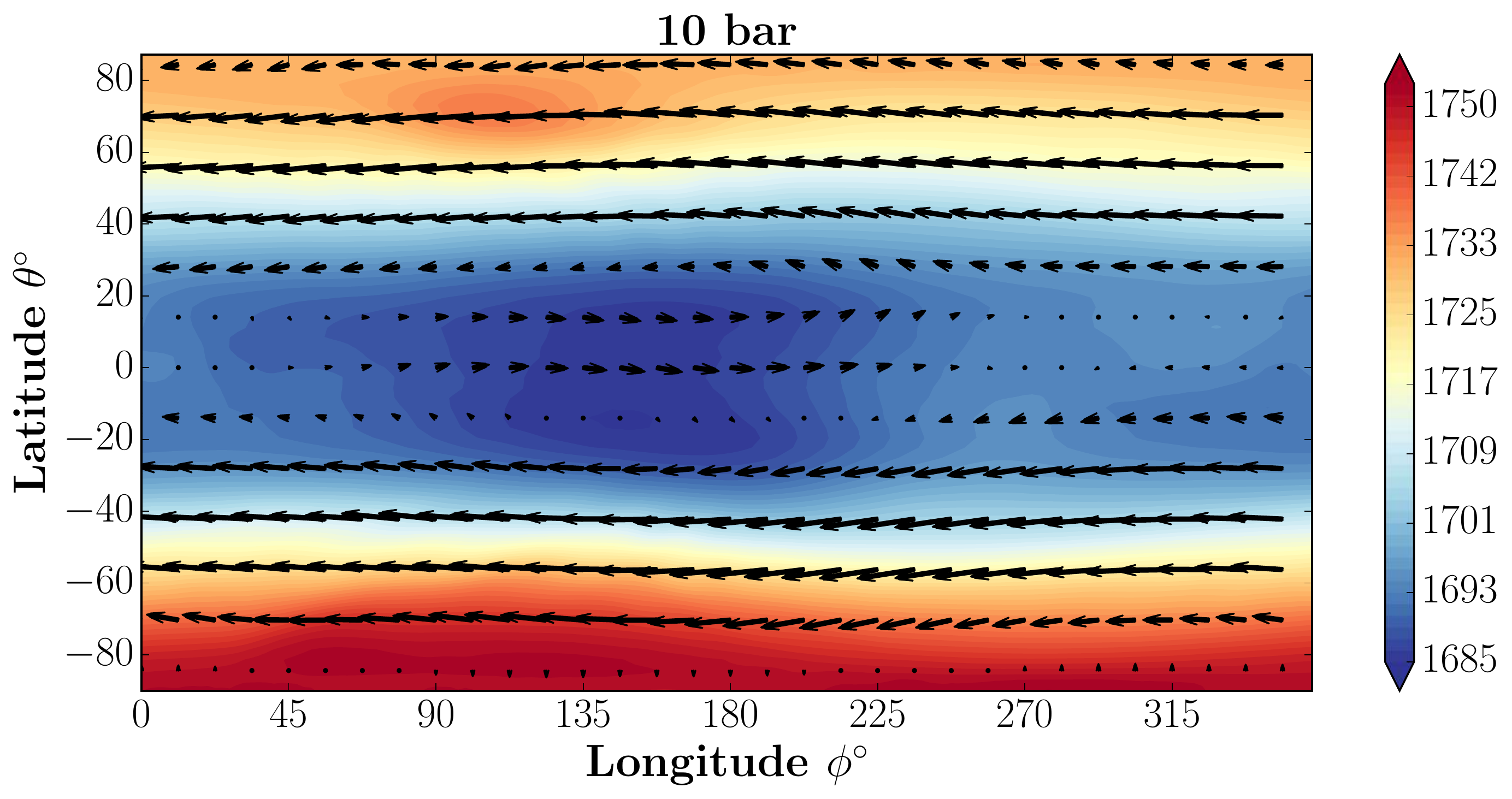}
\caption{Each panel shows T$_{\rm gas}$[K]  (colour bar) and the velocity field (given by the velocity vector $\rvert {\bf u}\lvert$ = $\sqrt{{\bf u}_{\rm h}^{2} + {\bf u}_{\rm m}^{2}}$) (arrows) at different atmospheric pressure iso-bars at p$_{\rm gas}$ =  0.1, 1, 10, 100 mbar and 1, 10 bar for different $\phi$ (longitudes) and $\theta$ (latitude).
Note: the colour bar scale is different for each plot.
The sub-stellar point is located at $\phi$ = 0\degr, $\theta$ = 0\degr.}
\label{fig:Tstruc}
\end{center}
\end{figure*}

This section presents our results regarding the combined modelling of cloud formation and radiative hydrodynamics for the giant gas planet HD 189733b.
We use snapshot results of our simulation at 65 Earth days to illustrate the global cloud formation structures.
We split our results into two broad areas, the gas phase properties such as temperature profiles (Sect. \ref{sec:tprof}) and horizontal velocity (Sect. \ref{sec:vprof});  
and the cloud properties such as cloud particle number density structures (Sect. \ref{sec:CloudNumberDensity}), mean cloud particle grain sizes (Sect. \ref{sec:GrainSizes}) and material composition (Sect.  \ref{sec:MaterialComposition}).  
Section \ref{sec:eldep} presents the depletion/replenishment of gas phase elements due to cloud formation processes and examines the hydrodynamic transport of elements from dayside to nightside.
A global summary of the results and cloud formation physics is presented in Sect. \ref{ref:ressummary}.
Section \ref{sec:CloudOpacity} presents the band by band gas and dust wavelength dependent opacity of the model in order to examine the radiative effects of cloud opacity on the temperature structure of the simulation.

\subsection{Global temperature profiles}
\label{sec:tprof}

\begin{figure*}
\begin{center}
\includegraphics[width=0.49\textwidth]{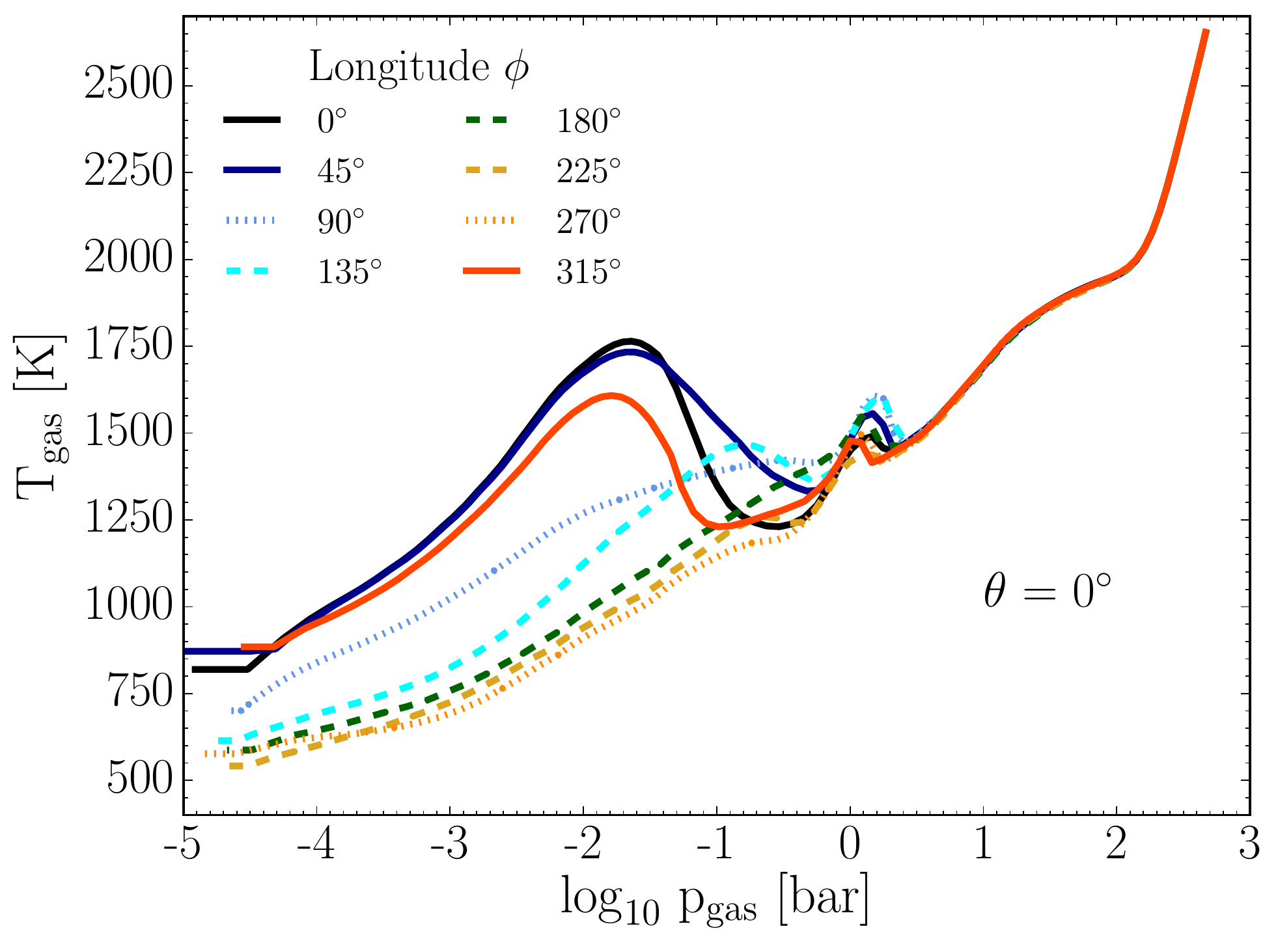}
\includegraphics[width=0.49\textwidth]{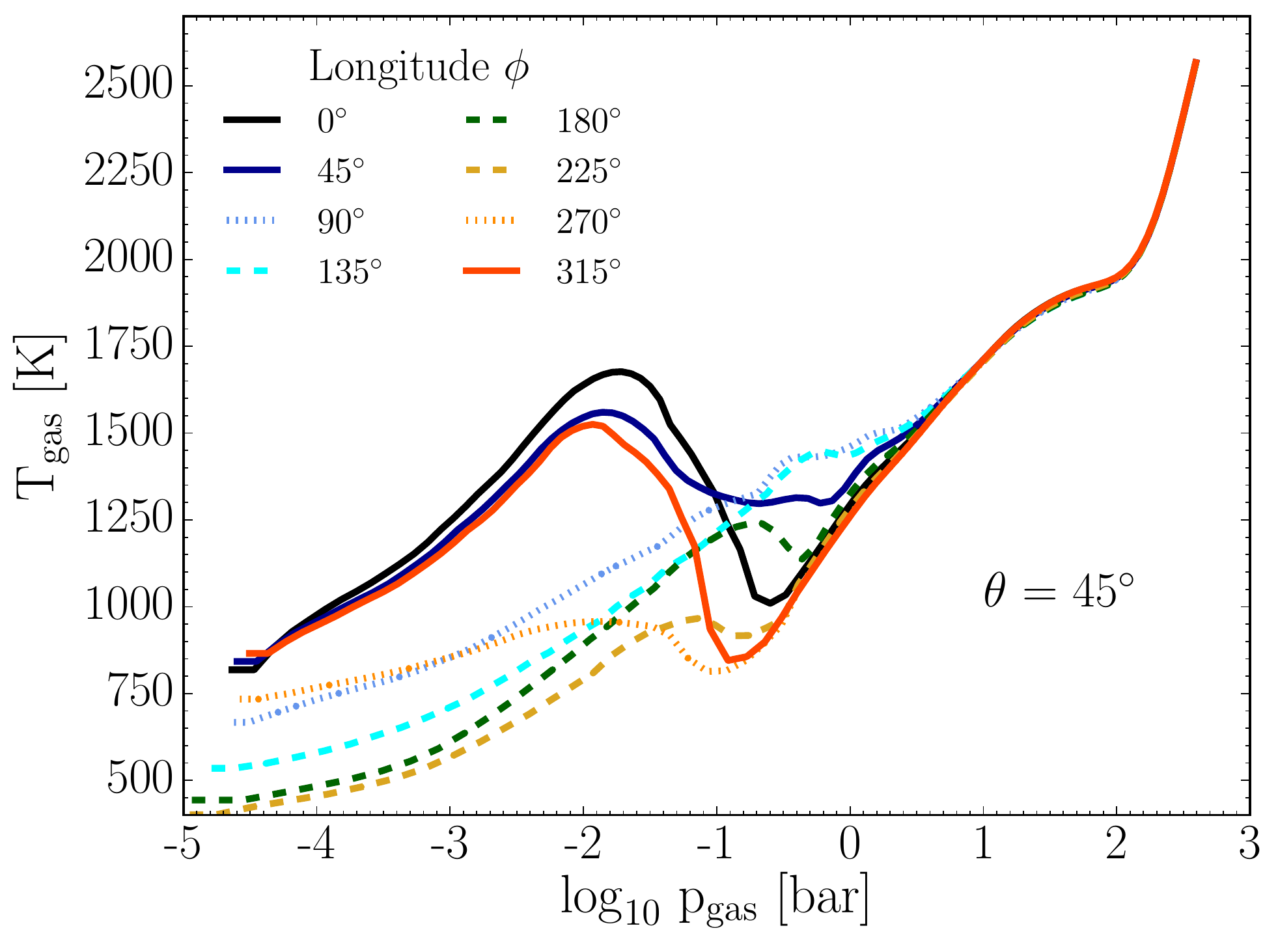}
\includegraphics[width=0.49\textwidth]{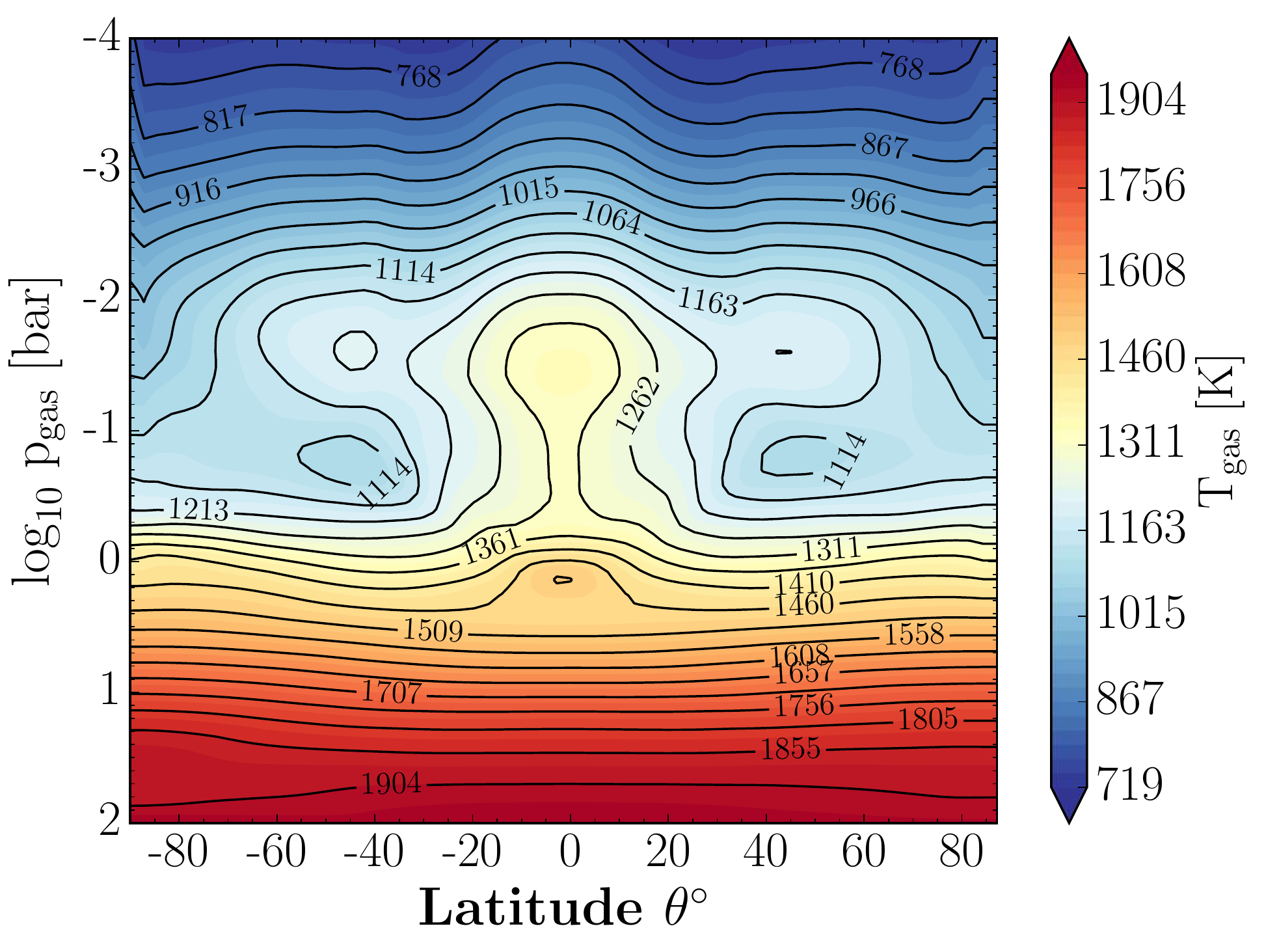}
\includegraphics[width=0.49\textwidth]{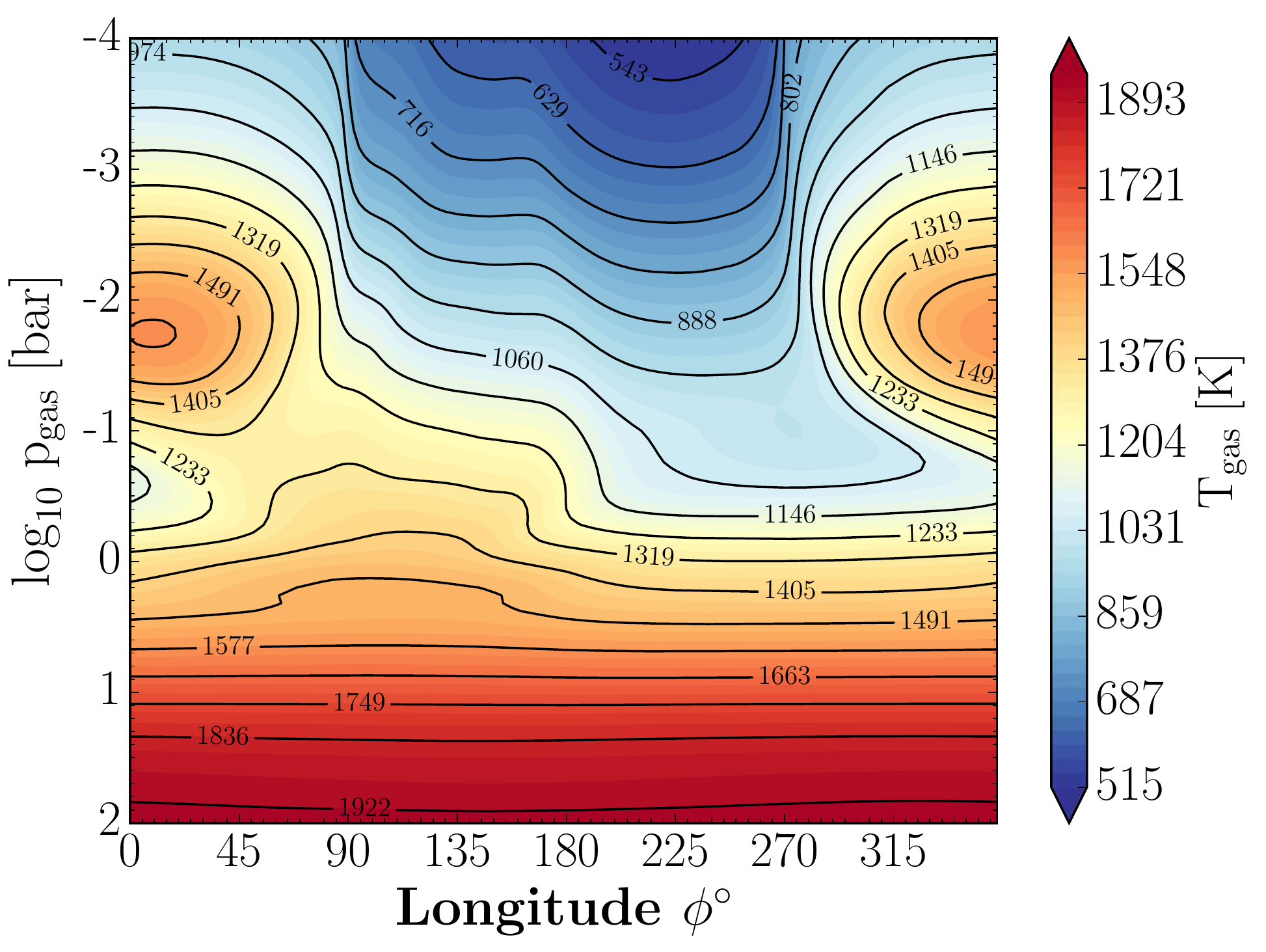}
\caption{
\textbf{Top:} 1D (T$_{\rm gas}$, p$_{\rm gas}$) trajectories at latitudes $\theta$ = 0\degr (left) , 45\degr (right). 
Dayside profiles (solid) show steep temperature inversions at $\sim$10 mbar, especially at higher latitudes.
Nightside (dashed) and terminator (dotted) profiles have smaller inversions.
\textbf{Bottom:} Zonal (left) mean T$_{\rm gas}$ [K] as a function of atmospheric pressure and meridional (right) mean T$_{\rm gas}$ [K] as a function of atmospheric pressure. 
The largest differences in latitudinal temperature contrasts occur from 10mbar to 1bar. 
The temperature is generally isothermal at atmospheric regions at pressures  $>$5 bar.}
\label{fig:Tstrucmean}
\end{center}
\end{figure*}

The local thermodynamic quantities like gas temperature and gas pressure, the local element abundances, determine the local cloud formation processes. 
The resulting cloud particles affects the local temperatures through their opacity, which in turn is coupled to the local pressure and density through the equation of state and Navier-Stokes equations. 
We therefore study the local gas temperature which will allow us to develop a global picture of the atmospheric temperature of hot Jupiters like HD 189733b under the influence of cloud formation.

A variation between dayside and nightside is present, with largest gradients in temperatures typically occurring near the terminator regions, most apparent at 0.1, 1 and 10 mbar (Fig. \ref{fig:Tstruc}).
The larger hydrodynamic velocities (super-sonic jet streams) at equatorial regions advects energy density Eastward, resulting in a longitude offset of the temperature maximum by $\phi$ $\sim$ 20-40\degr\ East compared to the sub-stellar point $\phi$ = 0\degr\ where the planet receives maximum irradiation from the host star.
This is most apparent at 100 mbar and 1 bar in Fig. \ref{fig:Tstruc}.
Differences in temperature ($>$100 K) between equatorial regions and mid-high latitudes are present in upper atmospheric regions.
At depths $>$5 bar the local gas temperature starts to become more uniform in longitude and latitude.
Hottest upper atmosphere regions (T$_{\rm gas}$ $>$ 1500 K) occur on the dayside at the peak of the temperature inversions ($\sim$ 10mbar).
The coolest regions occur at nightside mid-latitudes with temperatures of $\sim$400 K.
These low temperature regions correspond to the large scale, nightside vortex regions where global hydrodynamical motions do not efficiently transport dayside hot gas towards. 

Figure \ref{fig:Tstrucmean} displays the 1D (T$_{\rm gas}$, p$_{\rm gas}$) trajectories and shows that the atmosphere contains steep dayside temperature inversions at $\sim$10 mbar. 
A second, smaller temperature inversion occurs at higher gas pressures $\sim$1 bar for all longitudes at the equator ($\theta$ = 0\degr; Fig. \ref{fig:Tstrucmean}, top left). 
This temperature bump of 100$\,\ldots\,$200K bump initially develops on the dayside due to a back-warming effect of the larger cloud opacity at $\sim$1 bar (see also Sect. \ref{sec:CloudOpacity}).
Emission of radiation from hot gas at lower pressure is absorbed by the cloud layer at $\sim$1 bar, resulting in localised heating of the gas. 
The hot gas is then transported to the nightside by the horizontal winds at this pressure (Fig. \ref{fig:Vstruc}), resulting in a bump for all equatorial profiles at $\sim$1 bar. 

Figure \ref{fig:Tstrucmean} also shows the zonal and meridional mean gas temperature T$_{\rm gas}$ [K] as a function of depth.
Zonal mean temperature show how the global temperature structure is changing with latitude and depth.
This is useful in order to show global differences between equatorial and mid-high latitude regions for atmospheric properties.
Horizontal contours indicate a more uniform variation in temperature in latitude, while vertical contours indicate a greater variation with latitude.
Meridional mean temperatures show how the global temperature structure is changing with longitude and depth.
This is useful in order to show atmospheric differences between dayside and nightside regions.
Horizontal contours indicate a more uniform variation in temperature in longitude, while vertical contours indicate a grater variation with longitude.
The highest temperature regions at the upper atmosphere are concentrated at the equator, while a  larger ($>$100 K) difference occurs between equatorial and higher latitude regions.
From the meridional mean plot (Fig. \ref{fig:Tstrucmean}, bottom right), a stream of hotter gas is present past the $\phi$ = 90\degr\ terminator at 100 mbar due to hydrodynamic flows advecting energy to the nightside and into deeper regions of higher pressure.
The temperature becomes more uniform in longitude and latitude at deeper atmospheric regions $>$5 bar.

\subsection{Atmospheric velocity field}
\label{sec:vprof}

\begin{figure*}
\begin{center}
\includegraphics[width=0.49\textwidth]{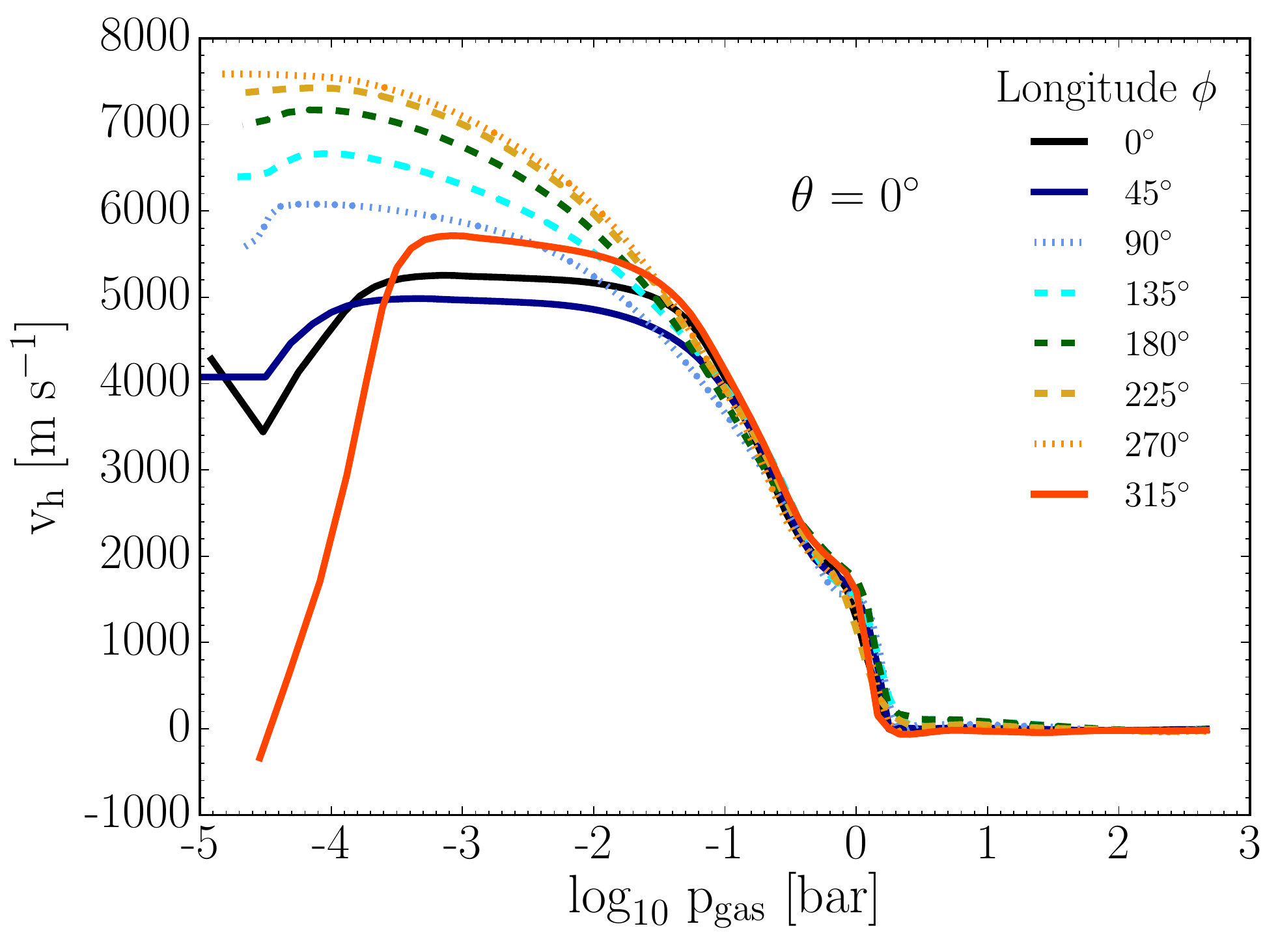}
\includegraphics[width=0.49\textwidth]{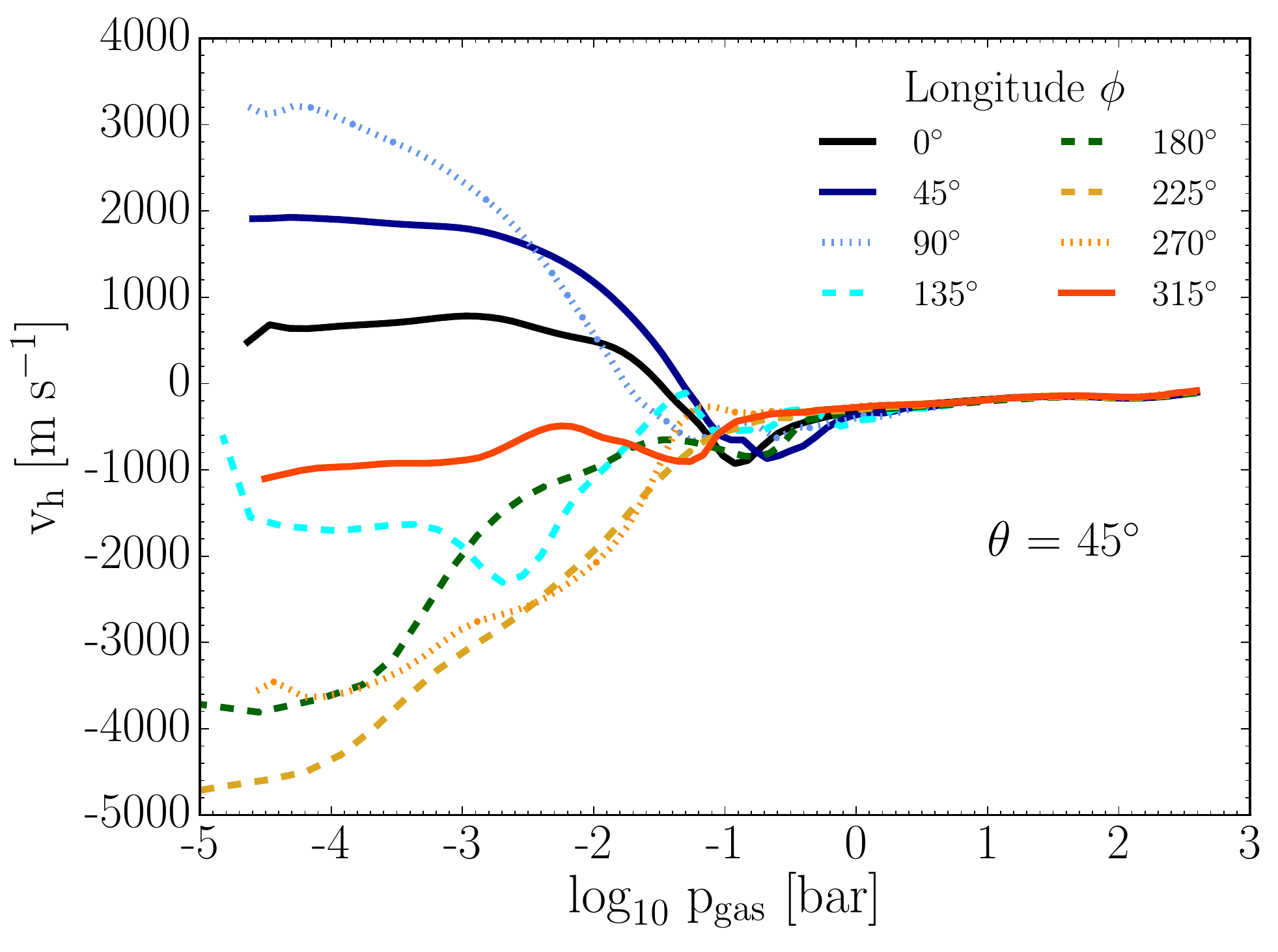}
\includegraphics[width=0.49\textwidth]{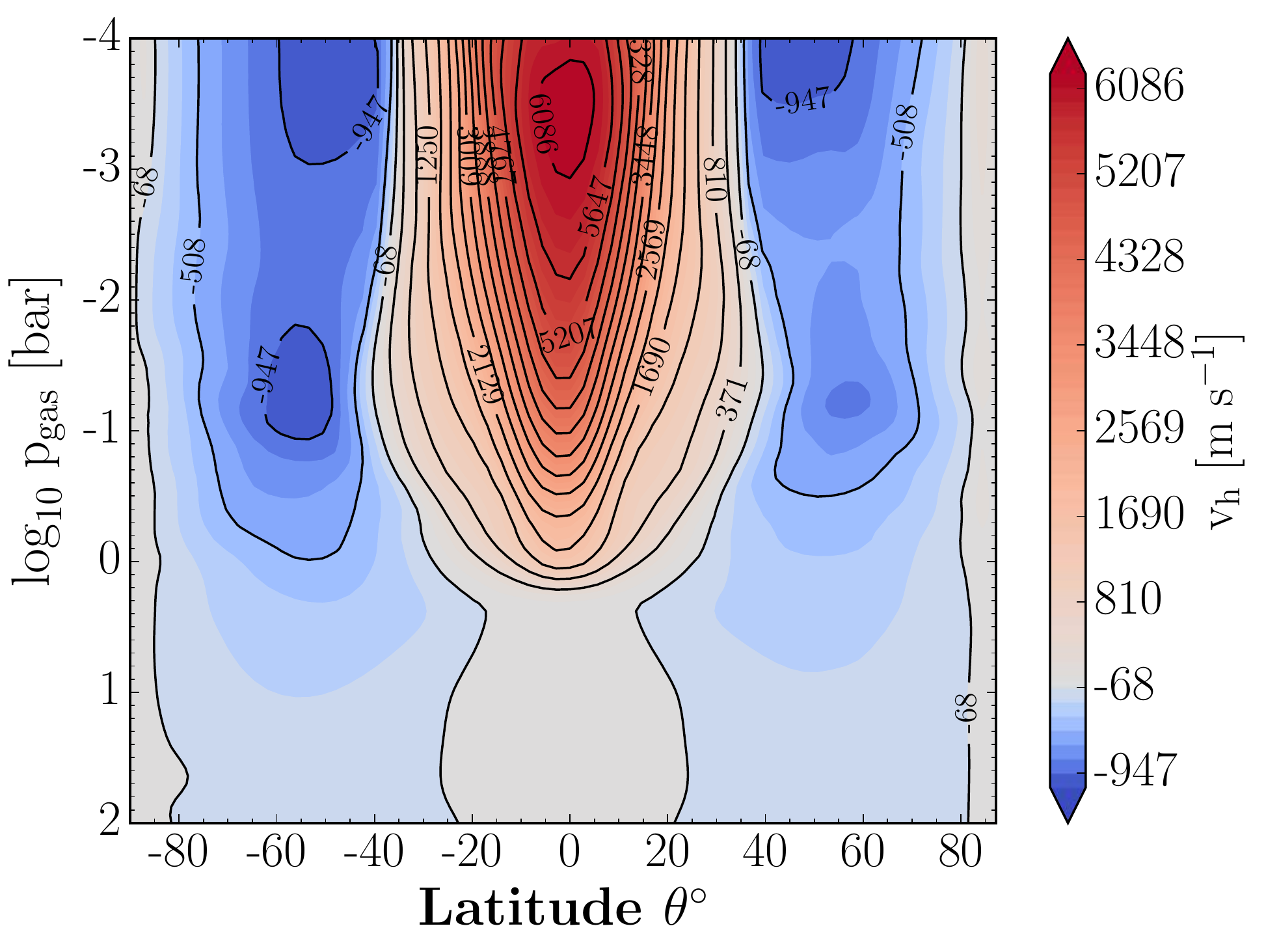}
\includegraphics[width=0.49\textwidth]{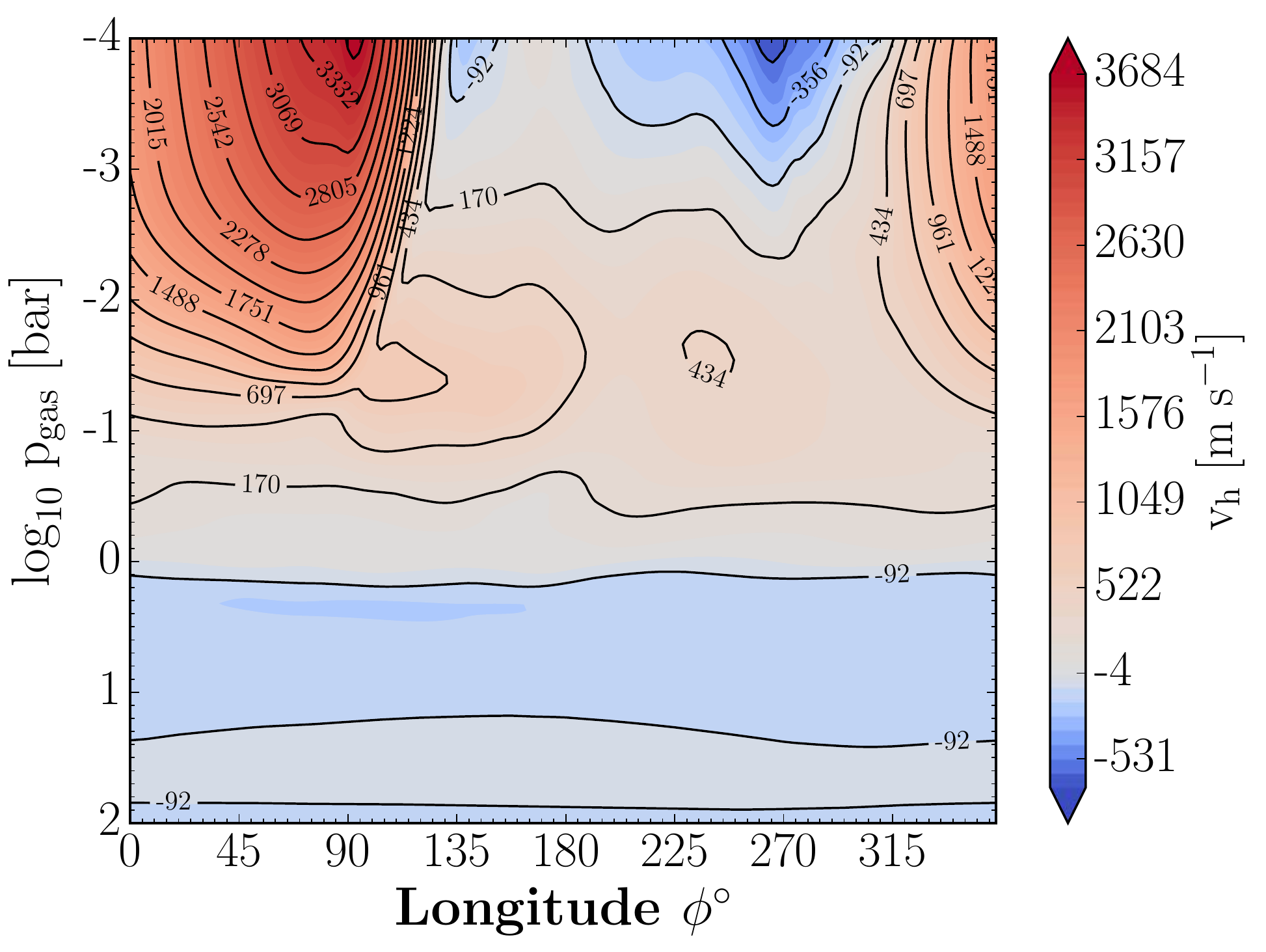}
\caption{\textbf{Top:} 1D zonal/horizontal velocity v$_{\rm h}$ [m s$^{-1}$] trajectories at latitudes $\theta$ = 0\degr (left) , 45\degr (right).
Equatorial regions show positive super-sonic flow confined to $\theta$ = $\pm$30\degr\ latitudes, with maximum velocities greater than 7000 m s$^{-1}$ at the upper nightside atmosphere.
Negative direction velocities occur at higher latitudes (|$\theta$| > 40\degr).
\textbf{Bottom:} Zonal (left) and meridional (right) mean horizontal velocity v$_{\rm h}$ [m s$^{-1}$] as a function of atmospheric gas pressure. 
Note: the colour scale bar has been normalised to 0 m s$^{-1}$.
The strongest zonal velocities occur at the equator.
Negative flows can be found at latitudes of $\theta$ $\pm$ 40$\ldots$80\degr.}
\label{fig:Vstruc}
\end{center}
\end{figure*}

3D RHD simulations provide information about the local and global hydrodynamical behaviour.
The dominant hydrodynamical feature of GCM/RHD models of hot Jupiter atmospheres is the formation of an equatorial jet.
The super-rotating, circumplanetary jet \citep{Tsai2014} efficiently advects energy density from day to night near the equatorial regions. 
This jet forms from the planetary rotation (Rossby waves) coupled with eddies which pump positive angular momentum toward the equator \citep{Showman2011a}.
This section studies the local and global velocity profiles of an atmosphere where cloud formation takes place.
1D profiles of zonal/horizontal velocity at the equator and mid-latitudes are presented in Fig. \ref{fig:Vstruc}. 
These show that an upper atmosphere super-sonic jet of velocity $>$ 4000 m s$^{-1}$ at equatorial regions.
A significant slow down of horizontal velocity at $\phi$$\sim$315\degr\ longitude occurs West of the sub-stellar ($\phi$ = 0\degr) point.
The maximum zonal velocities of $>$ 7000 m s$^{-1}$ occur on the nightside near the night-day $\phi$ = 315\degr\ terminator.
At mid-latitudes ($\theta$ $\sim$ 45\degr), nightside ($\phi$ = 135$\,\ldots\,$270\degr) regions contain super-sonic counter rotating flows with a velocity of $<$ $-$2000 m s$^{-1}$.
Figure \ref{fig:Vstruc} also shows the zonal mean gas velocity v$_{\rm h}$ [m s$^{-1}$] as at different latitudes and atmospheric pressures.
This shows that there is supersonic jet flow at the equator in the West-East direction.
Counter rotating flows occur at mid-latitudes with lower horizontal velocities than equatorial regions.
Below $\sim$1 bar, the horizontal motions are slower and longitude, latitude uniform until reaching the inner boundary of our computational domain.
The overall structural features remain similar to \citet{Dobbs-Dixon2013}.

\section{Dynamic mineral clouds in HD 189733b}

\begin{figure*}
\begin{center}
\includegraphics[width=0.49\textwidth]{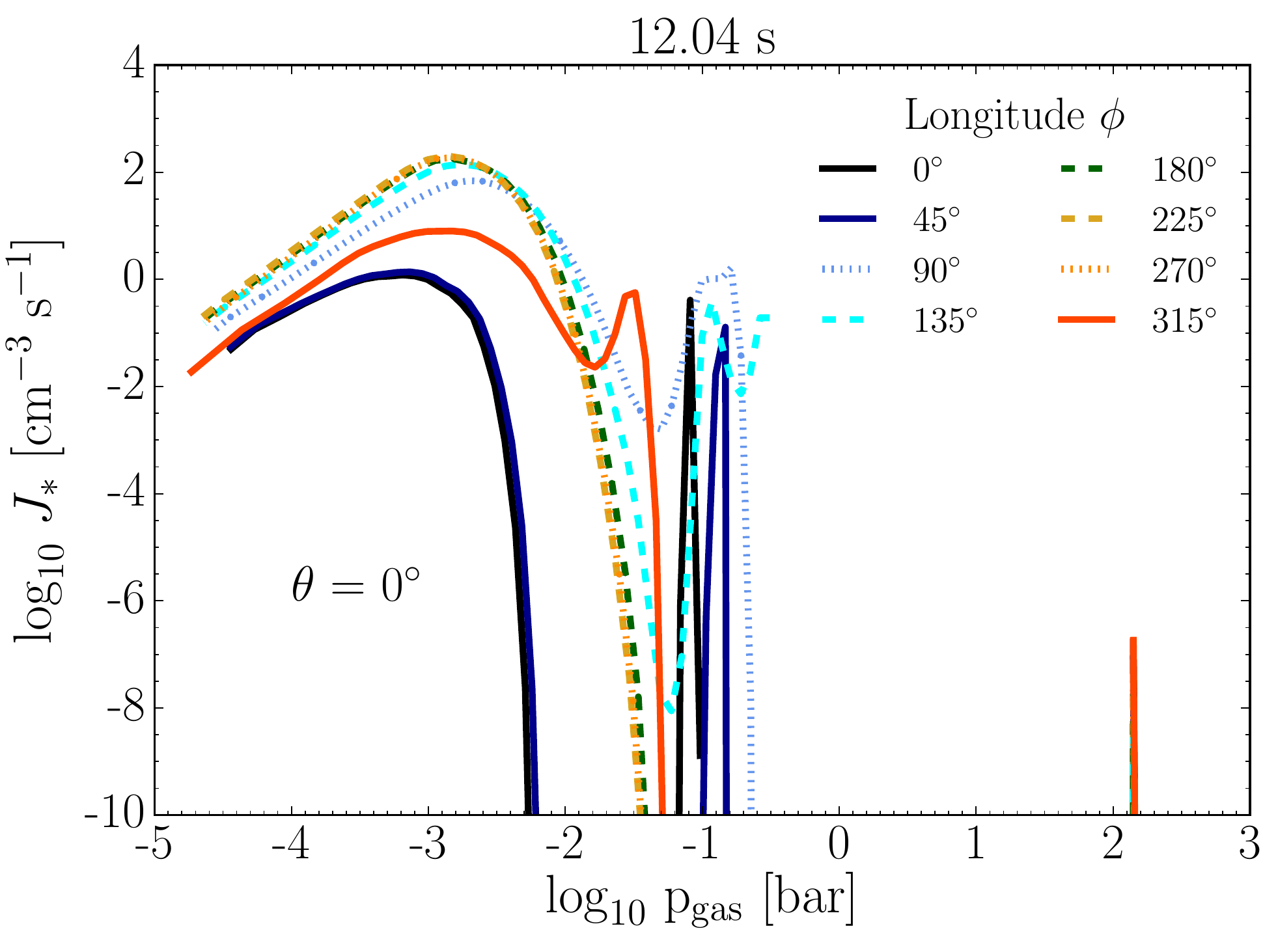}
\includegraphics[width=0.49\textwidth]{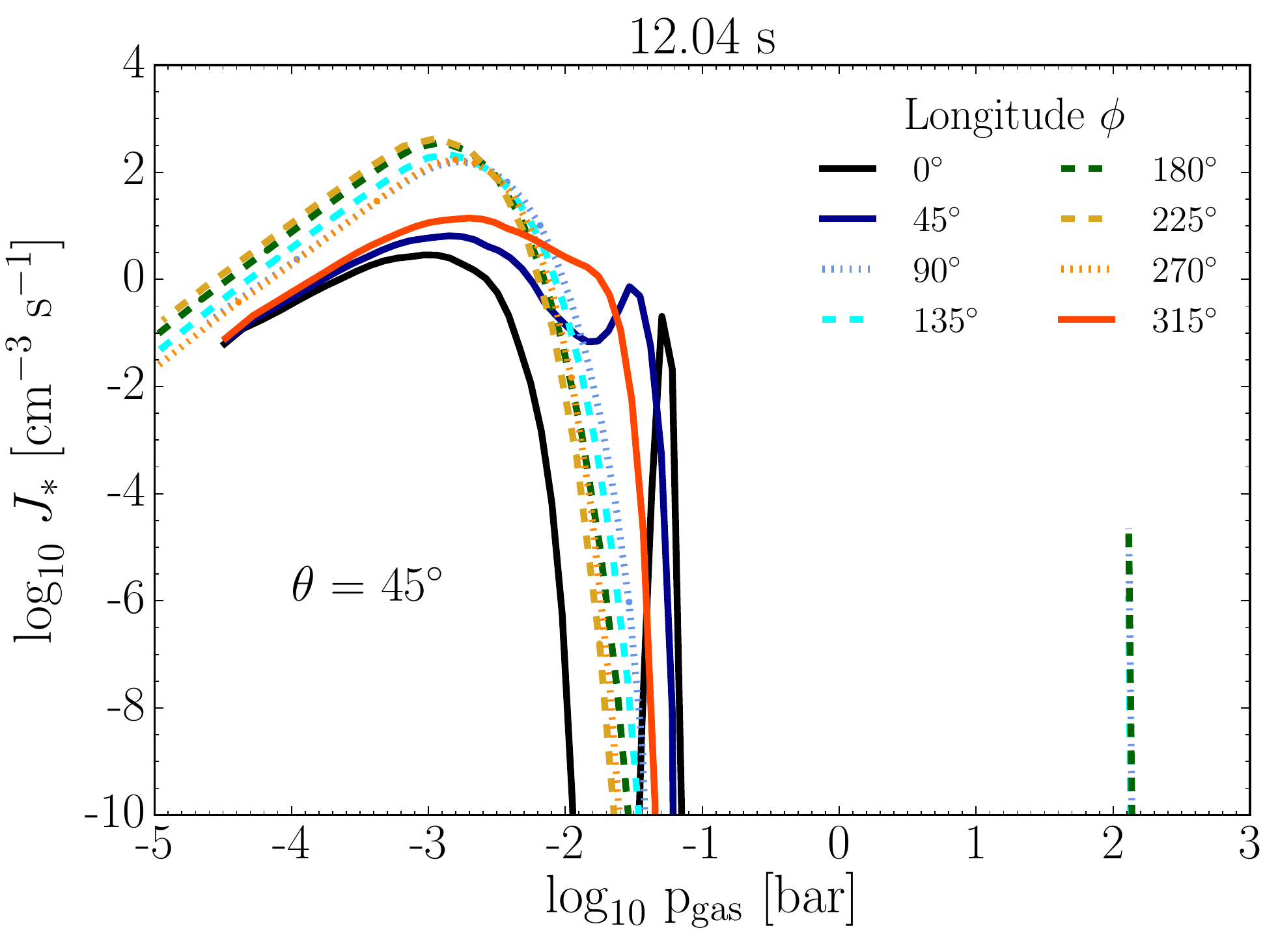}
\caption{\textbf{Top:} 1D nucleation rate $\log_{10}$ $J_{*}$ [cm$^{-3}$ s$^{-1}$] trajectories $\sim$ 12 seconds into the simulation for latitudes $\theta$ = 0\degr (left) , 45\degr (right).
The initial, most efficient nucleation regions occur at p$_{\rm gas}$ $<$ 100 mbar for all atmospheric profiles. 
Dayside equatorial profiles at $\phi$ = 0\degr, 45\degr\ have no nucleation occurring from $\sim$ 10-100 mbar, where the gas temperature is too high for the nucleation process.
The greatest magnitude of nucleation of seed particles is at nightside, mid-high latitude regions at $\sim$ 1 mbar.
}
\label{fig:zonal_J}
\end{center}
\end{figure*}

Giant gas planets like HD 189733b form clouds in their atmospheres from a chemically very rich gas phase. 
\citet{Lee2015b} have shown that the local thermodynamic conditions suggest that clouds form throughout the whole atmosphere of HD 189733b, although this result is limited by the non-global, 1D mixing approach.
A similar conclusions was reach for HD 209458b in a comparison study of both planets \citep{Helling2016}.
While \citet{Lee2015b} and \citet{Helling2016} present their results for stationary cloud structures, we now discuss the formation of clouds in a dynamic, time-dependent atmosphere in combination with the 3D atmospheric temperature and velocity fields. 
The following section shows how cloud properties like number density of cloud particles (Sect. \ref{sec:CloudNumberDensity}), cloud particle sizes (Sect. \ref{sec:GrainSizes}) and the material composition (Sect. \ref{sec:MaterialComposition}) develop in and form a dynamic cloud structure in an atmospheres with hydrodynamic jets and temperature inversions.

\subsection{Seed formation and cloud particle density}
\label{sec:CloudNumberDensity}

\begin{figure*}
\begin{center}
\includegraphics[width=0.49\textwidth]{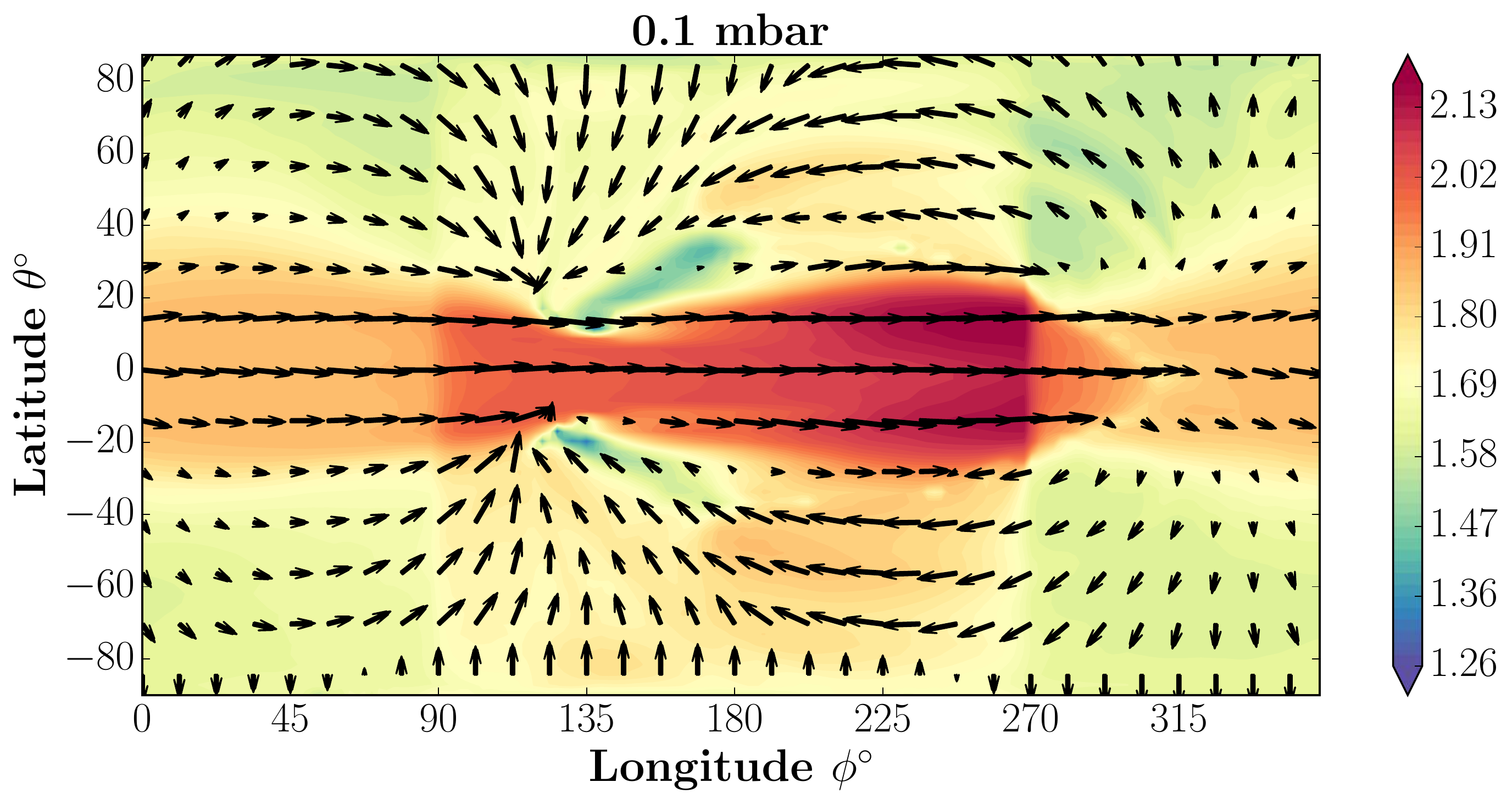}
\includegraphics[width=0.49\textwidth]{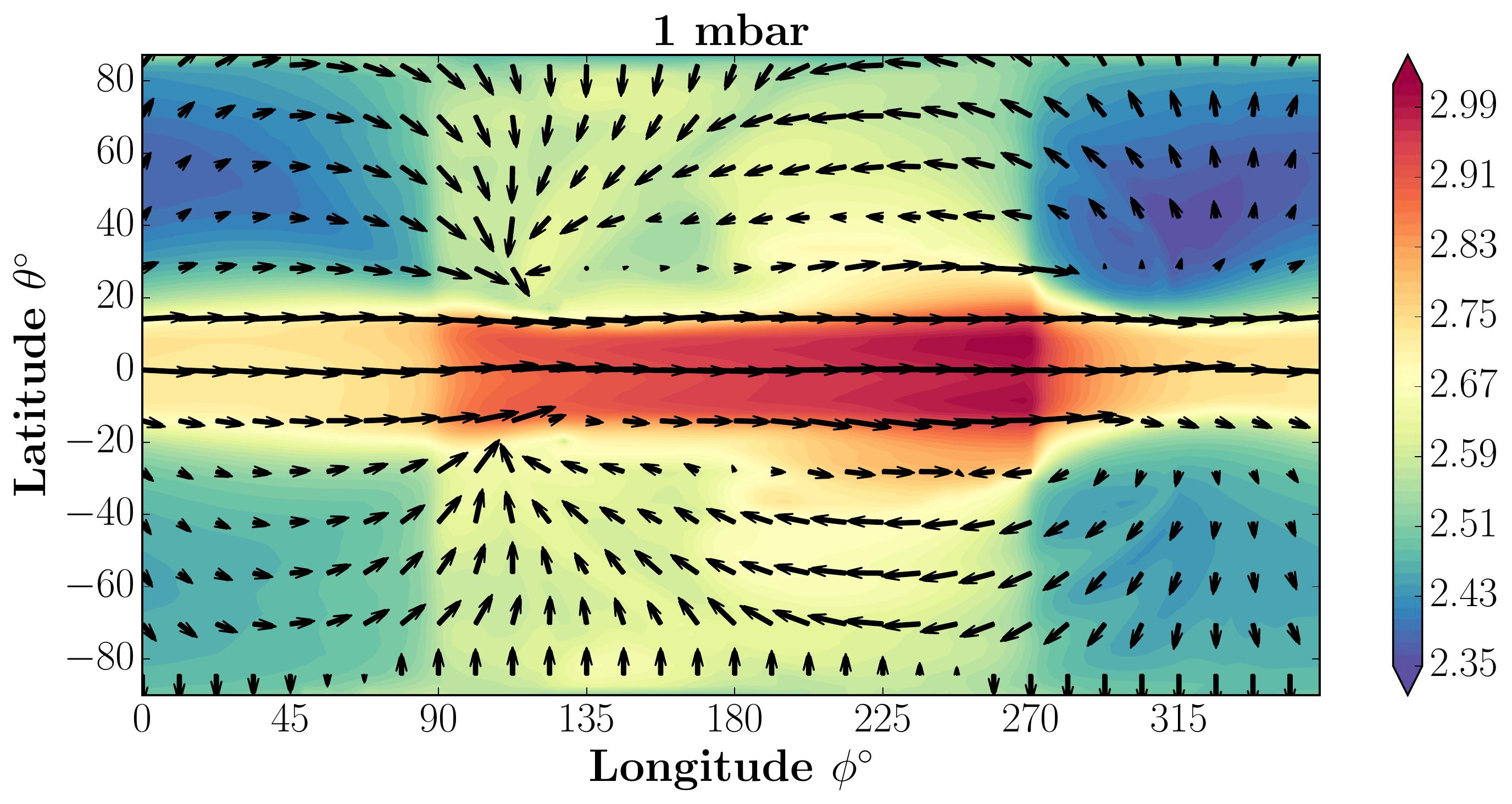}
\includegraphics[width=0.49\textwidth]{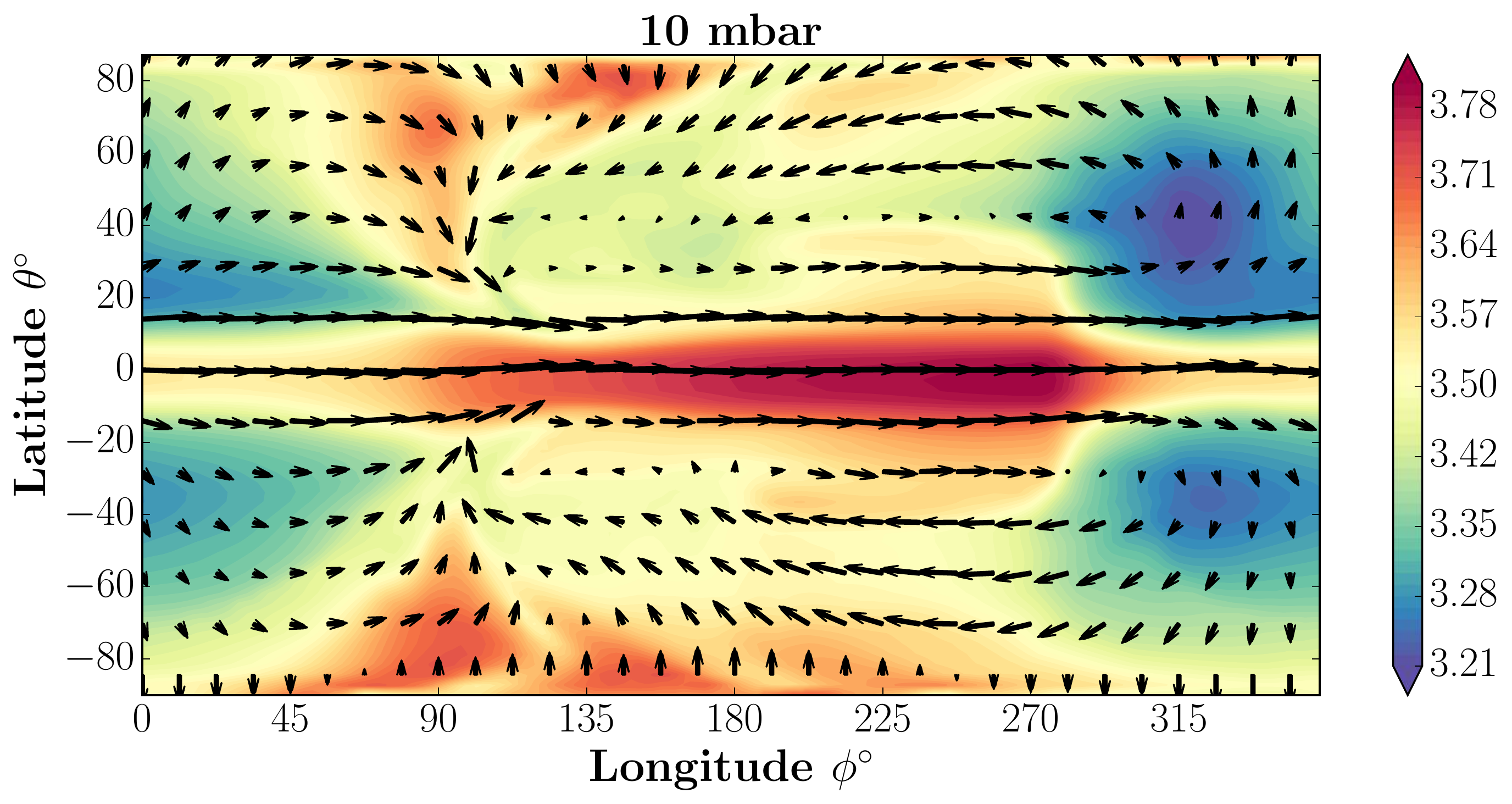}
\includegraphics[width=0.49\textwidth]{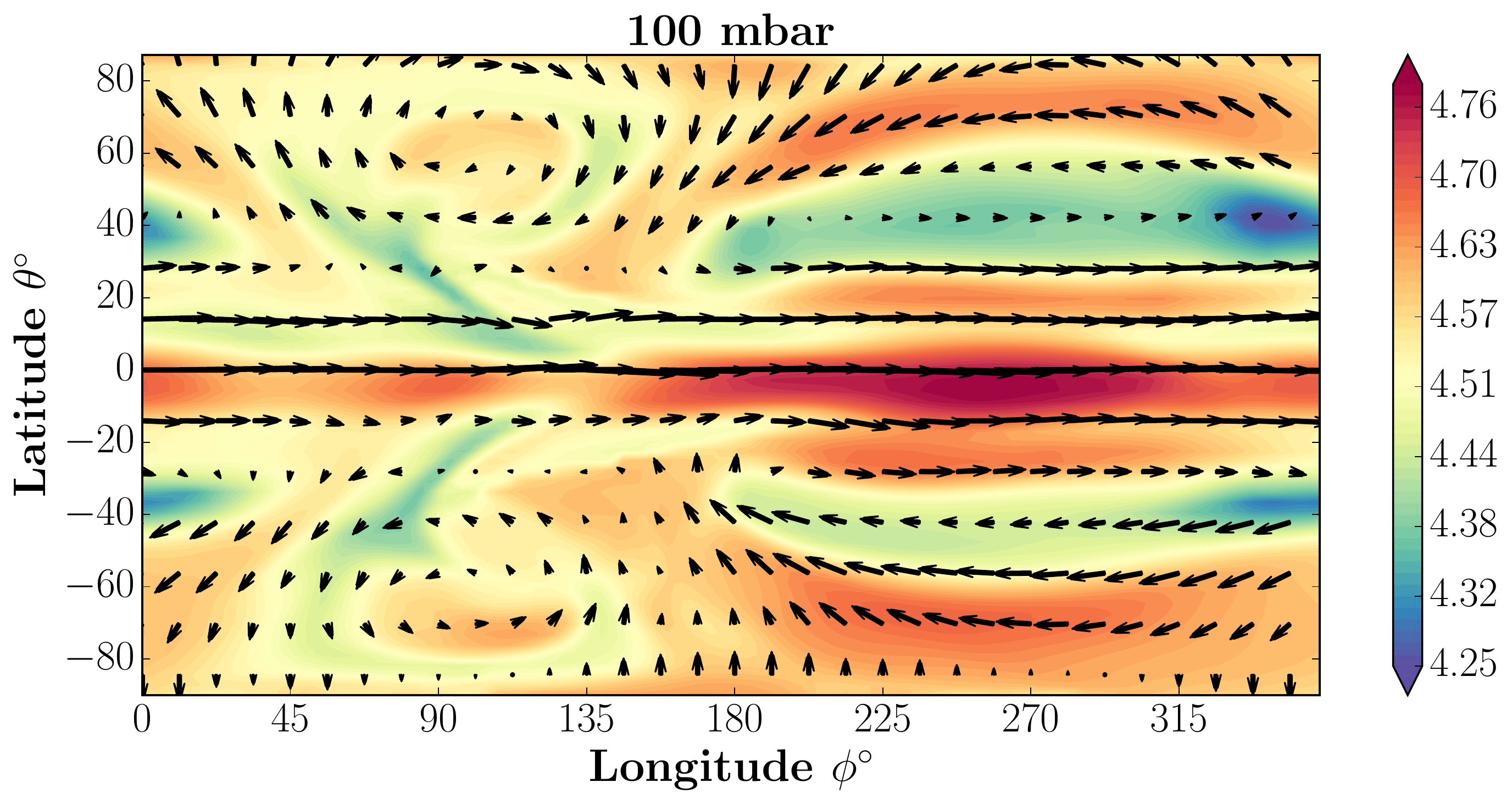}
\includegraphics[width=0.49\textwidth]{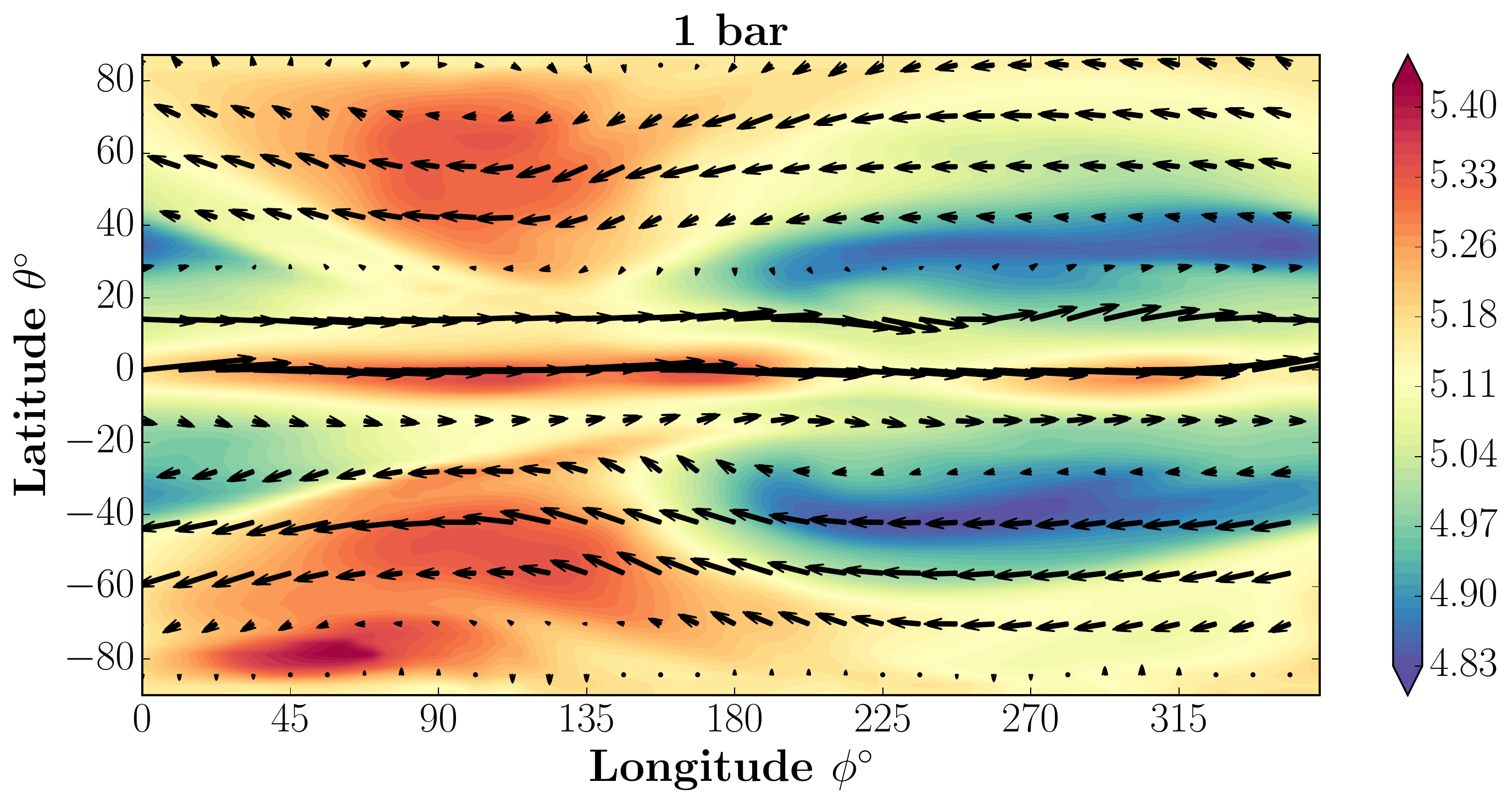}
\includegraphics[width=0.49\textwidth]{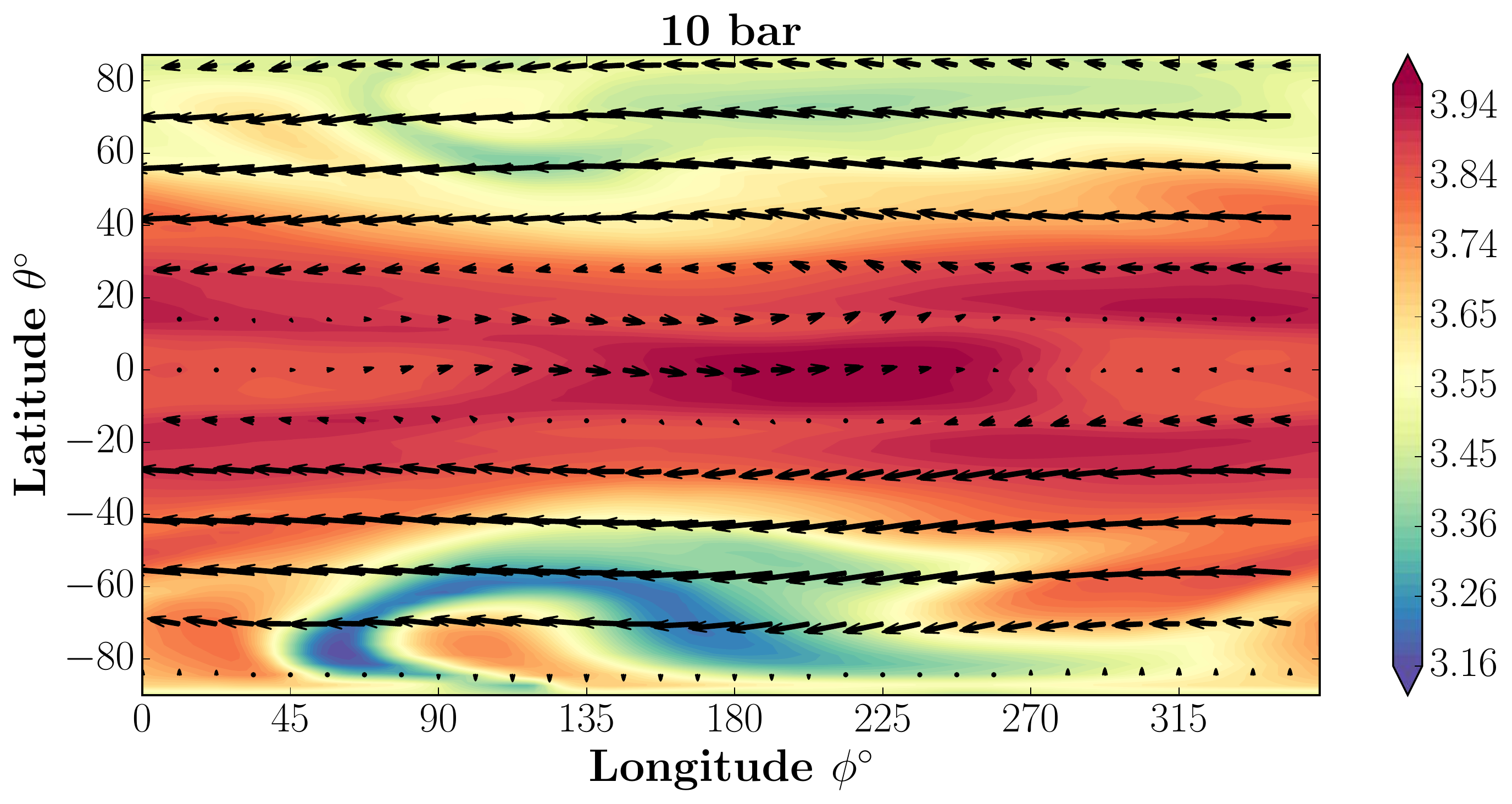}
\caption{Cloud particle number density of grains $\log_{10}$ n$_{\rm d}$ [cm$^{-3}$] (colour bar) and velocity field ($\rvert {\bf u}\lvert$ = $\sqrt{{\bf u}_{\rm h}^{2} + {\bf u}_{\rm m}^{2}}$) at 0.1, 1, 10, 100 mbar and 1, 10 bar for different $\phi$ (longitudes) and $\theta$ (latitude). 
Note: the colour bar scale is different for each plot.
The sub-stellar point is located at $\phi$ = 0\degr, $\theta$ = 0\degr.
Grains are typically more concentrated at equatorial nightside regions.
The number density increases until reaching a maximum near 1 bar, which then gradually falls until the lower computational boundary at $\sim$100 bar. 
The grains follow the flow patterns in the upper atmosphere, showing a preference to transport cloud particles to nightside equatorial regions.
Regions deeper than $\sim$1 bar show a more uniform distribution of cloud particles in latitude and longitude.}
\label{fig:ndstruc}
\end{center}
\end{figure*}

\begin{figure*}
\begin{center}
\includegraphics[width=0.49\textwidth]{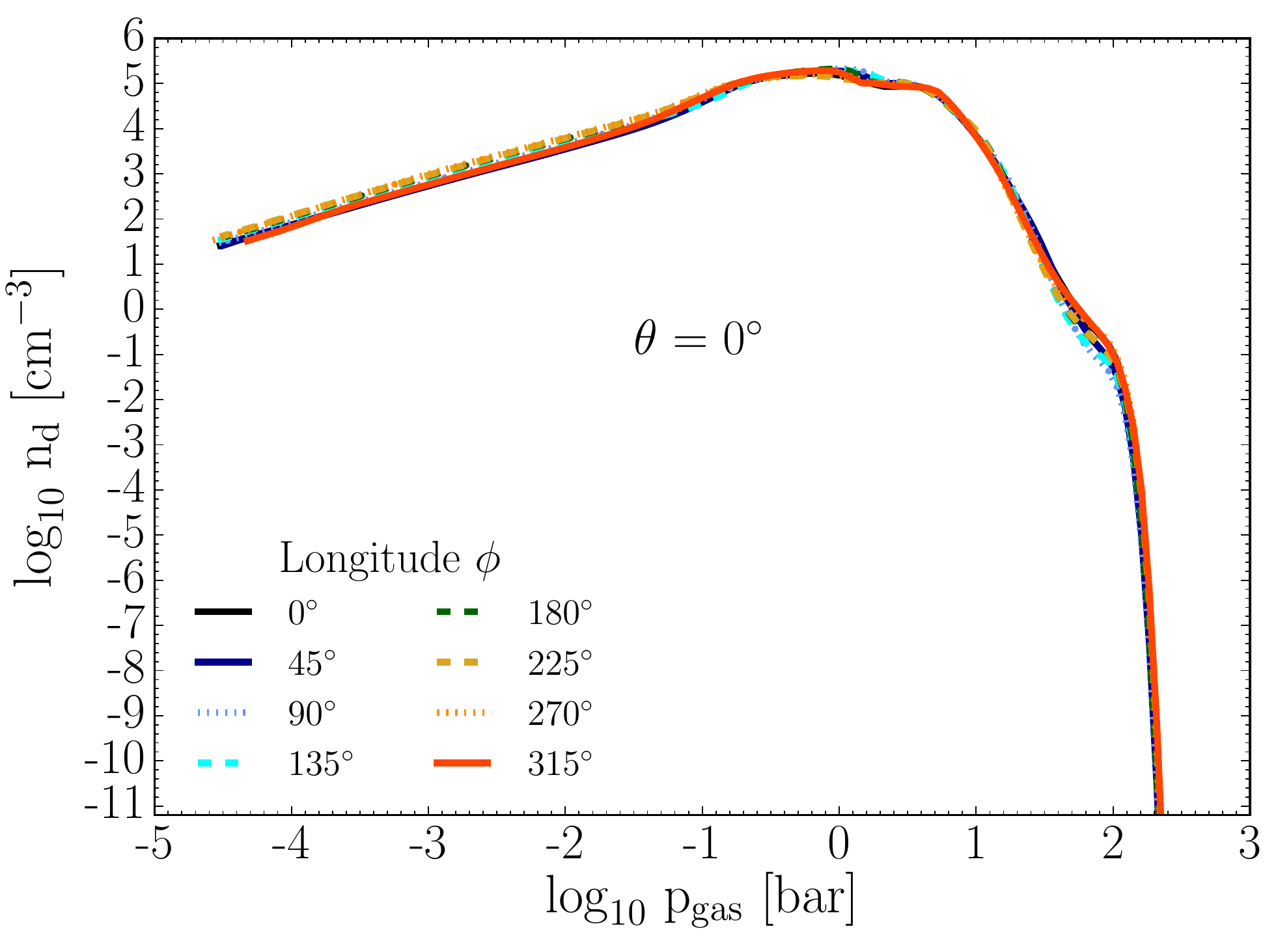}
\includegraphics[width=0.49\textwidth]{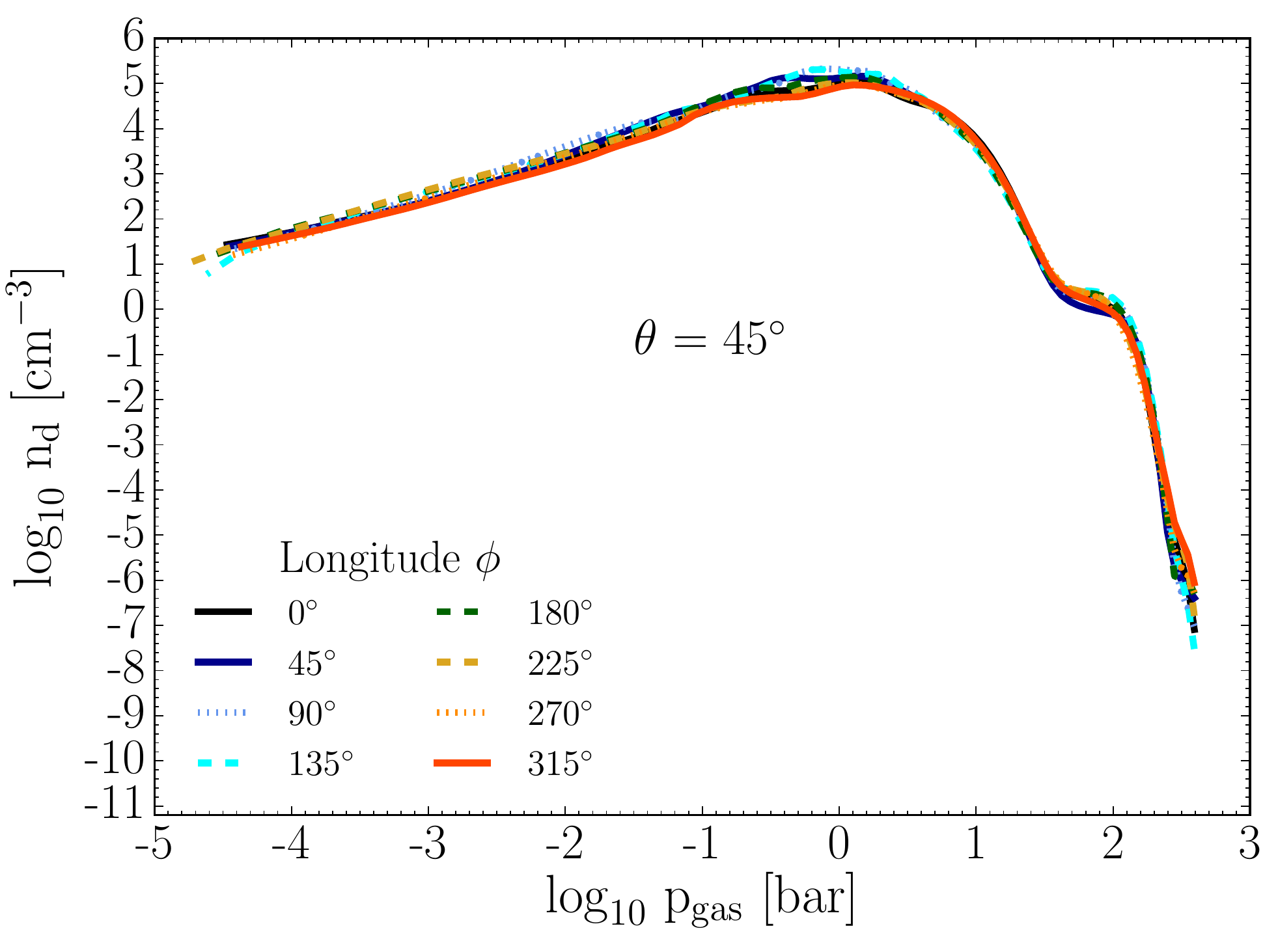}
\includegraphics[width=0.49\textwidth]{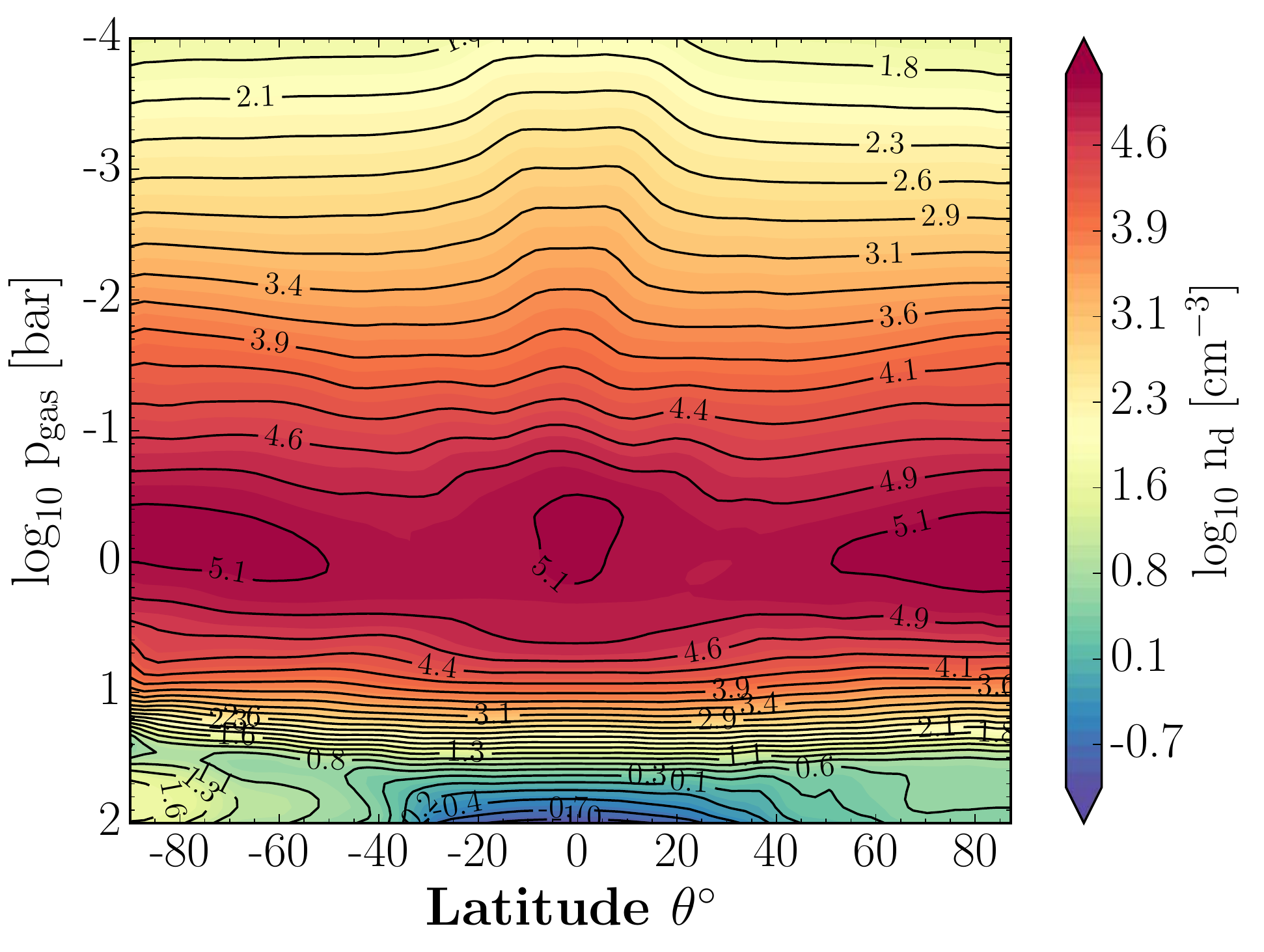}
\includegraphics[width=0.49\textwidth]{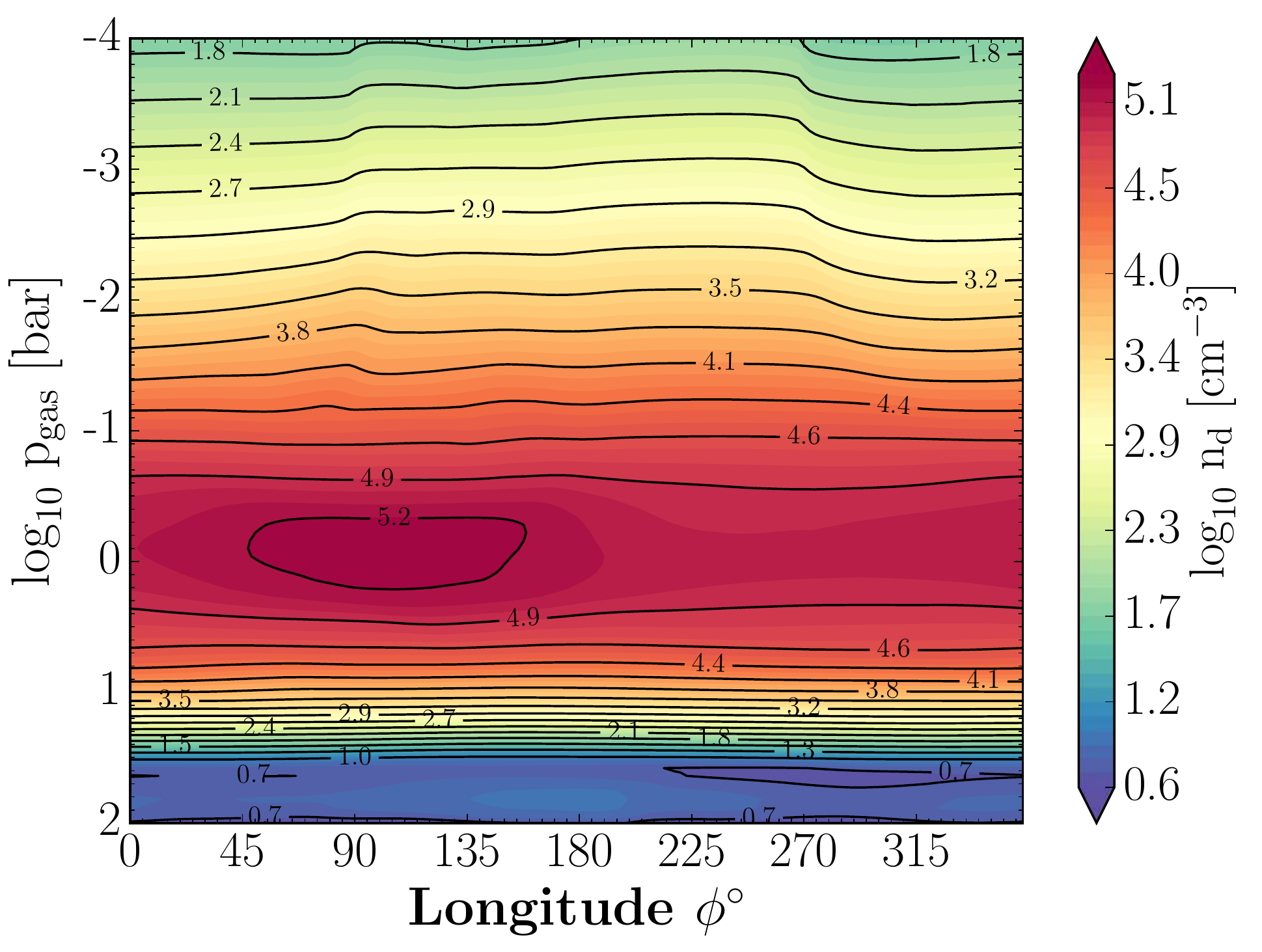}
\caption{\textbf{Top:} 1D cloud particle number density $\log_{10}$ n$_{\rm d}$ [cm$^{-3}$] trajectories at latitudes $\theta$ = 0\degr (left) , 45\degr (right).
The density structure is similar across the longitude and latitude range. 
Density rises to a maximum of $\sim$10$^{5}$ cm$^{-3}$ at $\sim$1 bar which contains the thickest and most opaque cloud regions.
\textbf{Bottom:} Zonal mean (left) and Meridional mean (right) of the number density of grains $\log_{10}$ n$_{\rm d}$ [cm$^{-3}$]. Note: the colour bar scale is different for each plot.
The most cloud dense region is from $\sim$100 mbar - 10 bar which is uniform across the globe. 
The thinest cloud layers are found at the simulation upper ($\sim$0.05 mbar) and lower ($\sim$500 bar) computational boundaries.}
\label{fig:zonal_nd}
\end{center}
\end{figure*}

The resultant number density structure of the cloud particles is a combination of the initial nucleation of seed particles and the hydrodynamic motions that transport cloud particles across the globe.
Early in the simulation, nucleation begins the cloud formation process with the most efficient nucleation occurring at the colder nightside mid-latitude regions (Fig. \ref{fig:zonal_J}).
Nucleation is a quick processes, and a few minutes/hours into the simulation atomic Ti is too depleted, limiting the nucleation of further cloud particles.
The nucleation source term, $J_{*}(\vec{r})$, for the dust moment equations presented in Sect. \ref{sec:Moments} becomes negligible across the globe.
Further evolution of the number density structure of cloud particles is then determined by the hydrodynamical and particle settling motions, rather than further nucleation.
Hotter dayside temperature regions at $\sim$10 mbar do not nucleate cloud particles at any point in time during the presently simulated epoch, but hydrodynamic motions transport cloud particles into these regions.
Seed particles remain thermally stable in these regions throughout the whole time-span of the present simulation.

Figure \ref{fig:ndstruc} shows the number density $n_{\rm d}$ [cm$^{-3}$] of cloud particles after the $\sim$60 Earth simulated days at isobars from 0.1 mbar - 10 bar.
The vertical frictional coupling between dust particles and gas is large enough that cloud particles move with the 3D gas flow efficiently.
This means, because we assume horizontal coupling, the highest cloud particle number density occurs near and at the equatorial regions for all atmospheric pressures. 
Drift velocities are small (\textbf{v}$_{\rm dr}$ $\sim$10$^{-4}$$\,\ldots\,$0.3 m s$^{-1}$) throughout the atmosphere (Fig. \ref{fig:convergence}), generally $<$10 \% the local vertical gas velocity.
This is a purely hydrodynamical effect and implies that the local grain sizes are not large enough to cause a significant de-coupling in the vertical direction.
Cloud particle motion is therefore dominated by the horizontal gas velocity (i.e. \textbf{v}$_{\rm dr}$ $\ll$ \textbf{u}$_{\rm gas, v}$ $<$ \textbf{u}$_{\rm gas, h}$).
The cloud structure predominantly follows the horizontal velocity structures in the atmosphere.
For example, cloud particles entering the equatorial jet stream will typically spend a longer time circulating in these regions due to the lower flux of particle out from the central jet, either from meridional motions or vertical settling.
Therefore, after a few days of simulation, after seed particle nucleation has become inefficient, equatorial regions are typically denser by $\sim$0.5 magnitudes compared to mid-high latitudes.
This is despite the majority of the seed particle nucleation taking place on nightside high-latitude regions, where the flow speed is typically slower, in the first hours of the simulation.
There is a build up of material on approach to the $\phi$ = 270\degr\ terminator corresponding to regions of highest horizontal velocity (u$_{\rm h}$ $\sim$ 7000 m s$^{-1}$).
This is due to the equatorial jet transporting material quickly around to the nightside and slowing down significantly (Fig. \ref{fig:Vstruc}) when reaching the night-day terminator.
This build up of cloud particles means that the overall flux of particles entering the dayside regions is reduced.
Additionally, particles entering the equatorial dayside regions are also transported quickly back onto the nightside by the equatorial jet, increasing the flux of particles towards the nightside.
This leads to differences in clouds number density, n$_{\rm d}$, between the dayside and nightside.
The cloud number density at 0.1, 1 and 10 mbar is also slightly ($\sim$1 \%) reduced by downward flows dragging the cloud particles to deeper depths near the day-night $\phi$ = 90\degr\ terminator and replenished by similar amounts at the upwelling night-day $\phi$ =  270\degr\ terminator.
At deeper regions ($>$100 mbar), due to the lower vertical gas velocities, the effect of vertical velocity on the number density structure is negligible.
At depths greater than 100 mbar there is less difference in number density between nightside-dayside and equatorial-high latitude regions.
Figure \ref{fig:zonal_nd} show 1D trajectory plots and the zonal and meridional mean number density as a function of pressure.
These clearly show a thicker cloud layer from $\sim$100 mbar - 10 bar which is relatively uniform throughout the planet which contains micron sized grains (Fig. \ref{fig:zonal_amean}).
On average, the atmosphere is fairly uniform in number density with 10-20$\%$ differences between equatorial and mid-high latitude regions and comparable difference between nightside and dayside regions.
This finding has implications for the cloud opacity which will therefore mainly be affected by the size of the cloud particles and their mixed material composition.

The large scale hydrodynamical motions explain the variety of the cloud number density seen on the dayside/nightside and show that cloud particle structures closely follow the horizontal, meridional and vertical gas dynamics at each atmospheric layer.
Efficient nucleation of seed particles occurs at mid-high latitudes on the nightside early in the simulation.
The gas dynamics then transports them, over time, to the equatorial regions where most of the cloud particles can be found by the end of the simulation.
This result may change if frictional coupling of the cloud particles with the atmospheric gas is altered by a force that specifically acts on cloud particles and causes them to move with \textbf{u}$_{\rm d}$ $\not =$ \textbf{u}$_{\rm gas}$  in horizontal/meridional  directions.

\subsection{Cloud particle sizes}
\label{sec:GrainSizes}

\begin{figure*}
\begin{center}
\includegraphics[width=0.49\textwidth]{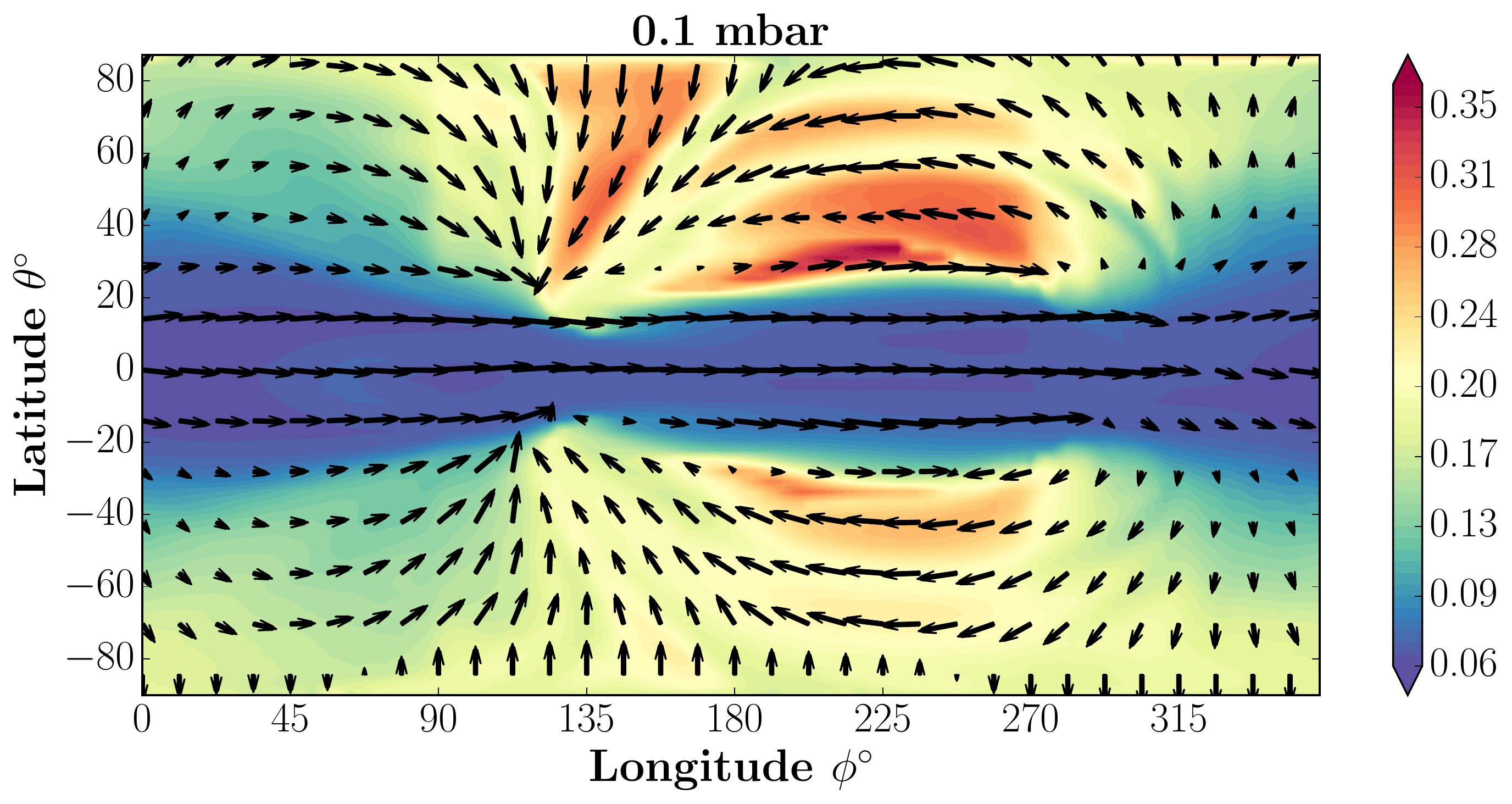}
\includegraphics[width=0.49\textwidth]{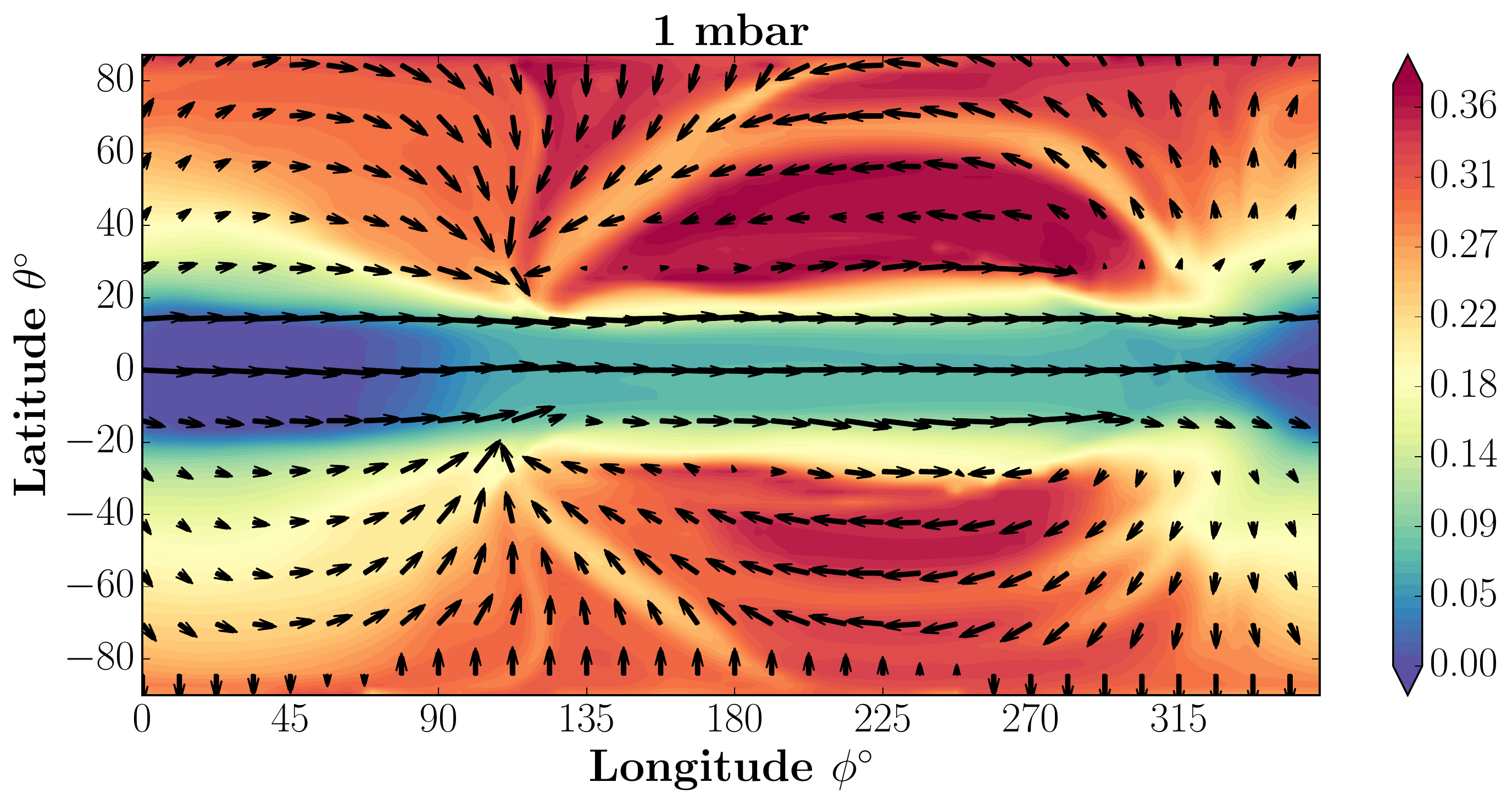}
\includegraphics[width=0.49\textwidth]{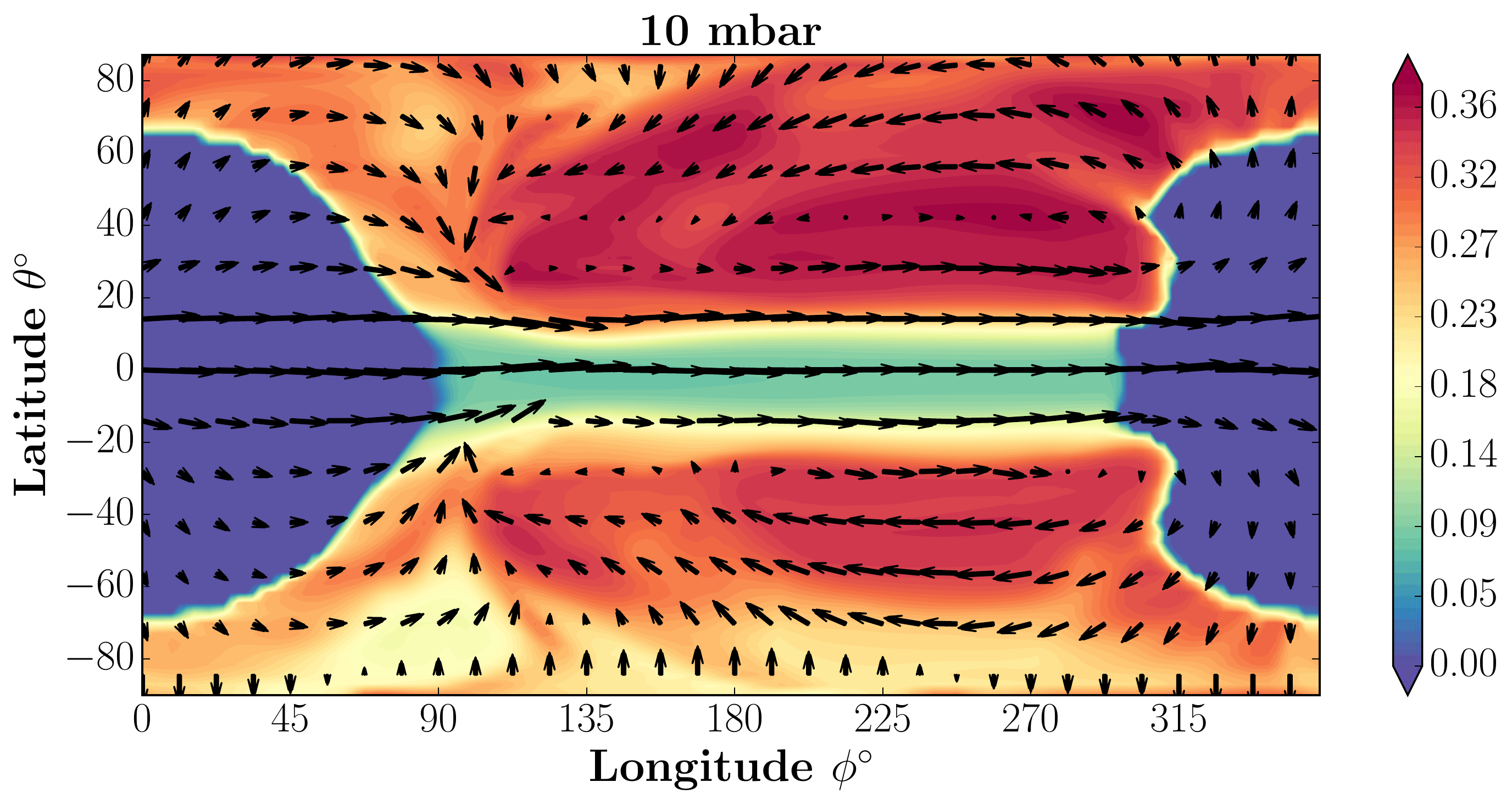}
\includegraphics[width=0.49\textwidth]{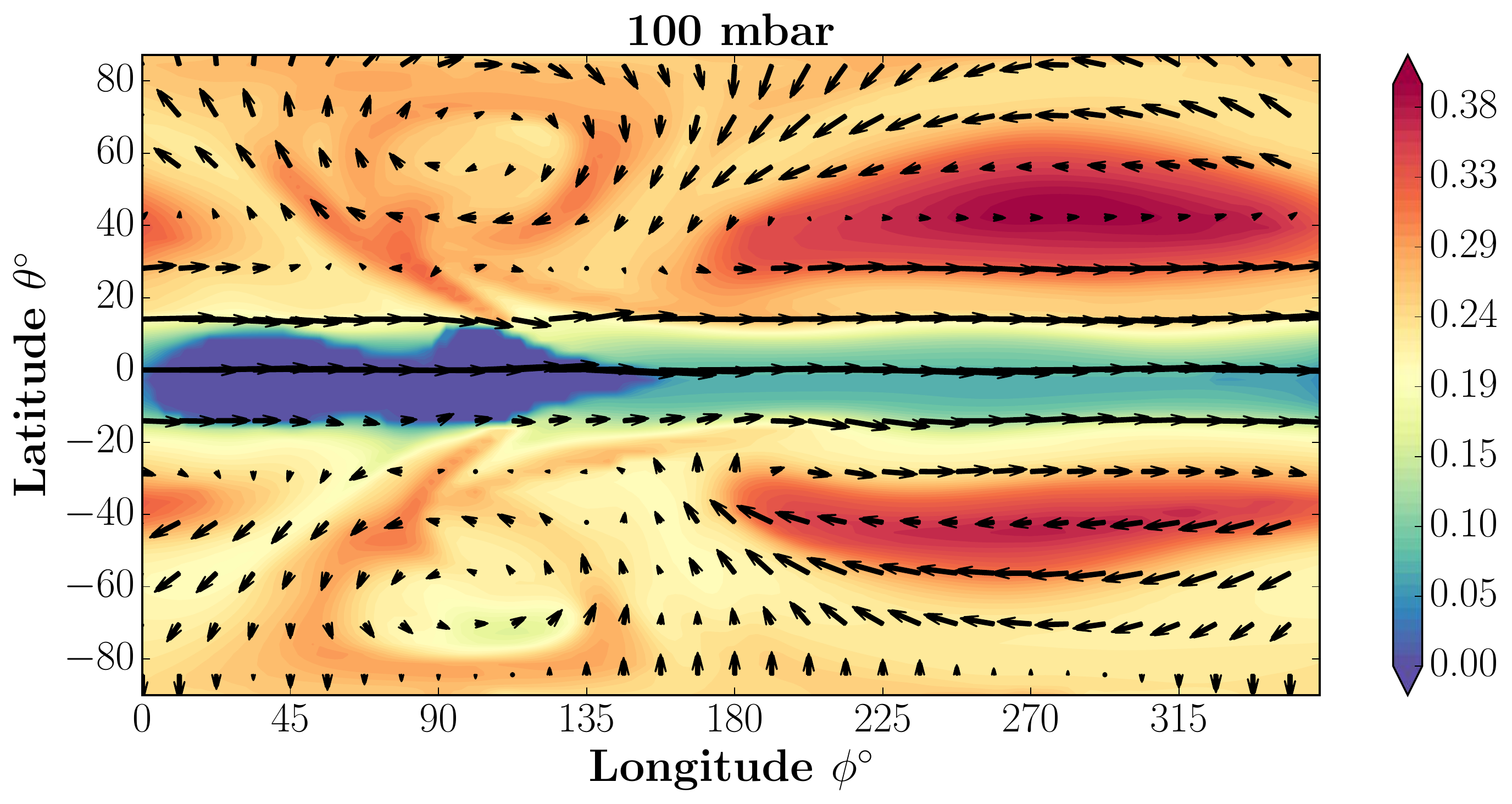}
\includegraphics[width=0.49\textwidth]{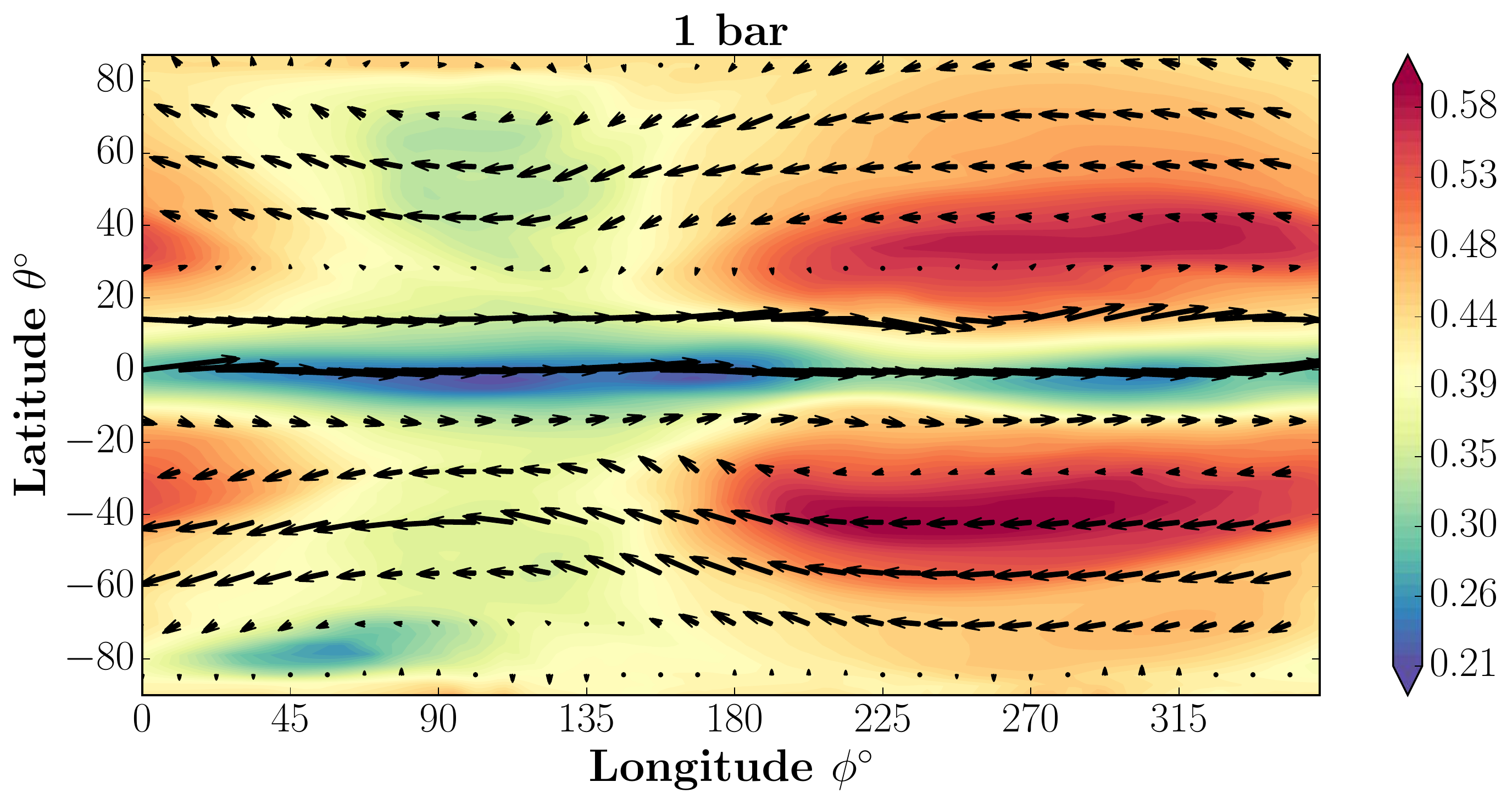}
\includegraphics[width=0.49\textwidth]{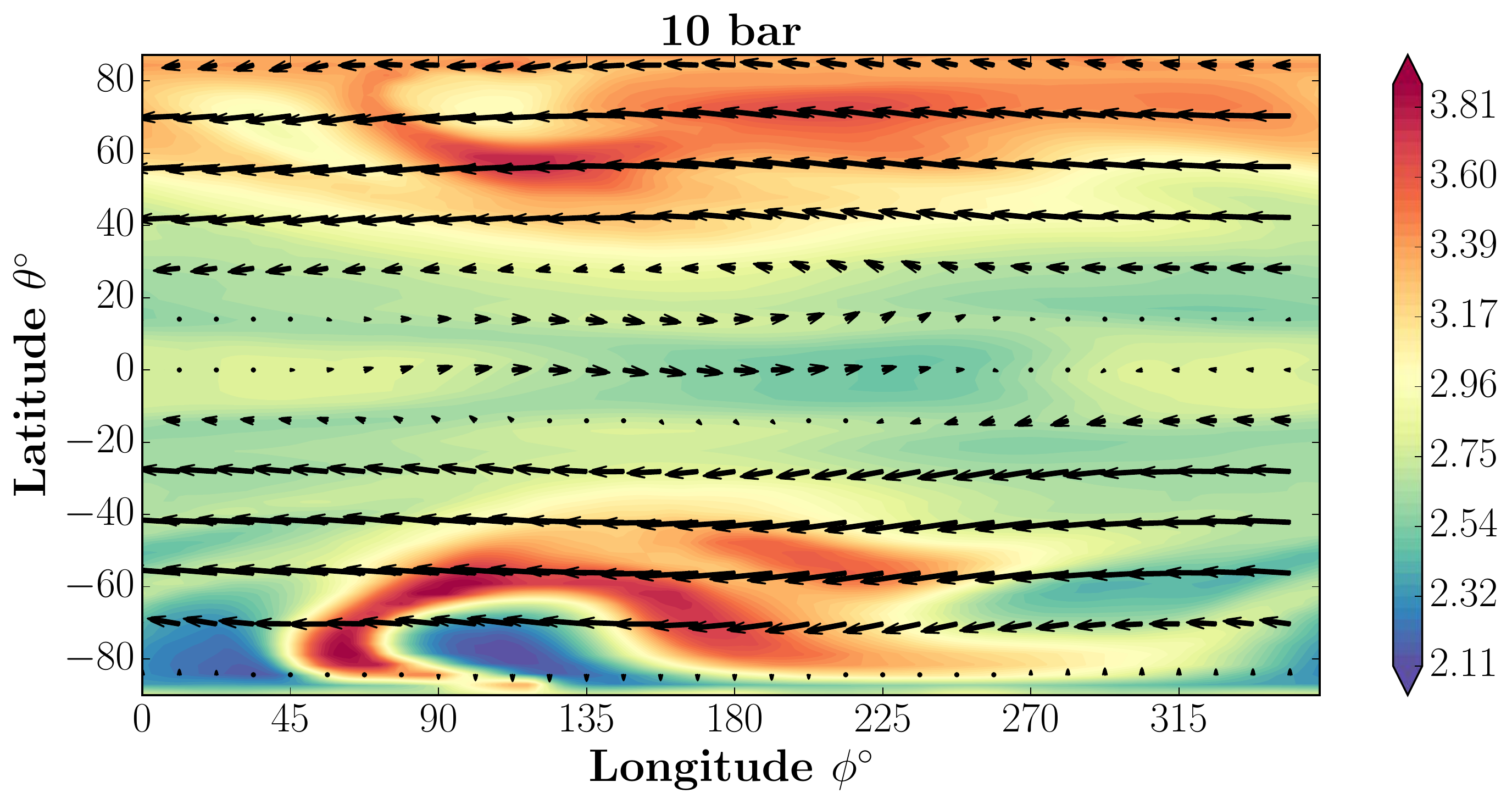}

\caption{Mean grain size $\langle$a$\rangle$ [$\mu$m] (colour bar) and velocity field ($\rvert {\bf u}\lvert$ = $\sqrt{{\bf u}_{\rm h}^{2} + {\bf u}_{\rm m}^{2}}$) at 0.1, 1, 10, 100 mbar and 1, 10 bar for different $\phi$ (longitudes) and $\theta$ (latitude).
Note: the colour bar scale is different for each plot.
The sub-stellar point is located at $\phi$ = 0\degr, $\theta$ = 0\degr.
The smallest grains at each layer typically reside at the equatorial regions. 
The largest grains are typically found on the nightside and at higher latitudes.
Deep purple/blue coloured regions at 1, 10 and 100 mbar contain seed particles of sizes $\sim$0.001 $\mu$m}
\label{fig:ameanstruc}
\end{center}
\end{figure*}

\begin{figure*}
\begin{center}
\includegraphics[width=0.49\textwidth]{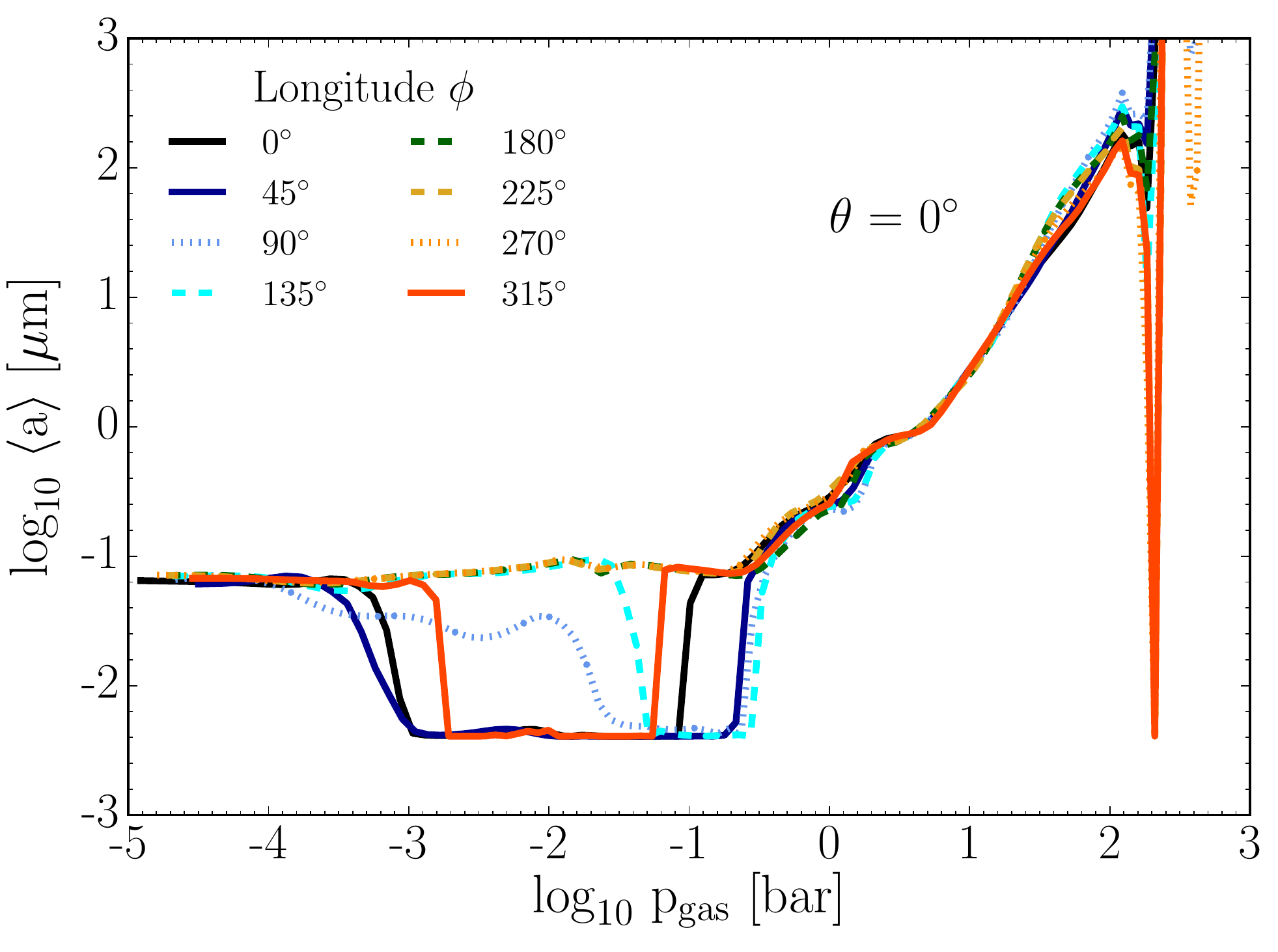}
\includegraphics[width=0.49\textwidth]{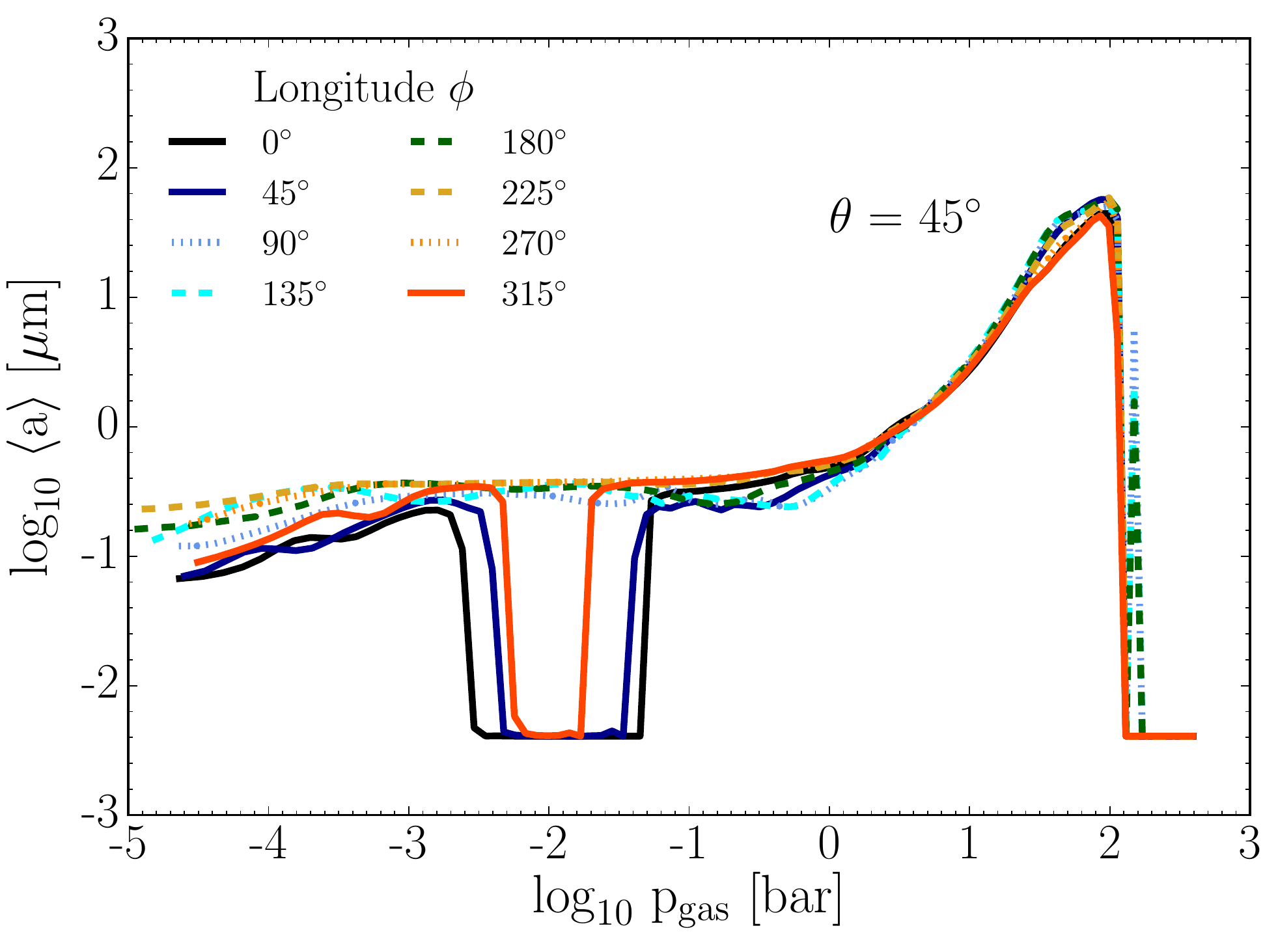}
\includegraphics[width=0.49\textwidth]{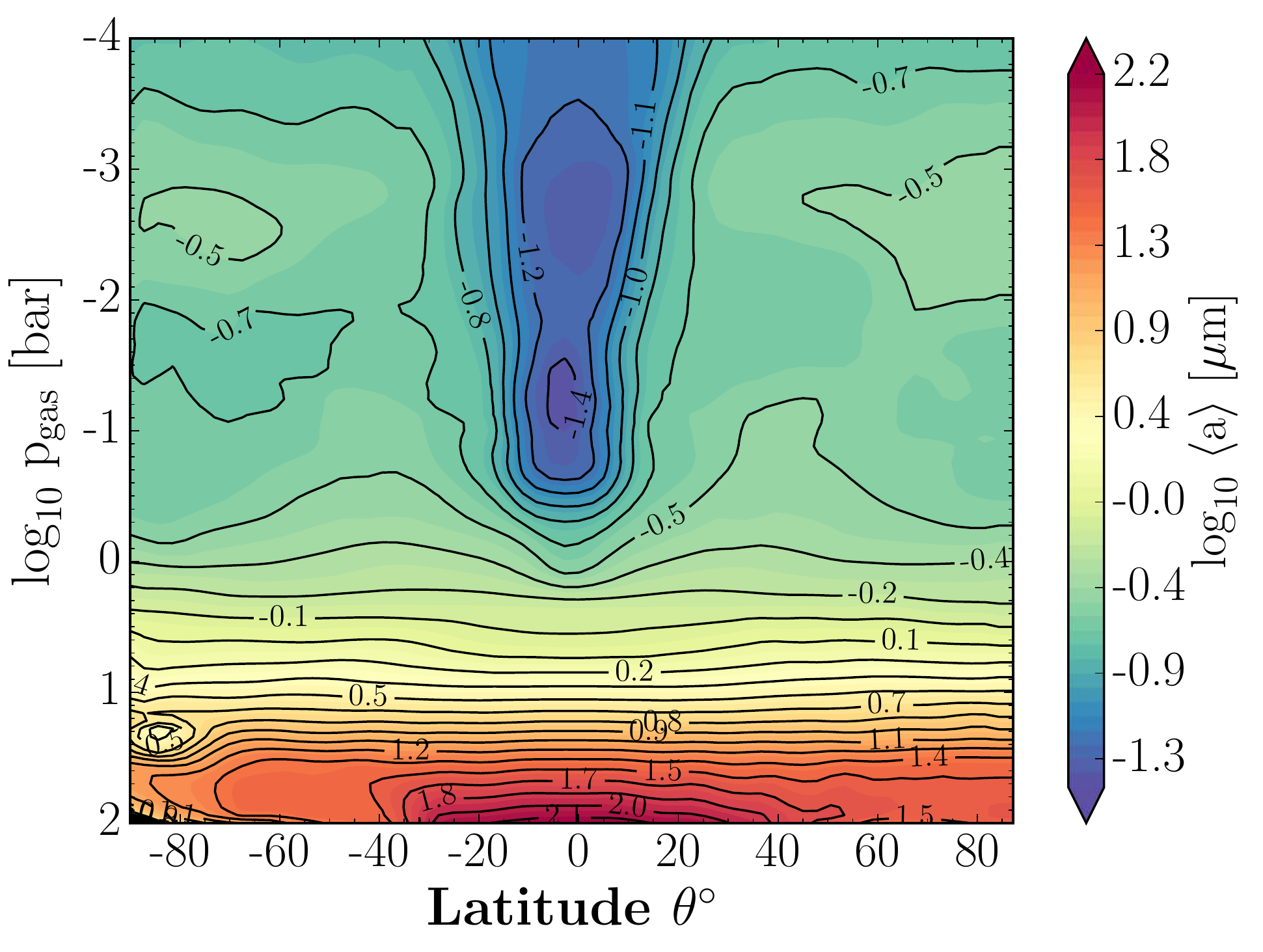}
\includegraphics[width=0.49\textwidth]{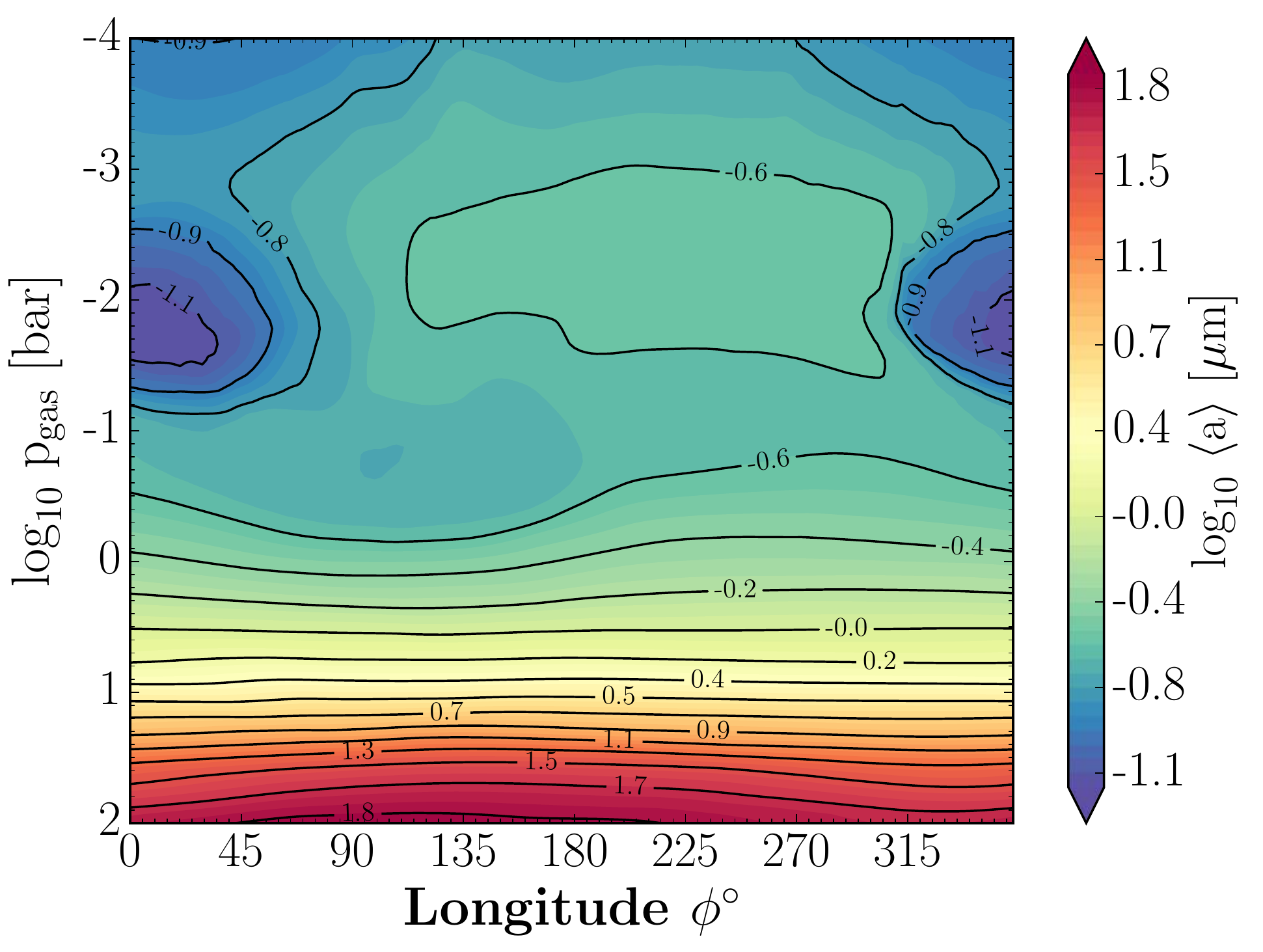}
\caption{\textbf{Top:} 1D mean grain size $\log_{10}$ $\langle$a$\rangle$ [$\mu$m] trajectories at latitudes $\theta$ = 0\degr (left) , 45\degr (right).
Cloud particles are typically $\sim$0.1 $\mu$m from 0.01-1 mbar. 
Seed particles are present between 1-100 mbar for many profiles.
\textbf{Bottom:} Zonal mean (left) and Meridional mean (right) average grain size $\log_{10}$ $\langle$a$\rangle$ [$\mu$m] as a function of pressure.
The smallest grains are found on the dayside at pressures 1-100 mbar at the equatorial regions.
The grain sizes are typically sub-micron above $\sim$5 bar and micron sizes below this pressure level.
}
\label{fig:zonal_amean}
\end{center}
\end{figure*}

Cloud particle sizes are a direct result of our cloud formation model (Sect. \ref{sec:Moments}).
Each growth species local growth/evaporation rate (Eq. \ref{eq:chinet}) determines the local grain size, which is a function of the local number of elements and temperature.
It also depends on the sinking/settling of cloud particles of different sizes over time as larger grains sink faster to higher pressure regions.
Figure \ref{fig:ameanstruc} shows the mean grain size $\langle$a$\rangle$ [$\mu$m] of cloud particles at pressure isobars from 0.1 mbar - 10 bar.
Dayside regions from 0.1 mbar - 1 bar contain smaller grains while the nightside contains larger grains.
This is most evident from the 0.1 and 10 mbar plots in Fig. \ref{fig:ameanstruc} where larger grains ($> $0.1 $\mu$m) reside on the nightside while smaller grains ($< $0.1 $\mu$m) are found on the dayside.
Asymmetry in grain size between equatorial and mid-high latitudes is also seen, with equatorial regions containing the smallest grains at any given atmospheric pressure and larger grains supported at higher latitudes.
At $\sim$10 mbar seed particles ($\sim$0.001 $\mu$m) reside on the dayside, corresponding to the highest upper atmosphere temperature regions where all other growth species are evaporated.
At depths $\sim$ 1 bar particles grow to 1 $\mu$m sizes or larger.
At such high densities, the frictional coupling to the gas phase is almost complete, resulting in very small drift velocities (v$_{\rm dr}$ $<$ 0.001 m s$^{-1}$).

In Fig. \ref{fig:zonal_amean} the 1D trajectories at equatorial and mid-latitudes show a varied depth dependent grain size.
Nightside and mid-latitude terminator regions regions typically contain grain sizes of $\sim$0.1 $\mu$m or less down to 1 bar where they grow to micron sizes and above.
Dayside and the $\phi$ = 90\degr, 135\degr\ equatorial profiles show the presence of seed particles from $\sim$0.1-100 mbar.
Figure \ref{fig:zonal_amean} shows the zonal and meridional averaged mean particle size (note: log scale) as a function of pressure.
The atmosphere typically contains sub-micron particles down to a pressure level of $\sim$5 bar.
The equatorial dayside regions contain the smallest particulates from 0.1-100 mbar.
The nightside mean particle size is also $\sim$0.5-1 magnitudes larger than the dayside grains but remain sub-micron at these pressure levels.
The largest particles ($\sim$0.1-1 mm) reside at the most dense parts of the atmosphere at gas pressures $>$ 10 bar.
Gradients (up to 1 order of magnitude) in cloud particle size occur near the $\phi$ = 90\degr\ terminator, while grain sizes at the $\phi$ = 270\degr\ terminator remain relatively homogenous with longitude.
This suggests that transit spectroscopy \citep{Pont2013} would sample a variety of cloud particle sizes.
The lower temperature regions at the center of vortex regions on the upper atmosphere nightside contain the largest cloud particles at each pressure level, suggesting that these vortex regions can efficiency trap and grow larger particles.

In summary, our model produces a variety of cloud particle sizes dependent on the local thermo-chemical conditions of the atmosphere.
A large portion of the hot equatorial dayside contains seed particles of nm size in contrast to the cooler, nightside and mid-latitude regions where grain sizes can be $>$ 0.1 $\mu$m.
We, however, note that the present results are limited to the growth of TiO$_{2}$[s], SiO[s], SiO$_{2}$[s], MgSiO$_{3}$[s], Mg$_{2}$SiO$_{4}$[s] materials, which suggest that the present grain sizes to be  a lower limit (Sect. \ref{sec:Discussion}).

\subsection{Cloud material composition}
\label{sec:MaterialComposition}

\begin{figure*}
\begin{center}
\includegraphics[width=0.32\textwidth]{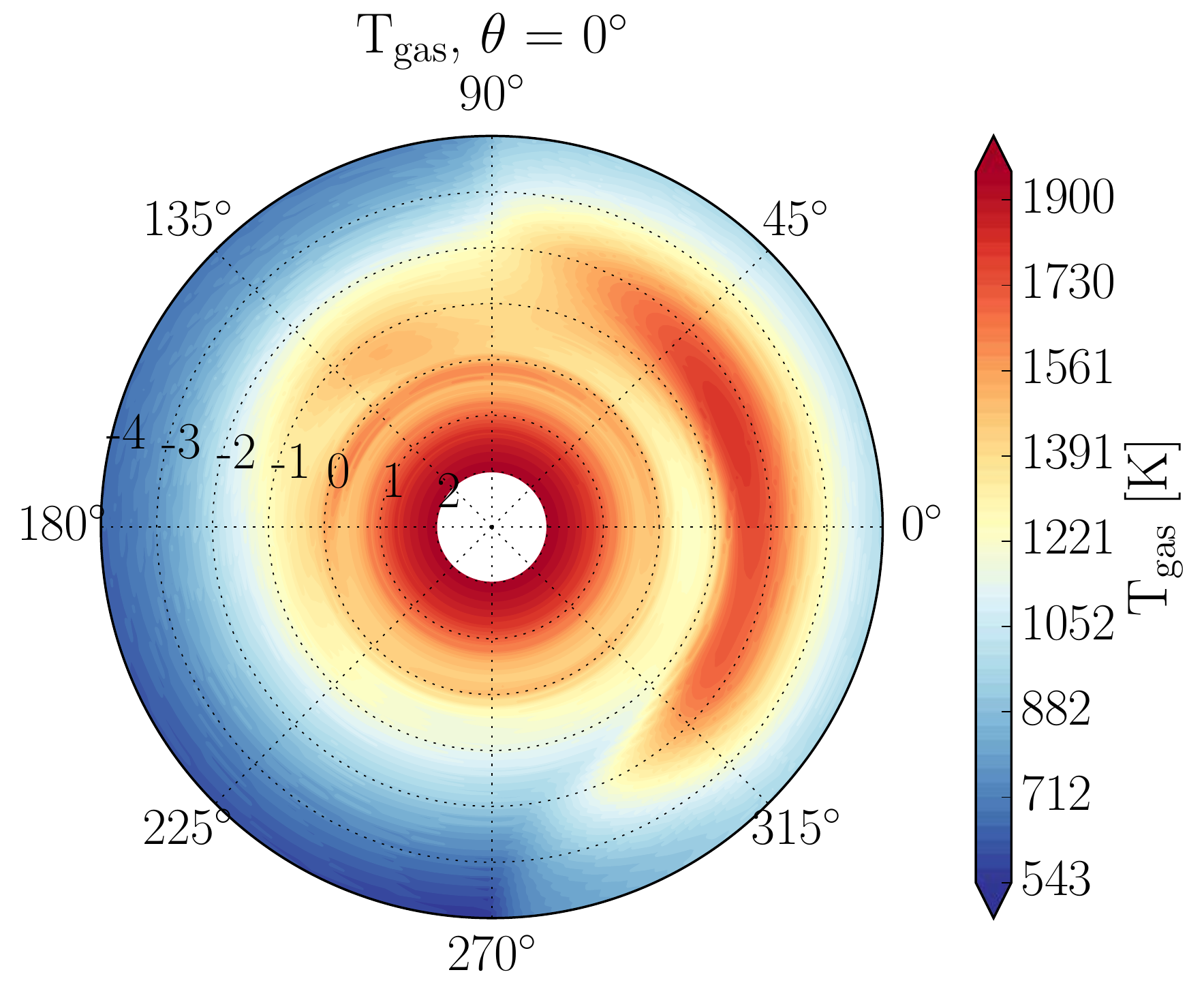}
\includegraphics[width=0.32\textwidth]{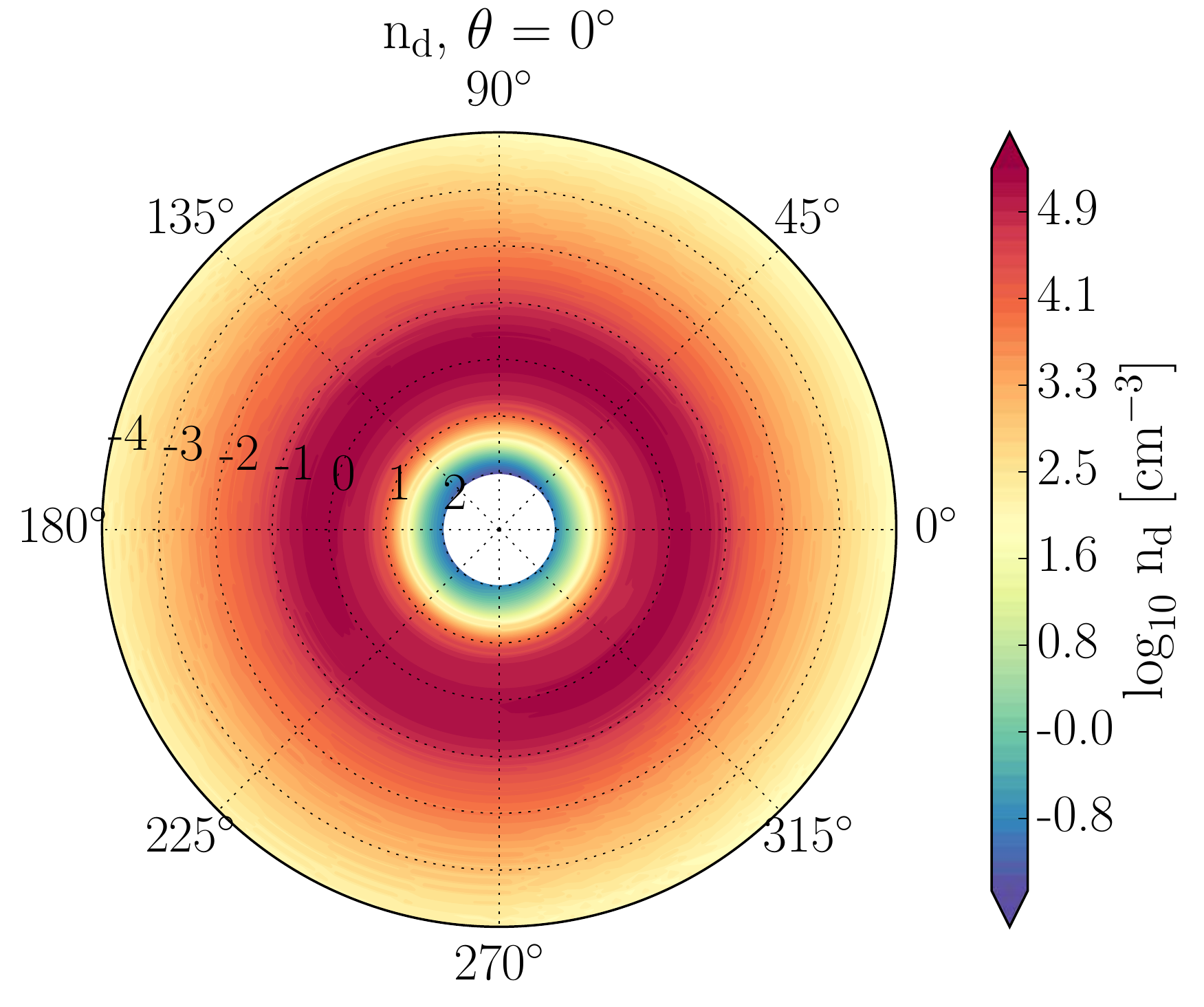}
\includegraphics[width=0.32\textwidth]{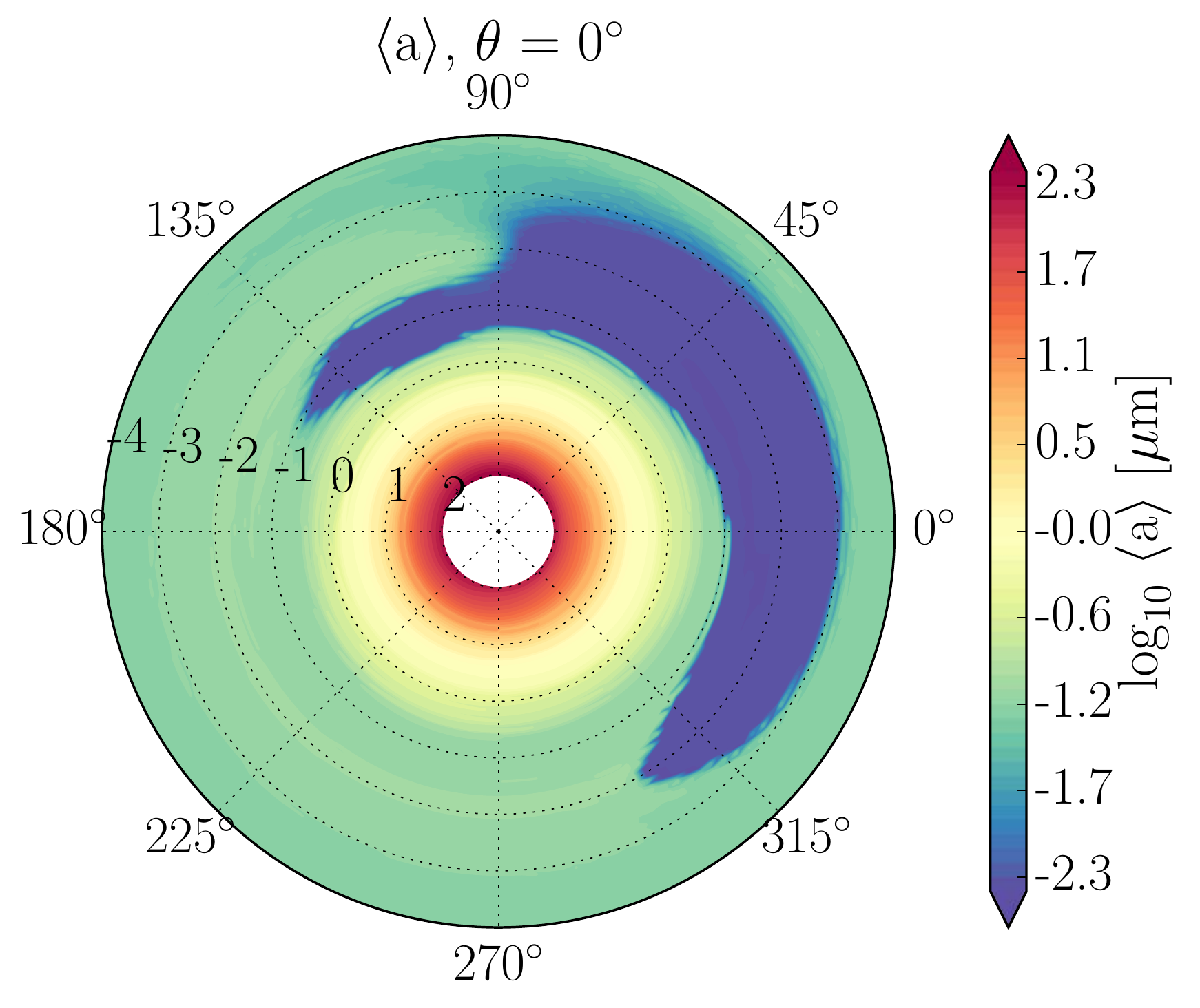}
\includegraphics[width=0.32\textwidth]{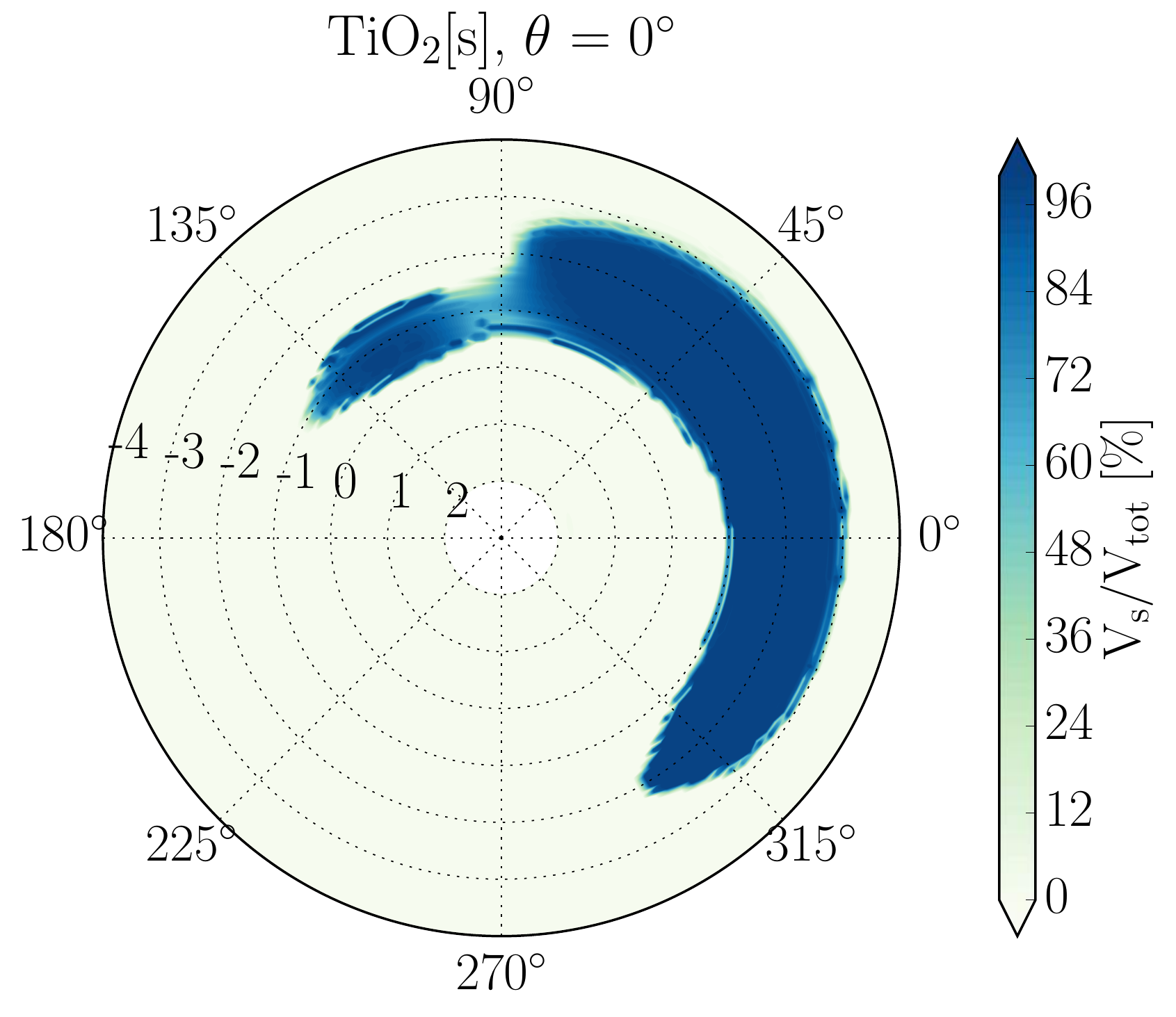}
\includegraphics[width=0.32\textwidth]{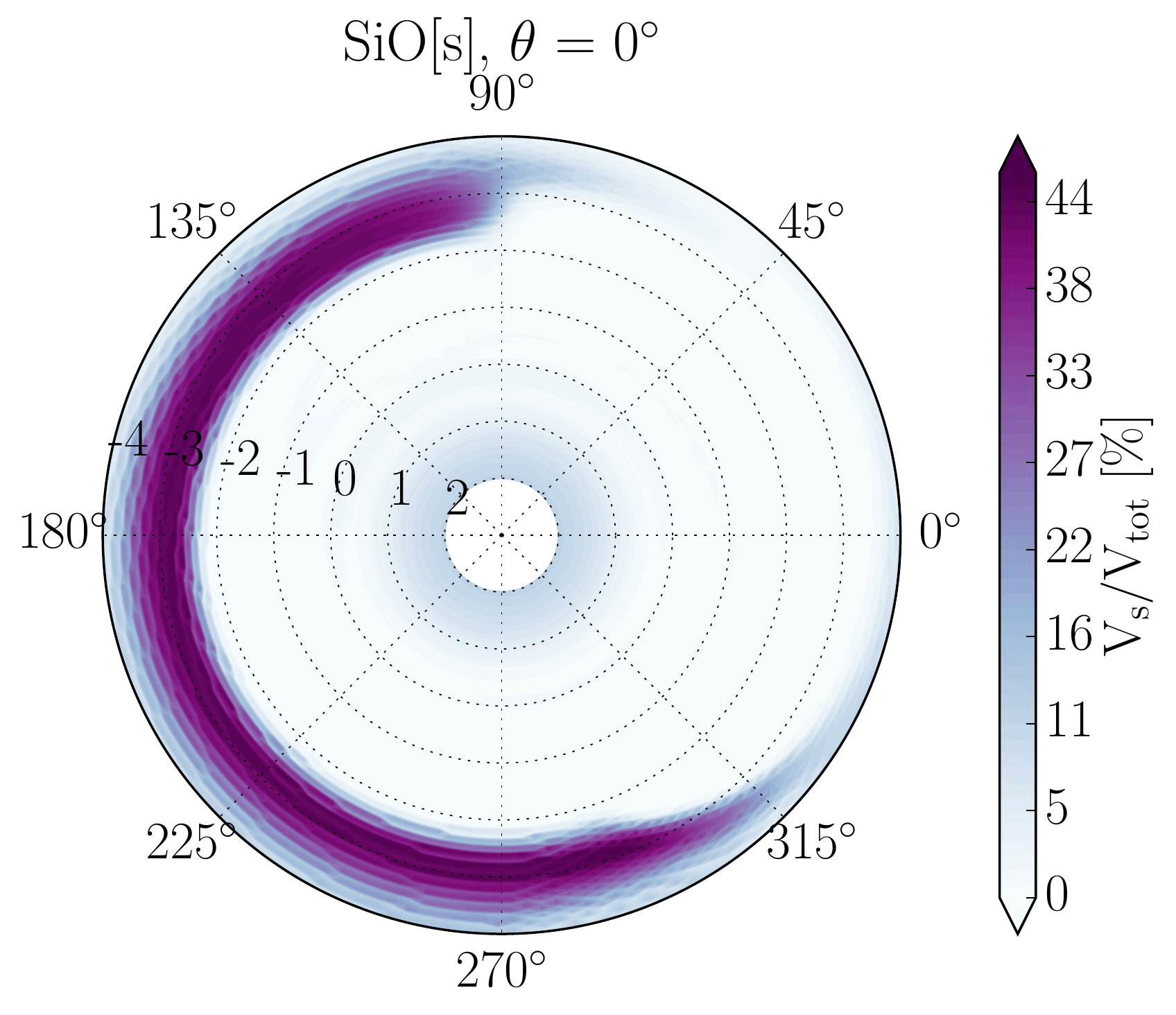}
\includegraphics[width=0.32\textwidth]{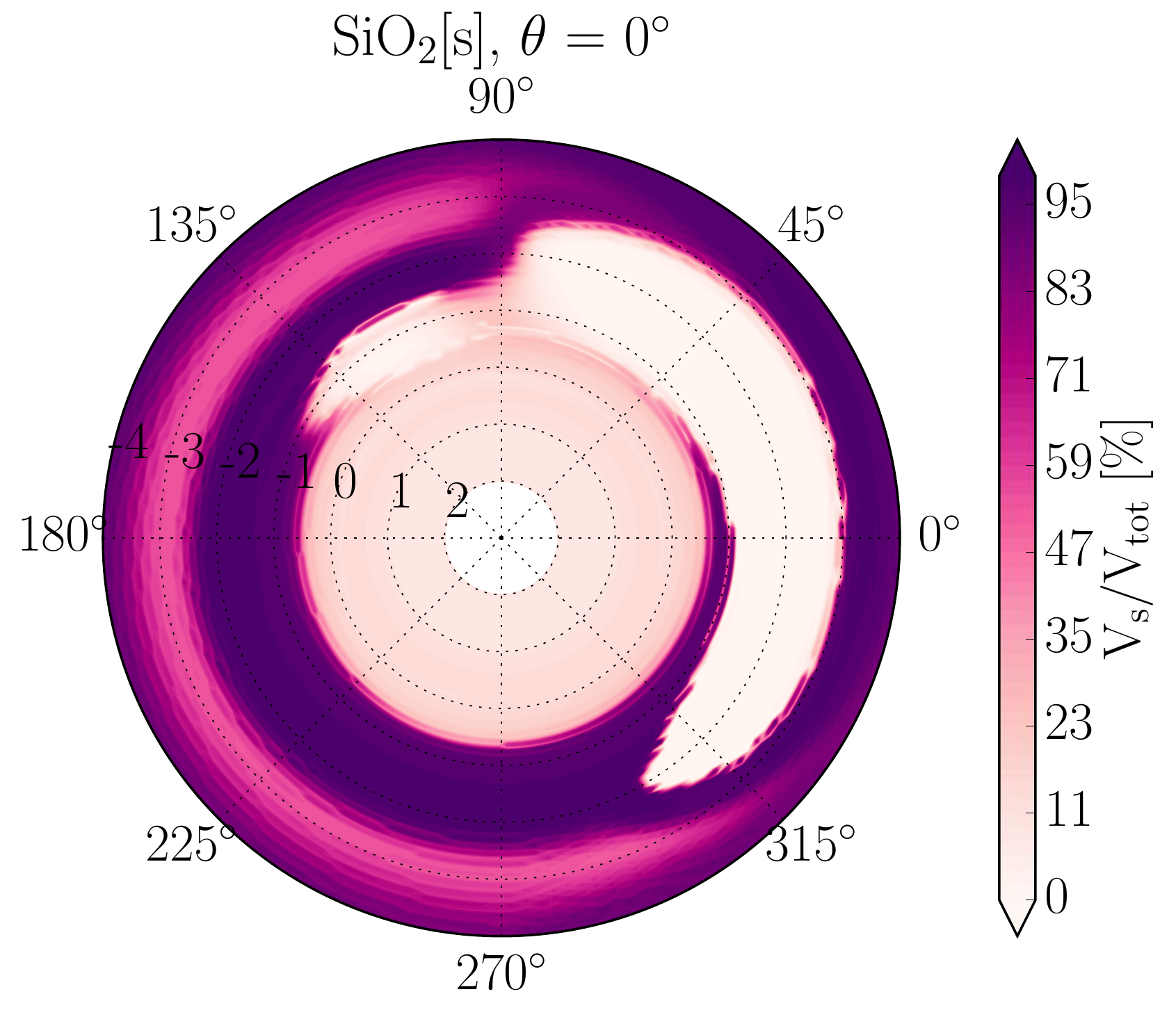}
\includegraphics[width=0.32\textwidth]{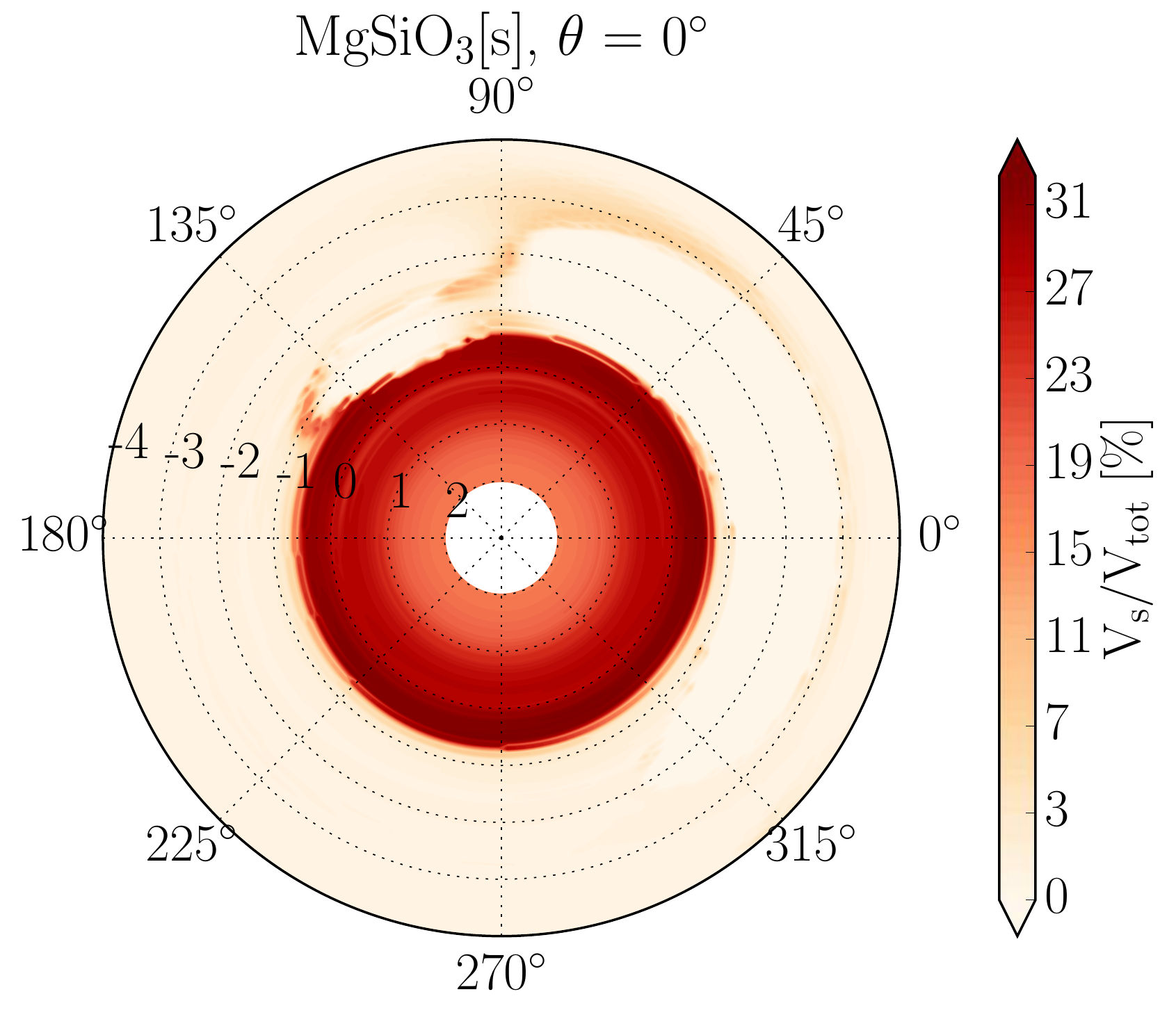}
\includegraphics[width=0.32\textwidth]{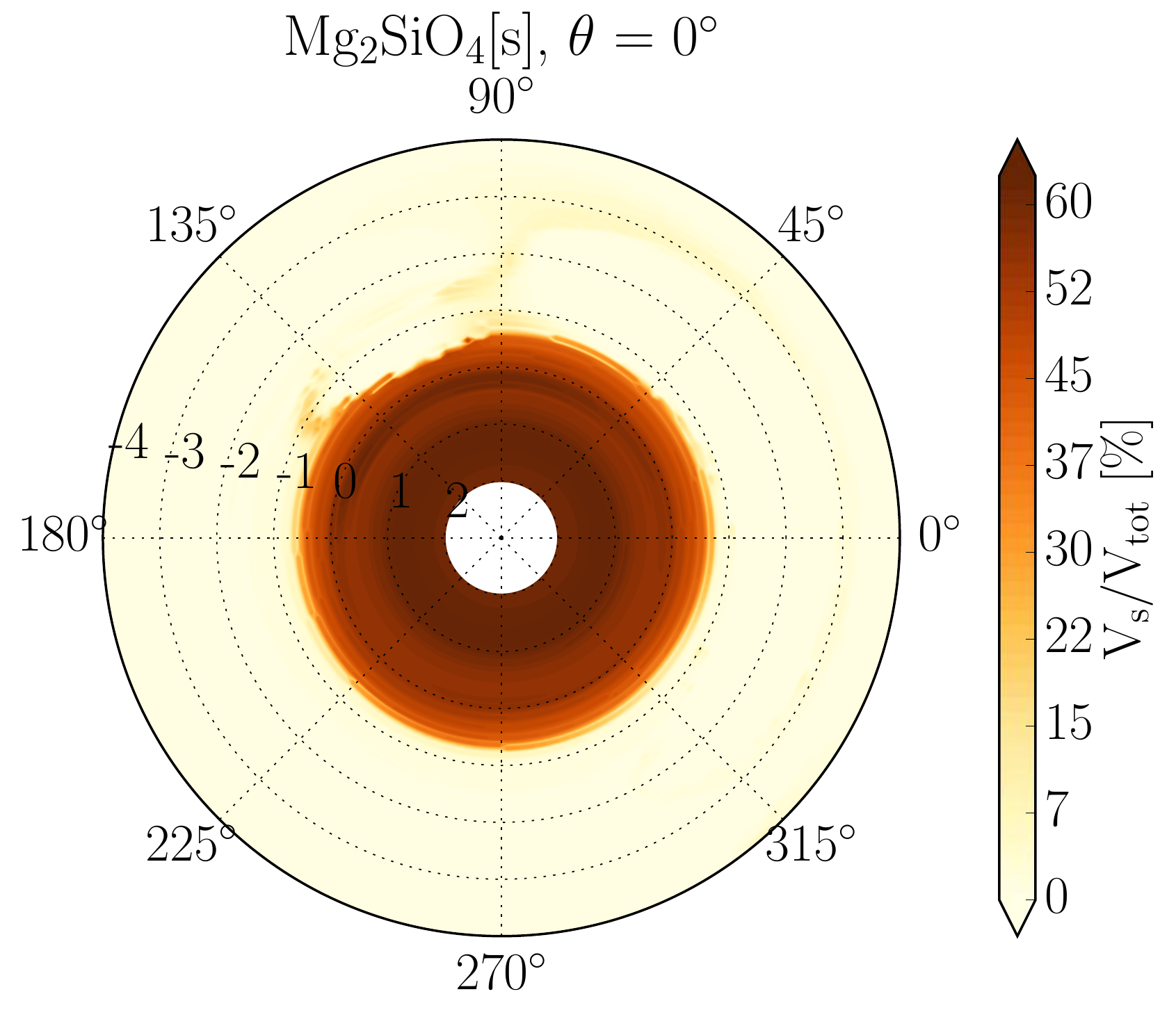}
\caption{Meridional polar slices of atmospheric cloud properties at the equator ($\theta$ = 0$^{\circ}$).
\textbf{Top:} Local gas temperature (T$_{\rm gas}$ [K]), cloud particle number density ($\log_{10}$ n$_{\rm d}$ [cm$^{-3}$]) and mean cloud particle grain size ($\log_{10}$ $\langle$a$\rangle$ [$\mu$m]).
\textbf{Middle:} Volume fraction (V$_{\rm s}$/V$_{\rm tot}$ [$\%$]) of the cloud particle composition containing TiO$_{2}$[s], SiO[s] and SiO$_{2}$[s].
\textbf{Bottom:} Volume fraction (V$_{\rm s}$/V$_{\rm tot}$ [$\%$]) of the cloud particle composition containing MgSiO$_{3}$[s] and Mg$_{2}$SiO$_{4}$[s].
Outer circular values denote longitude at intervals of $\phi$ = 45\degr\ from the sub-stellar point ($\phi$ = 0\degr). Radial values indicate $\log_{10}$ p$_{\rm gas}$ isobars from 0.1 mbar - 100 bar.
The globe is irradiated from the direction of the colour bars.
Note: the size of the annulus is not scaled to planetary radius.}
\label{fig:classical_meri_0}
\end{center}
\end{figure*}

\begin{figure*}
\begin{center}
\includegraphics[width=0.32\textwidth]{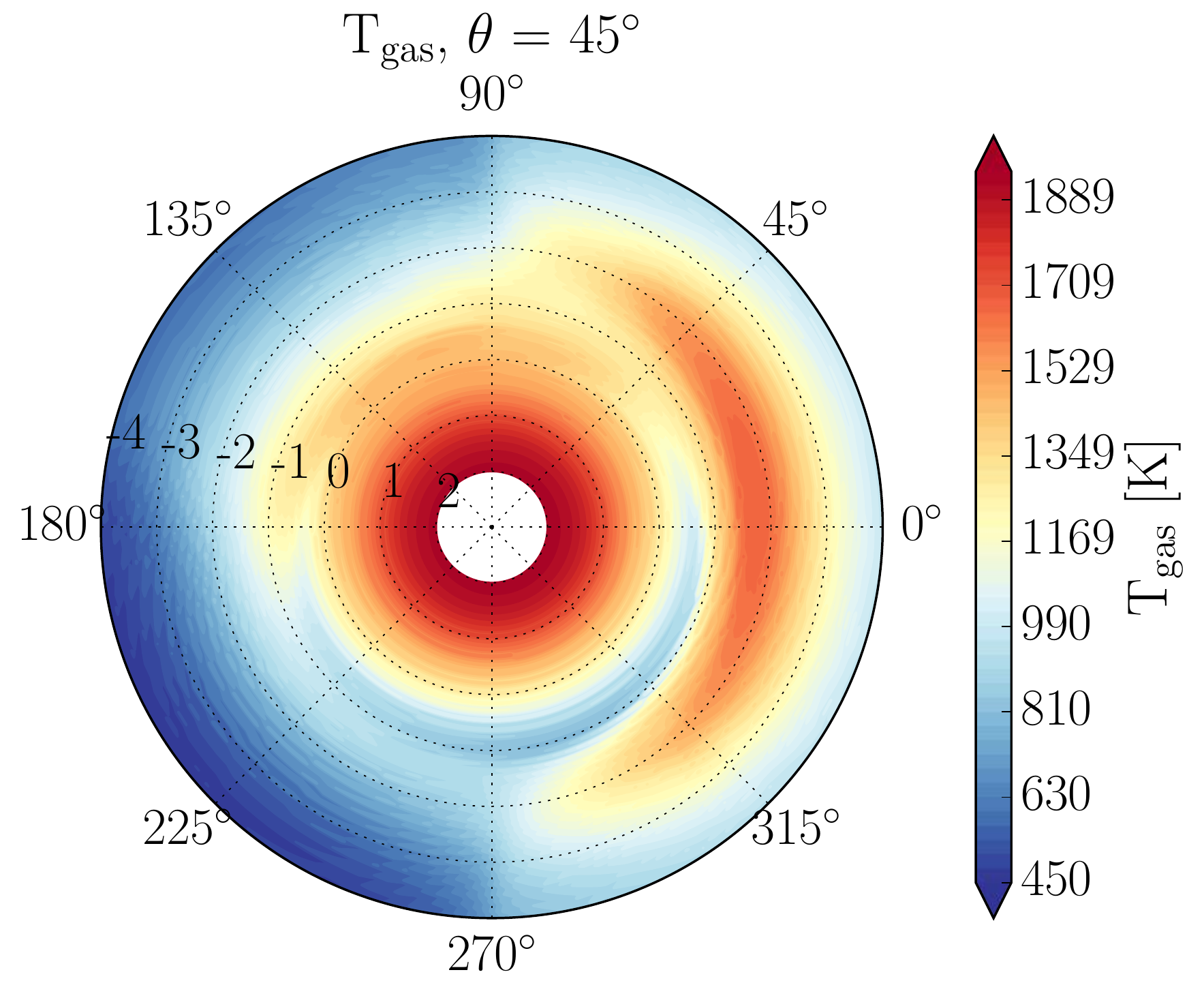}
\includegraphics[width=0.32\textwidth]{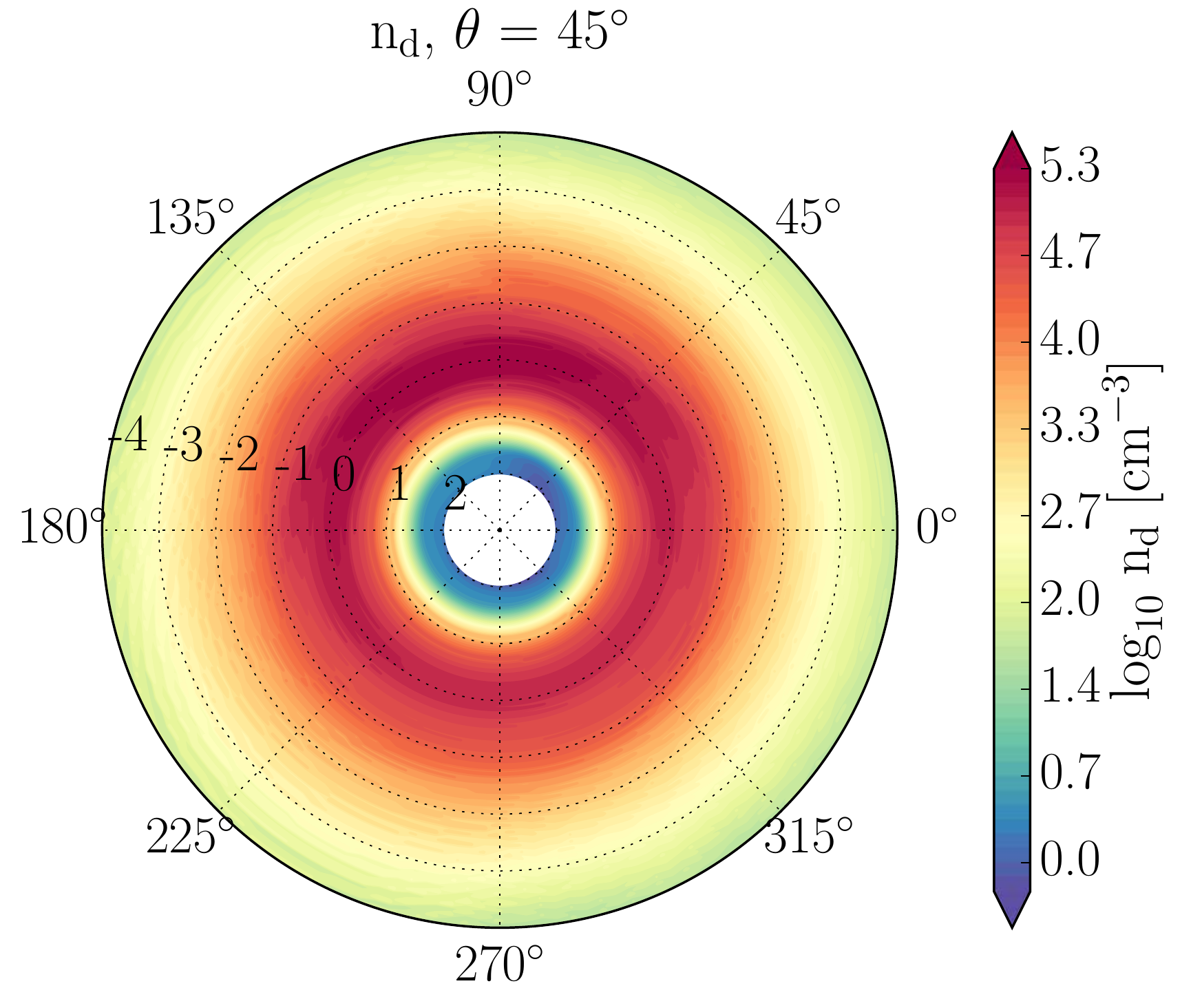}
\includegraphics[width=0.32\textwidth]{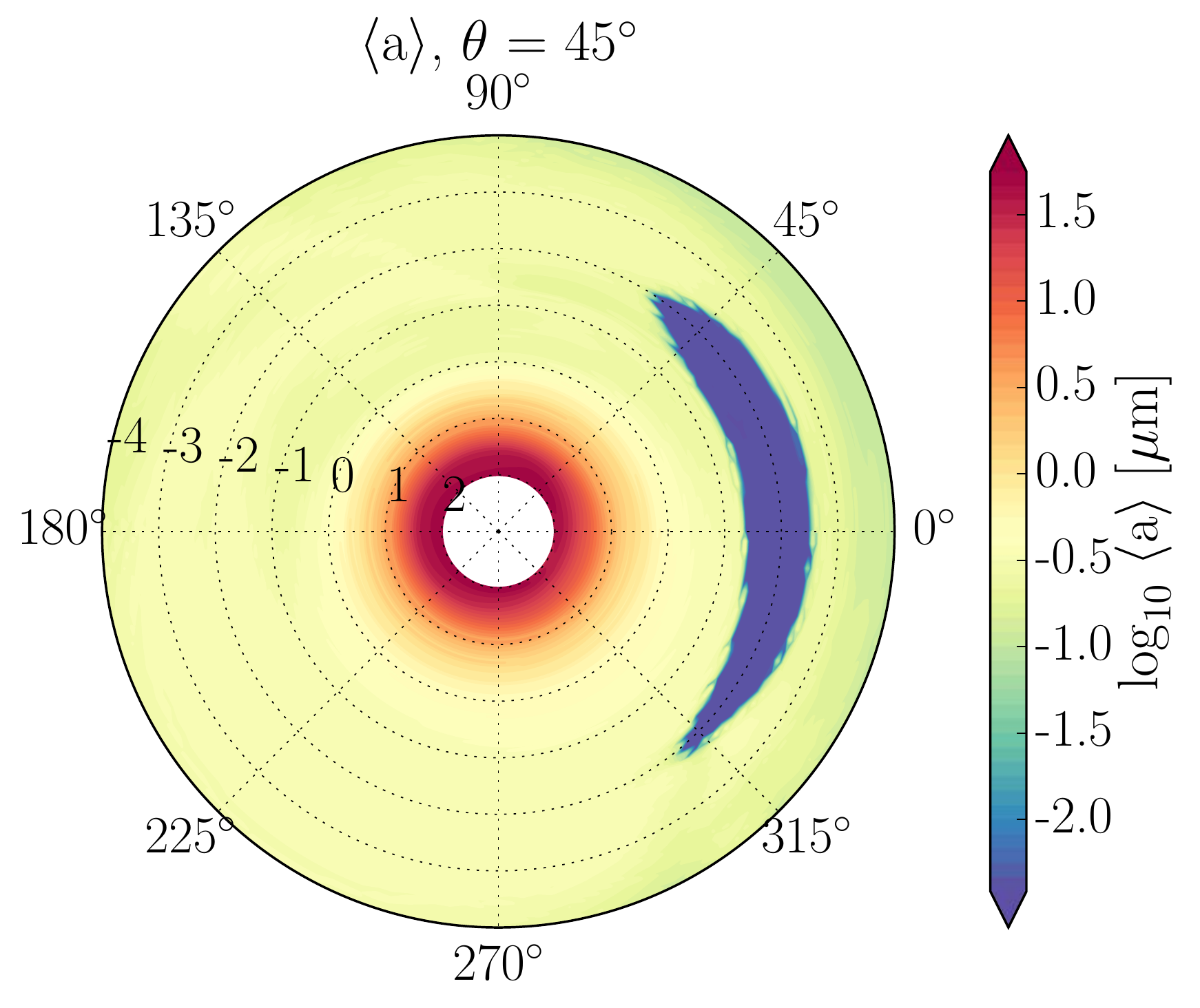}
\includegraphics[width=0.32\textwidth]{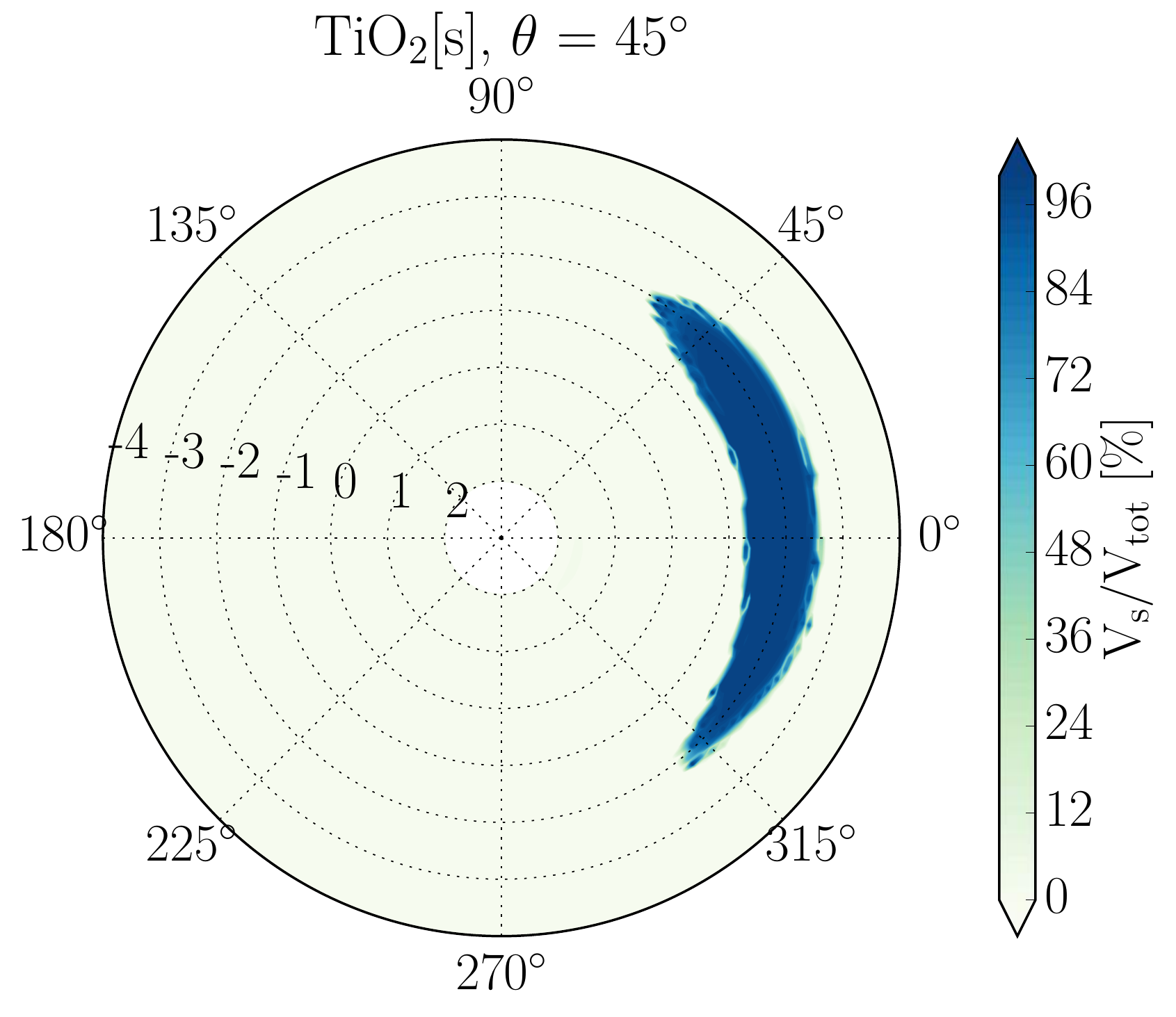}
\includegraphics[width=0.32\textwidth]{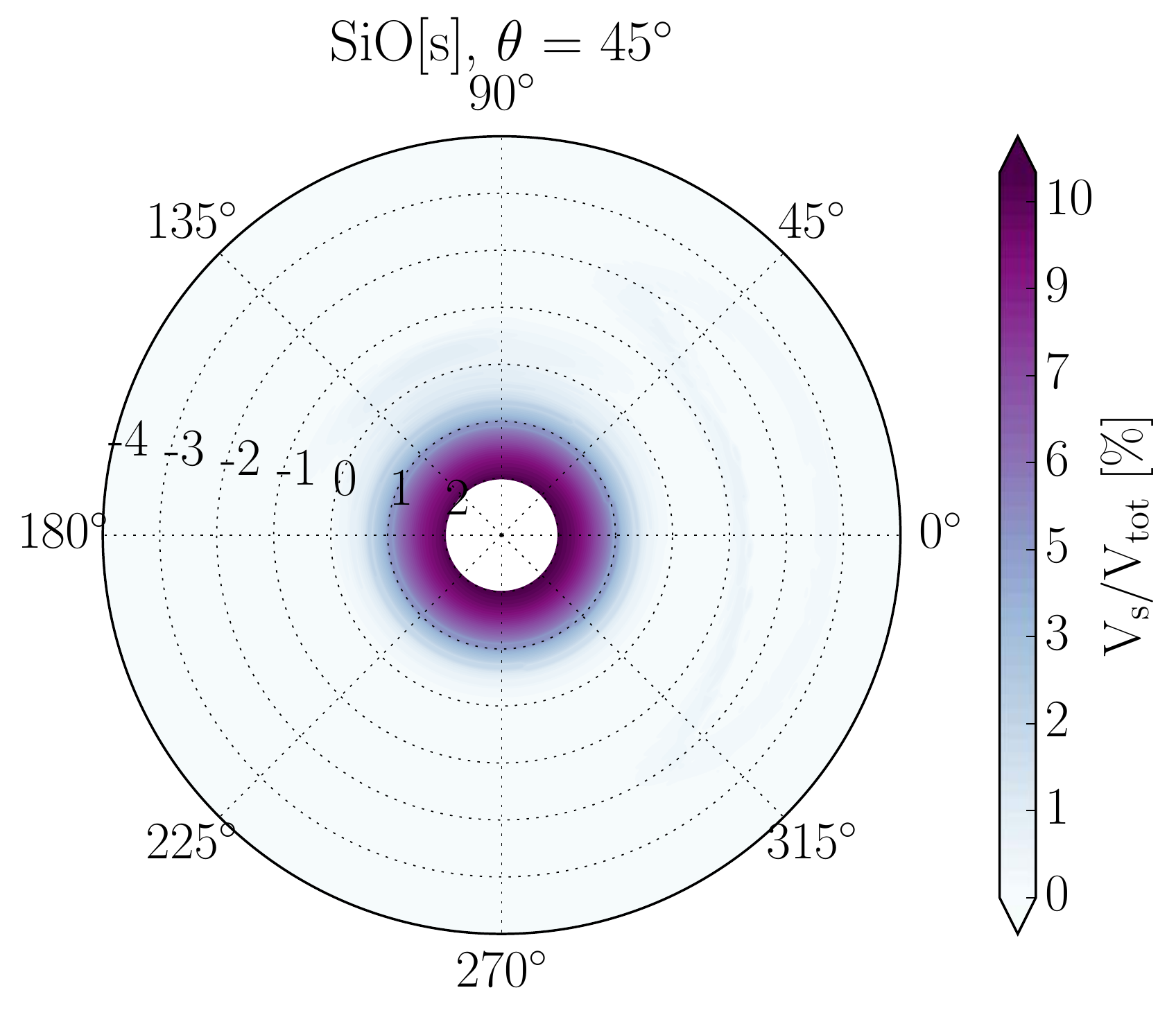}
\includegraphics[width=0.32\textwidth]{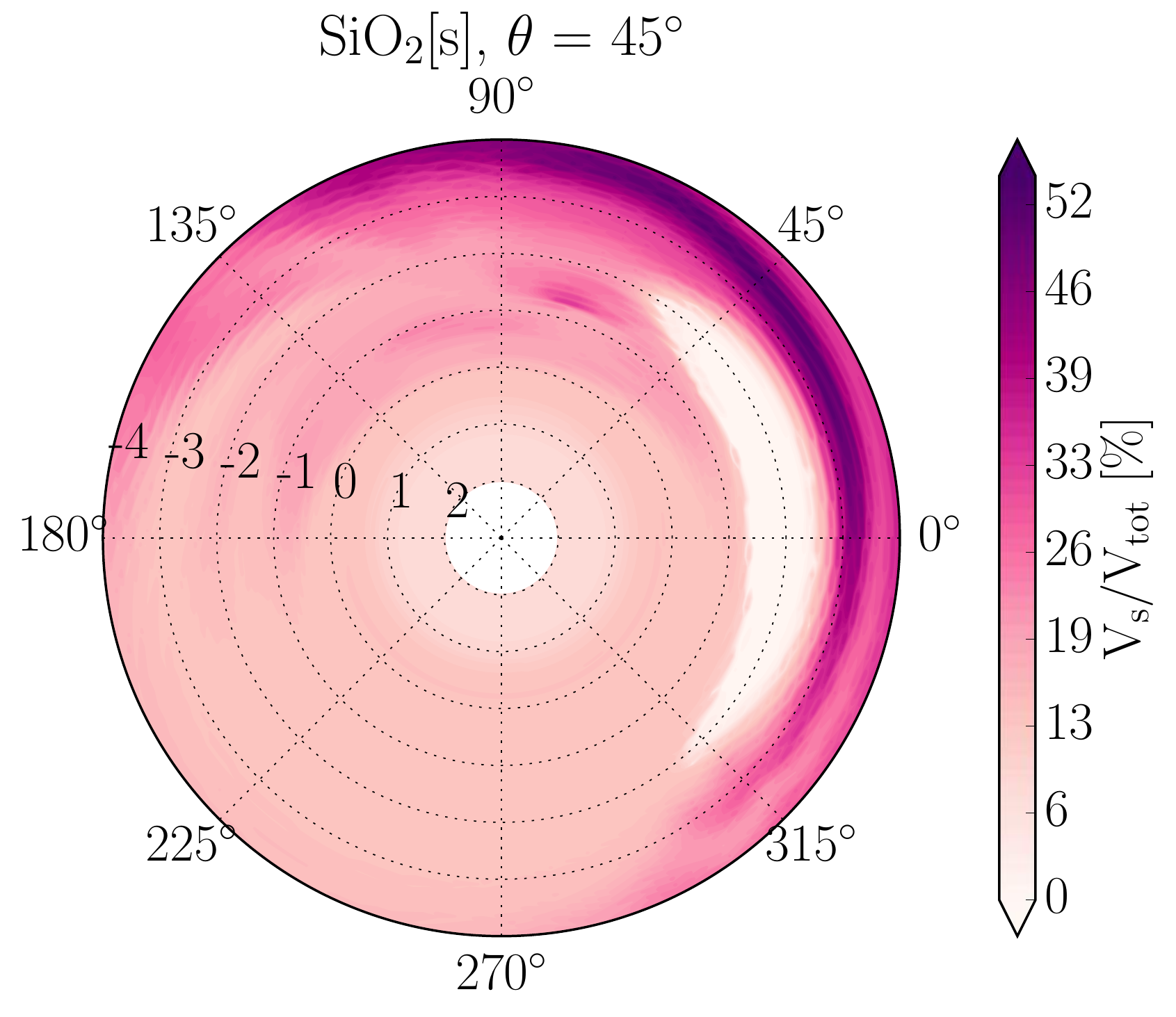}
\includegraphics[width=0.32\textwidth]{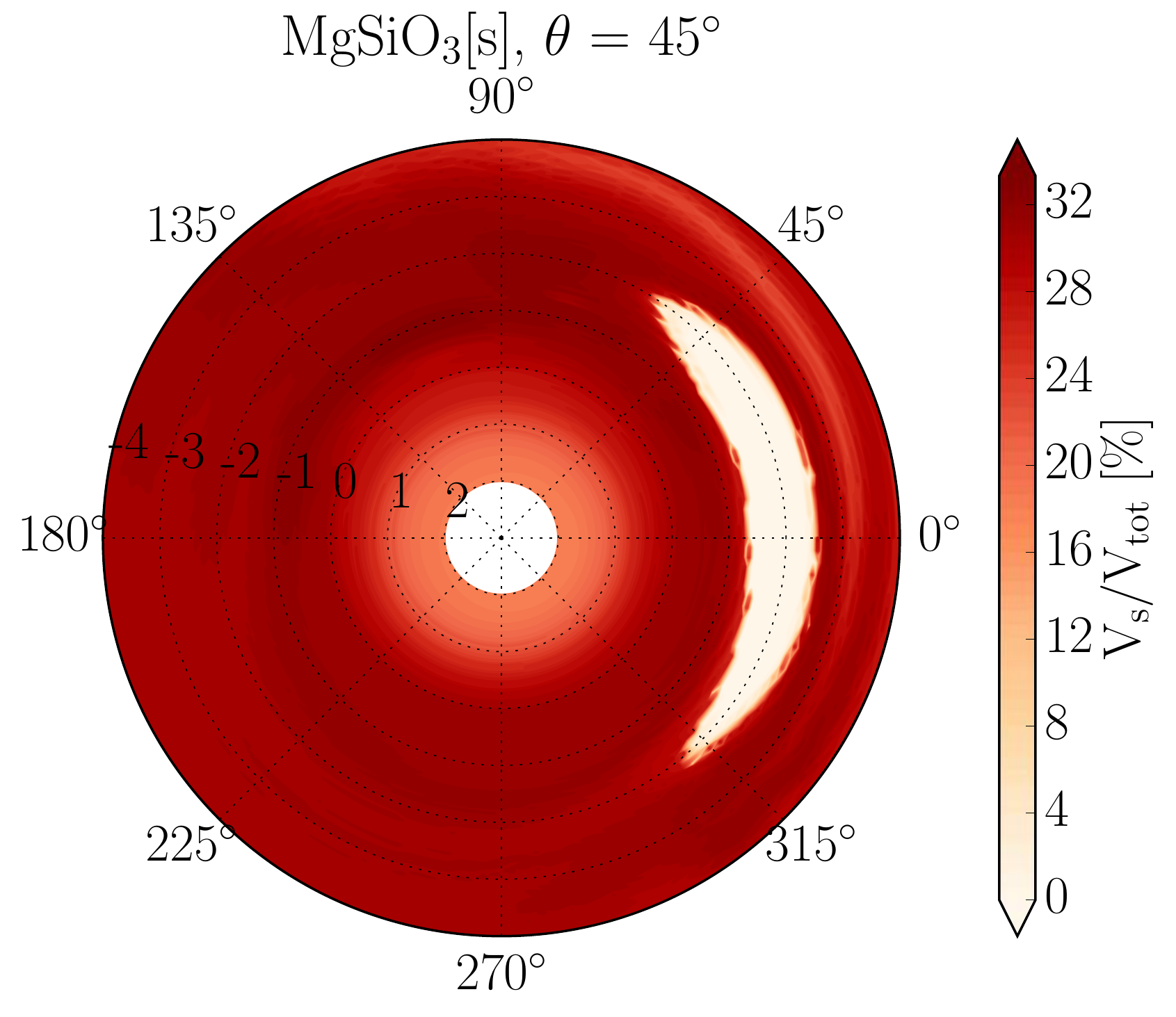}
\includegraphics[width=0.32\textwidth]{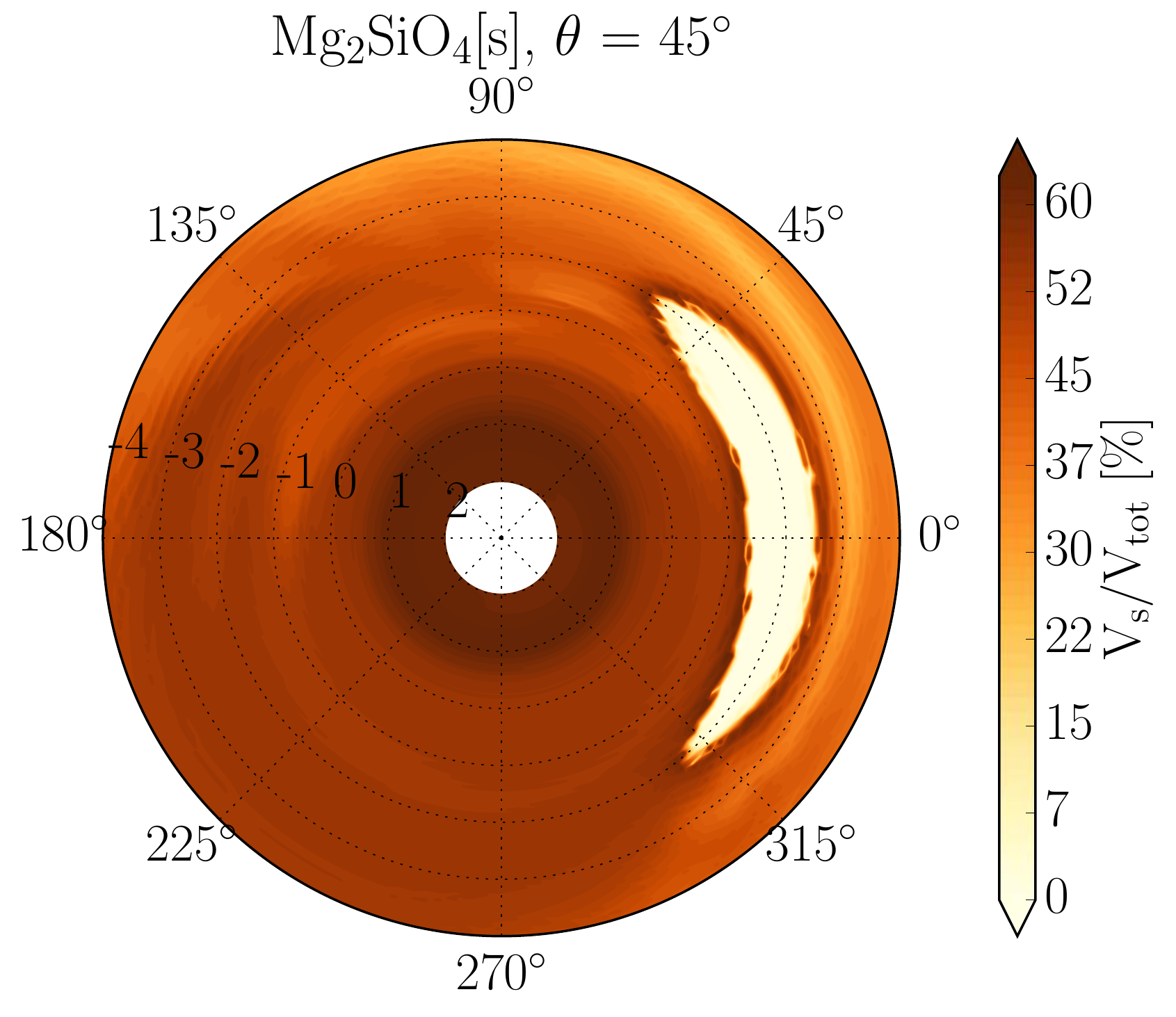}
\caption{Meridional polar slices of atmospheric cloud properties at the equator ($\theta$ = 45$^{\circ}$).
\textbf{Top:} Local gas temperature (T$_{\rm gas}$ [K]), cloud particle number density ($\log_{10}$ n$_{\rm d}$ [cm$^{-3}$]) and mean cloud particle grain size ($\log_{10}$ $\langle$a$\rangle$ [$\mu$m]).
\textbf{Middle:} Volume fraction (V$_{\rm s}$/V$_{\rm tot}$ [$\%$]) of the cloud particle composition containing TiO$_{2}$[s], SiO[s] and SiO$_{2}$[s].
\textbf{Bottom:} Volume fraction (V$_{\rm s}$/V$_{\rm tot}$ [$\%$]) of the cloud particle composition containing MgSiO$_{3}$[s] and Mg$_{2}$SiO$_{4}$[s].
Outer circular values denote longitude at intervals of $\phi$ = 45\degr\ from the sub-stellar point ($\phi$ = 0\degr). Radial values indicate $\log_{10}$ p$_{\rm gas}$ isobars from 0.1 mbar - 100 bar.
The globe is irradiated from the direction of the colour bars.
Note: the size of the annulus is not scaled to planetary radius.}
\label{fig:classical_meri_45}
\end{center}
\end{figure*}

Cloud particles form that are made of a mix of materials that are locally thermally stable. 
The composition of these material mixes changes depending on the local thermo-chemical conditions that a cloud particle may encounter when being advected due to the presence of a velocity field.
Depending on the local temperature and element abundance properties, different solid growth species may dominate the bulk composition compared to others.
The local composition of our mixed material cloud particles is therefore dependent on the local growth and evaporation rates as well as the transport of cloud particles and gas phase elements.
Figure \ref{fig:classical_meri_0} and Fig. \ref{fig:classical_meri_45} show meridional slices of temperature, cloud particle number density, mean grain size and composition at latitudes $\theta$ = 0\degr , 45\degr\ respectively.
These show a complicated, non-uniform composition structure depending on what material is thermo-chemically favourable at each local atmospheric regions.
These plots visualise the interplay between the gas temperature, number density, grain size and composition of the cloud particles.

TiO$_{2}$[s] is a high-temperature condensate which forms stable clusters that become subsequently more stable with size through homogenous nucleation. 
TiO$_{2}$[s] is therefore an efficient seed formation species and will also contribute to the material richness of the grain mantle by surface growth processes.
TiO$_{2}$[s] rich grains (V$_{\rm s}$/V$_{\rm tot}$ $\gtrsim$ 80$\%$) are generally found between pressures of 1-100 mbar on the dayside of planet, corresponding to the hottest regions of the upper atmosphere.
These regions primarily consist of near seed particle size ($\sim$0.001 $\mu$m) cloud particles due to the more volatile materials evaporating off the grain surface.
These seed particles also appear on the nightside of the planet from $\phi$ = 90\degr\ to $\phi$ = 135\degr\ at $\sim$100 mbar due to the equatorial jet efficiently circulating hot gas to the nightside and to greater depths.
At mid-latitudes, pure TiO$_{2}$[s] grains are only found in regions with the highest local gas temperatures at $\sim$10 mbar, also seed particle sized.
These seed particles are thermo-chemically stable.
Elsewhere in the atmosphere, TiO$_{2}$[s] constitutes less than 5$\%$ of the grain volume.
Other materials grow more efficiently due to the greater element abundance of their constituent elements.
Deep atmospheric regions near the lower computational boundary ($\sim$500 bar) contain pure TiO$_{2}$[s] seed particles where other material is thermally unstable.

SiO[s] is typically $<$ 5$\%$ of the volume fraction in most of the atmosphere.
However, it is found in significant volume fractions of $>$ 33$\%$ at the equatorial regions from  $\phi$ = 90\degr\ to $\phi$ = $\sim$315\degr\ at gas pressures of p$_{\rm gas}$ $\sim$ 0.1-10 mbar.
This corresponds to regions of lower gas temperature and density where Mg/Si-growth is unfavourable.
SiO[s] can also be found at the hotter and denser inner atmosphere from 10-100 bar where SiO[s] contributes 10 \% to the total volume of the cloud particles.

SiO$_{2}$[s] dominates (V$_{\rm s}$/V$_{\rm tot}$ $\gtrsim$ 33$\%$) the dayside equatorial upper atmospheric regions from 0.1-1 mbar. 
It is especially dominant at the upper hotter regions from 0.1-1 mbar where Mg/Si-materials are thermodynamically unfavourable with near 100$\%$ composition in some regions.
Grain sizes at these regions are $\sim$ 0.1 $\mu$m.
At mid-latitudes, SiO2[s] contributes with $>$10 \% to the total cloud particle volume at all longitudes and pressure levels, with large volume fractions $>$50 $\%$ at dayside pressures of 0.1-1 mbar.

MgSiO$_{3}$[s] is perhaps the most interesting species included our models since its optical properties have been used to fit transit spectra, Rayleigh slope observations.
We find that it comprises a large amount (V$_{\rm s}$/V$_{\rm tot}$ $>$ 20$\%$) of the grain composition at mid-high latitudes and at all depths, excluding seed particle regions.
However, at equatorial regions it can only be found at the deeper, denser parts of the atmosphere from 100 mbar.
At equatorial longitudes from $\phi$ = 45\degr-90\degr\ at $\sim$1 mbar it can be found to be 10-20 \% of grain volume.

Mg$_{2}$SiO$_{4}$[s] is found to be the most abundant Mg/Si material.
Mg$_{2}$SiO$_{4}$[s] and MgSiO$_{3}$[s] can be found in the same regions in the atmosphere, and follow similar trends for their thermal stability.
However, Mg$_{2}$SiO$_{4}$[s] contributes a larger volume fraction when both materials are thermally stable due to its larger monomer volume.
It is the most dominant material at pressures greater than $\sim$500 mbar with grain volumes over 50$\%$.

Overall, a complicated longitude, latitude and depth dependence of the cloud composition across the globe with the gas temperature playing a key role.
The thermal instability of the silicate materials at the dayside upper atmosphere regions leads to large volumes of the dayside containing thermally stable, nm-sized TiO$_{2}$[s] seed particles.
An equatorial belt of SiO$_{2}$[s] and SiO[s] forms due to the different thermo-chemical conditions between mid-latitudes and equatorial regions. 
Silicate materials such as SiO$_{2}$[s], MgSiO$_{3}$[s] and Mg$_{2}$SiO$_{4}$[s] are abundant at terminator regions ($\phi$ $\sim$ 90\degr, 270\degr) probed by transit spectroscopy.

\subsection{Non-uniform element abundances}
\label{sec:eldep}

\begin{figure*}
\begin{center}
\includegraphics[width=0.37\textwidth]{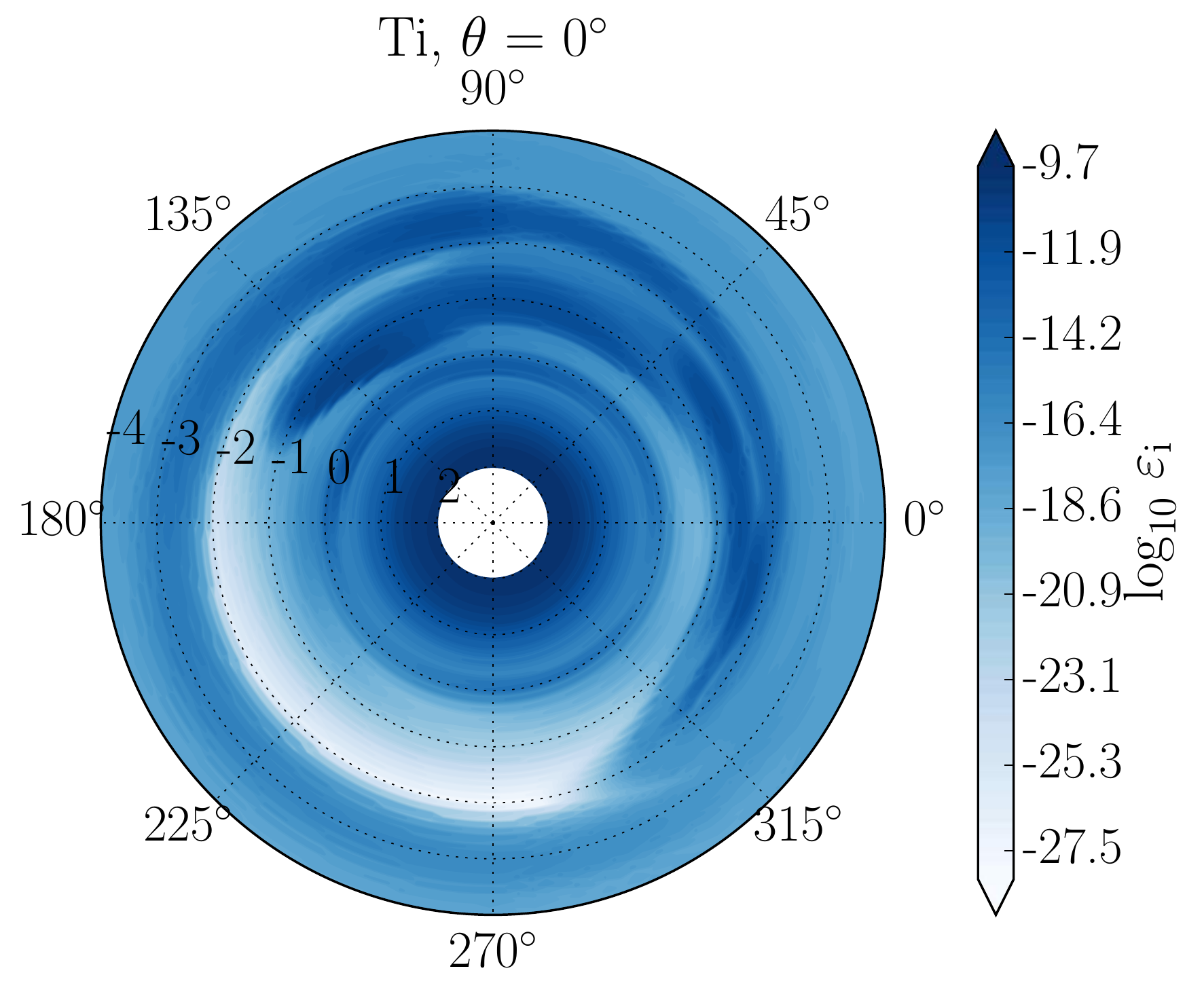}
\includegraphics[width=0.37\textwidth]{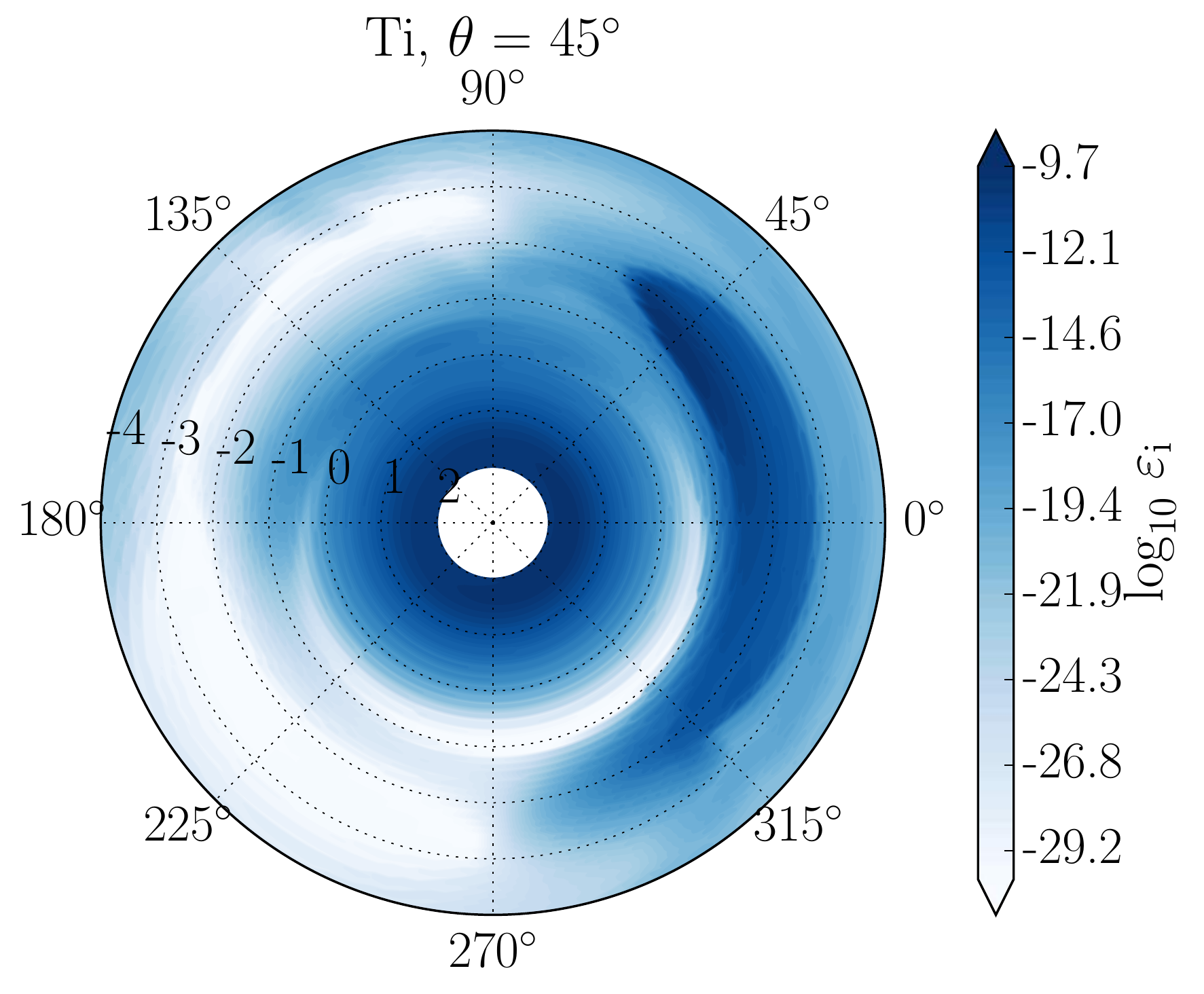}
\includegraphics[width=0.37\textwidth]{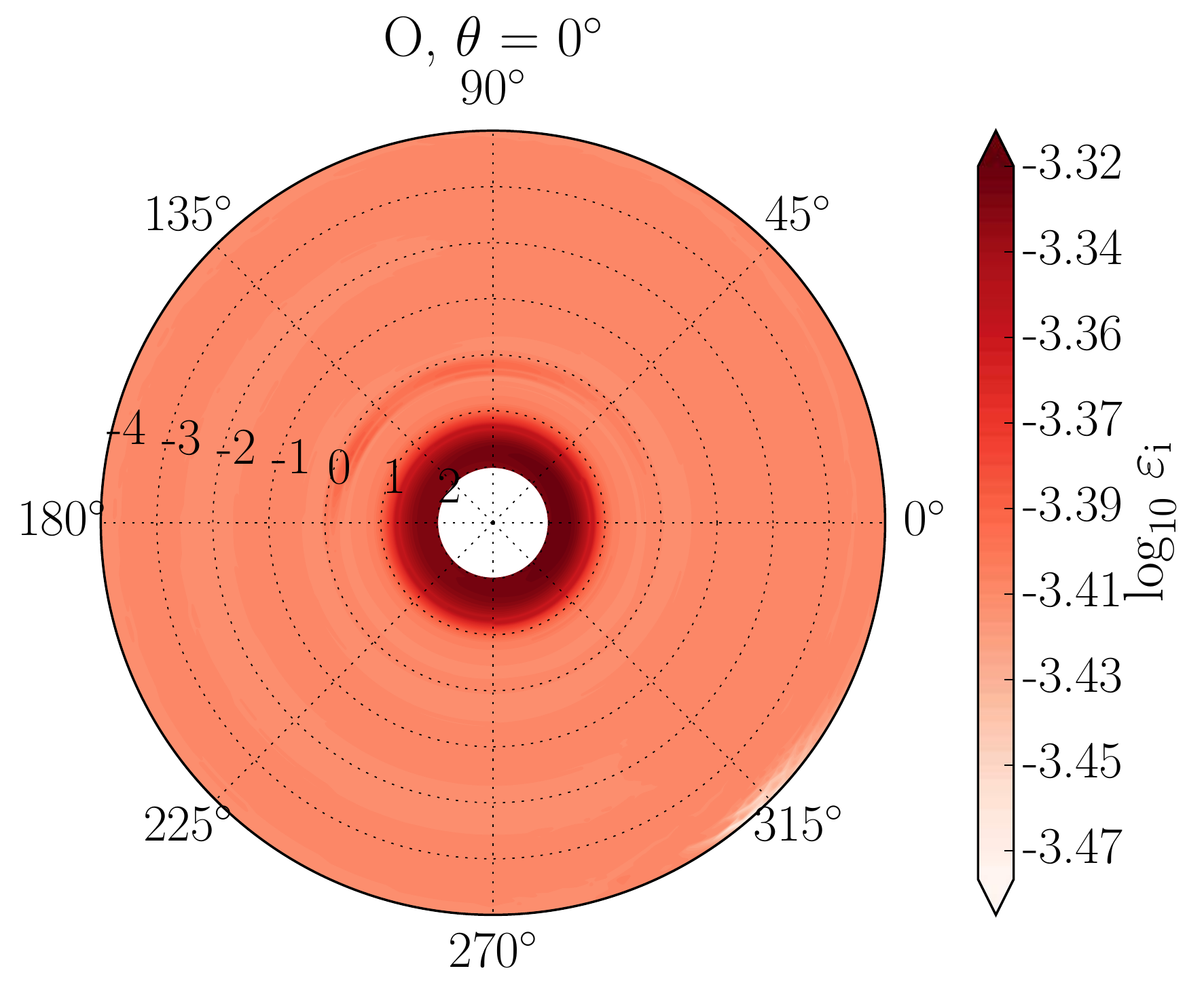}
\includegraphics[width=0.37\textwidth]{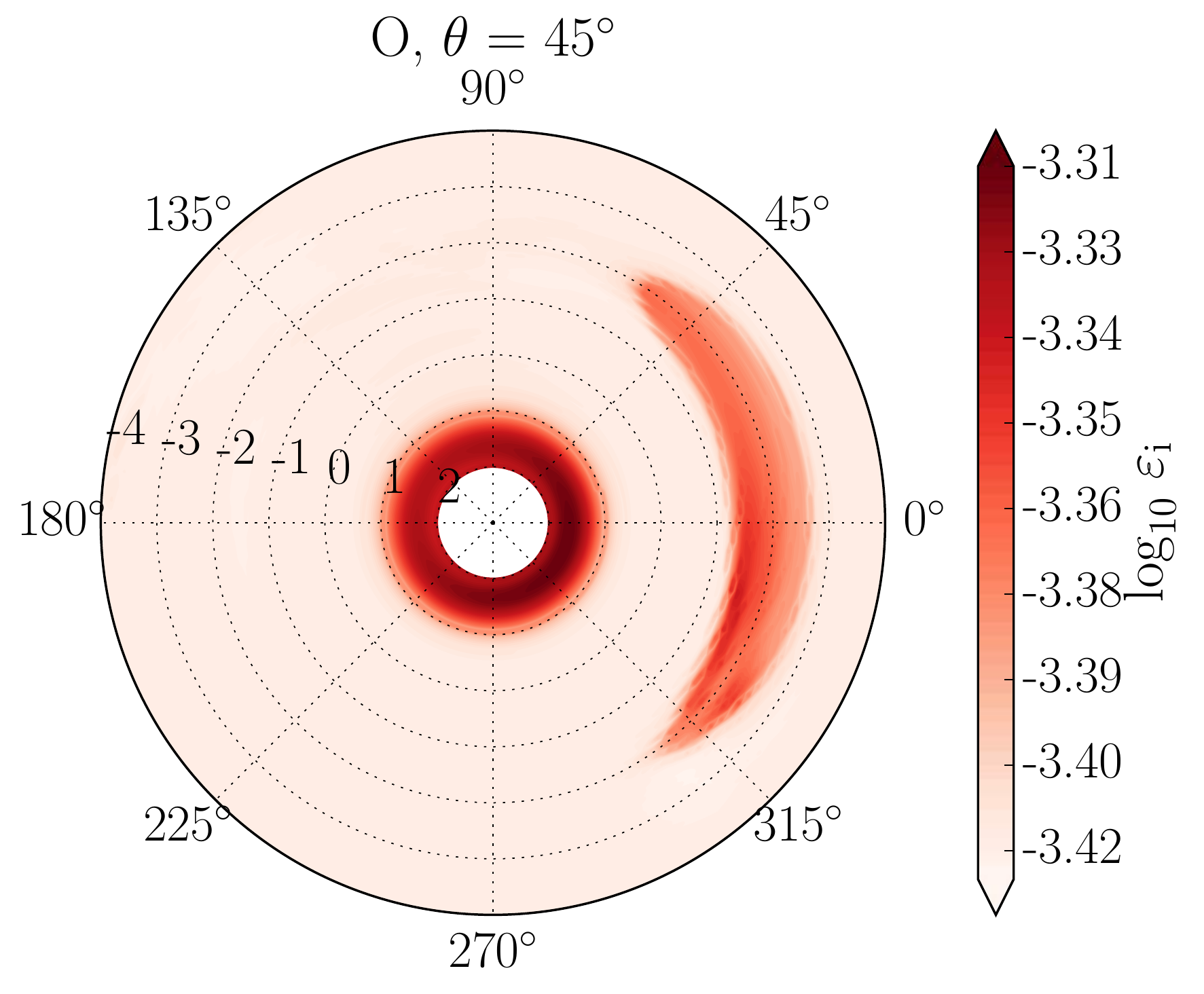}
\includegraphics[width=0.37\textwidth]{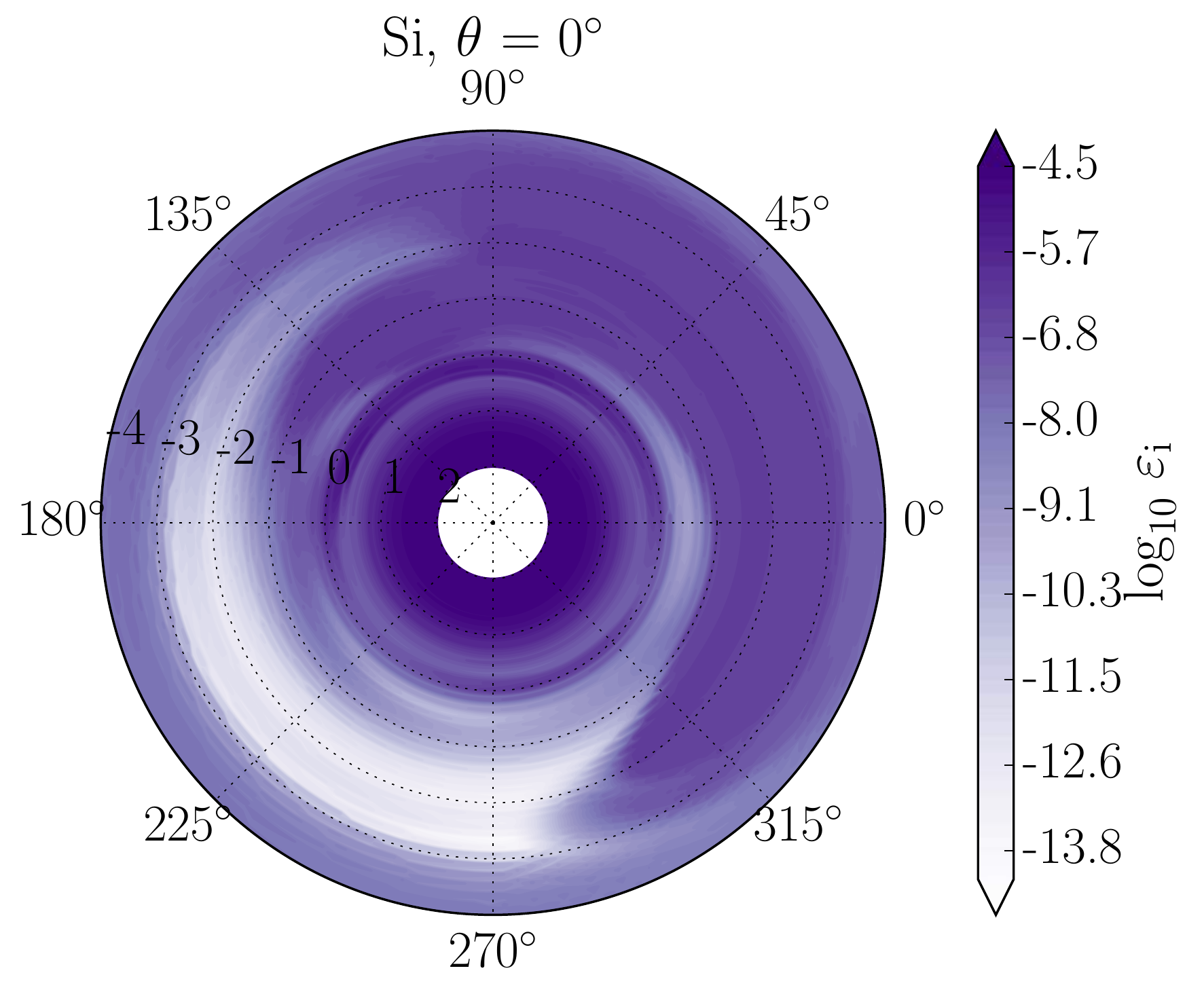}
\includegraphics[width=0.37\textwidth]{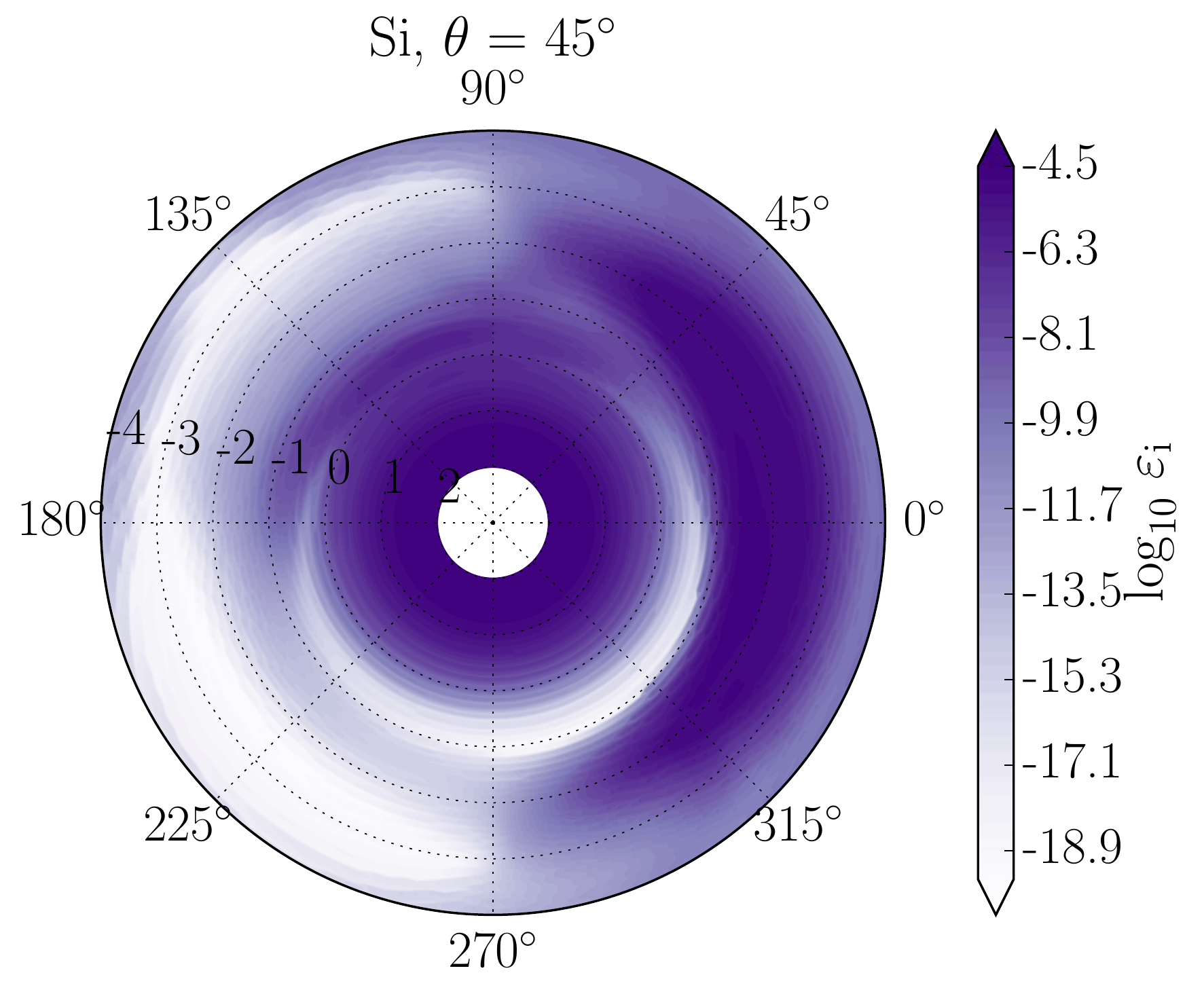}
\includegraphics[width=0.37\textwidth]{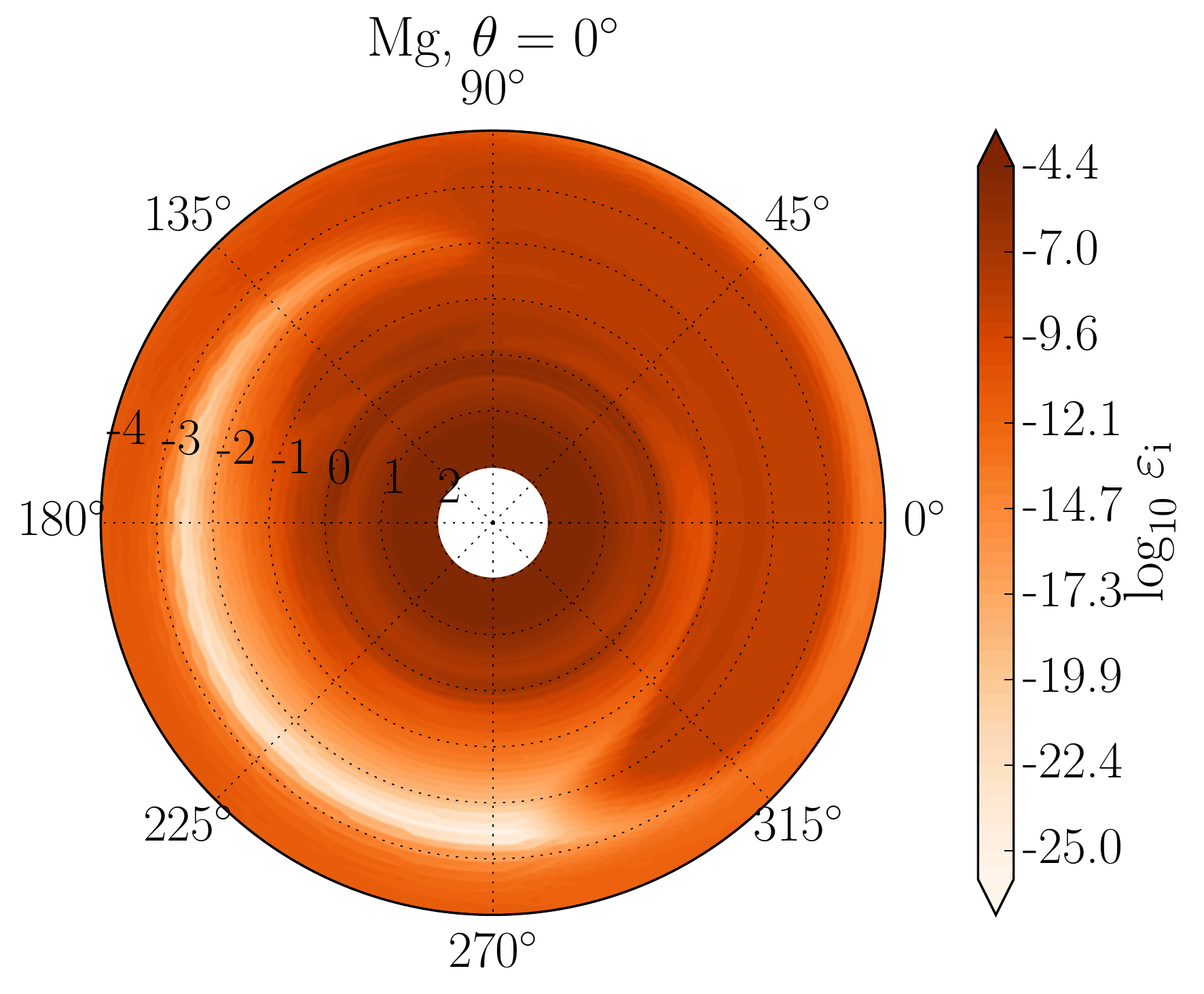}
\includegraphics[width=0.37\textwidth]{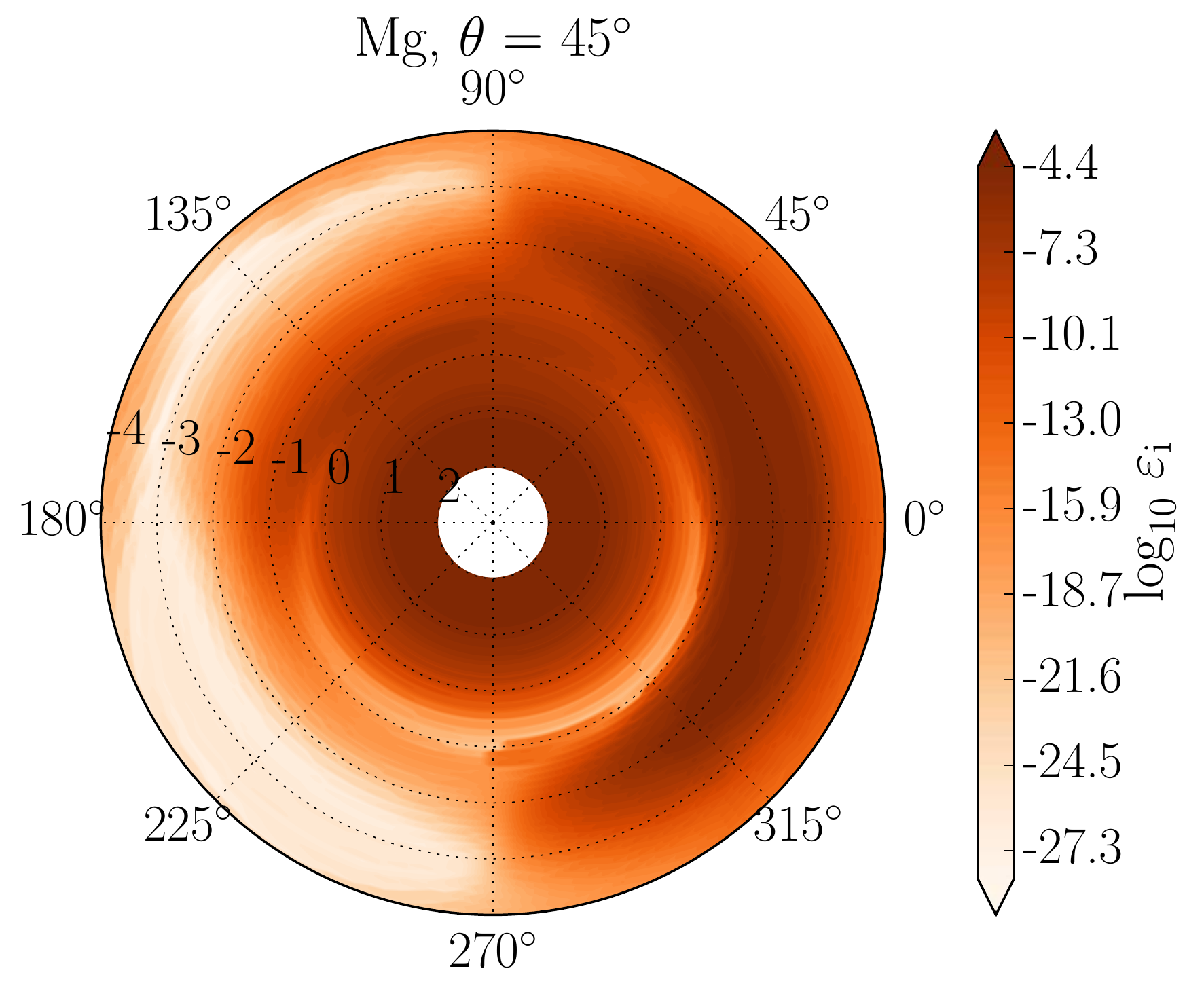}
\caption{Meridional polar slices of gas phase element abundance $\log_{10} \varepsilon_{i}$ = $n_{i}$/$n_{\langle H\rangle}$ (ratio $i$ to Hydrogen abundance) of
\textbf{Top:} Ti (blue), 
\textbf{Second row:} O (red),
\textbf{Third row:} Si (purple) and
\textbf{Fourth row:} Mg (orange/brown) at $\theta$ = 0$^{\circ}$ (left) and  $\theta$ = 45$^{\circ}$ (right), respectively.
For reference, solar metallicity $\log_{10} \varepsilon^{0}_{i}$ of the elements from \citet{Asplund2009} are, Ti: -7.05, O: -3.31, Si: -4.49,  Mg: -4.40.
Lighter coloured regions indicate a depletion of elements due to the cloud formation processes.
Darker coloured regions indicate a local element abundance closer to the initial solar values.
Outer circular values denote longitude at intervals of $\phi$ = 45\degr\ from the sub-stellar point ($\phi$ = 0\degr). Radial values indicate $\log_{10}$ p$_{\rm gas}$ isobars from 0.1 mbar - 100 bar.
The globe is irradiated from the direction of the colour bars.
Note: the size of the annulus is not scaled to planetary radius.}
\label{fig:eps_meri}
\end{center}
\end{figure*}

To complete our understanding of why certain mineral materials are thermally unstable at certain atmospheric regions, we look at the elemental abundances of elements involved in the cloud formation.
The local element abundances determine the gas phase chemical composition which are used to determine the composition of the cloud particles.
Several materials can be thermally stable at a particular T$_{\rm gas}$-p$_{\rm gas}$, meaning each of the condensing materials (S $>$ 1) must be considered when calculating the element depletion \citep[Appendix B,][]{Helling2006}. 
If one solid material has reached an equilibrium state (S = 1), this does not imply that all the constituent elements of the material have been condensed.
If the initial abundance of elements in the gas phase is low for a particular element (e.g. Ti), then any changes in the elemental abundance due to the cloud formation processes will be greater compared to more abundant elements. 
Elements are depleted from the gas phase where cloud particle formation has occurred and are replenished where the material that the cloud particles are composed of has evaporated from the grain bulk.
The "evaporation window" of a material marks the atmospheric longitude ($\phi$) location where a particular cloud particle material becomes thermally unstable.
As cloud particles are transported towards this longitude by circulating gas, the elements from the unstable material replenishes the gas phase.
More volatile materials evaporate further away from the sub-stellar point ($\phi$ = 0\degr), while more thermally stable material evaporates closer to the sub-stellar point. 
Figure \ref{fig:eps_meri} shows meridional slices of the gaseous elemental abundance of Ti, O, Si and Mg at $\theta$ = 0$^{\circ}$, 45$^{\circ}$.
This shows where elements have been depleted in the atmosphere by the formation of cloud particles or replenished by the evaporation of thermally unstable materials.

Ti is typically depleted by 2-20 orders of magnitude depending on the location in the atmosphere.
The magnitude of this severe depletion is due to the initial solar metallicity low abundance of Ti in the atmosphere. 
The highest abundance of gas phase Ti elements occurs at the dayside at $\sim$ 10mbar where Ti is replenished by the evaporation of TiO$_{2}$[s] in these regions.
Seed particles are not evaporated, however, indicating that the TiO$_{2}$[s] seed particles remain thermally stable in these regions.
The gas temperature is also too high to nucleate seed particles at these regions.
The atmosphere never returns to solar metallicity values due to the thermal stability of the TiO$_{2}$[s] seed particles throughout the 3D radial extent of the simulation boundaries.
The evaporation window for TiO$_{2}$[s] occurs at approximately $\phi$ =  315\degr\ longitude at 10 mbar.
At $\theta$ = 0\degr\ the equatorial jet carries uncondensed gas towards the nightside where the local temperatures are low enough to allow an efficient surface growth of TiO$_{2}$[s]. 
The result is a decreasing Ti-element abundance by $>$ 10 magnitudes on the nightside.
The growth efficiency of TiO$_{2}$[s] is less compared to other materials because less Ti is available to condense into TiO$_{2}$[s] (compare \citealt{Helling2006}).

O, the most abundant element considered here, is depleted by a maximum of 30$\%$ throughout the atmosphere. 
Replenishment of oxygen at $\theta$ = 45\degr, 10 mbar occurs due to the evaporation of O bearing silicate materials.

Si is the least depleted element on the nightside in comparison to Ti and Mg.
Upper cooler atmospheric regions where SiO[s] is present tend to be less depleted.
The depletion of Si at the equator, $\theta$ = 0\degr, is from SiO$_{2}$[s] growth.
The evaporation window for Si/O material is $\phi$ $\approx$ 300\degr\ longitude at 10 mbar.
Equatorial jets resupply the nightside with Si elements from evaporating dayside silicate materials.

Mg is typically more depleted on nightside regions indicating the efficient formation of Mg/Si-material.
Gas containing Mg is transported to the nightside by the equatorial jet, where the lower temperatures allow the Mg/Si-material to be thermally stable.
Severe depletion occurs at $\sim$1-10 mbar where Mg/Si-material surface growth occurs.
The evaporation window for Mg/Si-material is $\phi$ $\approx$ 280\degr\ longitude at 10 mbar.

Due to the global dynamics of the upper atmosphere, any replenished elements from evaporating material at mid-latitudes gets funnelled towards the equator.
This hydrodynamic preference is how the mid-latitude, nightside regions contain some of the most element depleted areas of the globe with $>$10 magnitude depletion.
Si/O materials are also dominant at the equator due to these hydrodynamic motions, compared to Mg/Si despite both elements have similar initial solar abundances.
At mid-latitude regions the efficient growth of Mg$_{2}$SiO$_{4}$[s] removes 1 Mg atom more from the gas phase compared to Si atoms for each Mg$_{2}$SiO$_{4}$[s] monomer condensed onto the cloud particle.
The meridional motions then funnel gas from these mid-latitudes regions towards the equatorial jet, leading to a greater gaseous abundance of Si at the equator compared to Mg.
This is seen in the different ranges of colour bar for Si and Mg in Fig. \ref{fig:eps_meri} where Si is typically more abundant by 5-10 magnitudes.
The supersaturation ratio (Eq. \ref{eq:sratio}) for SiO[s] and SiO$_{2}$[s] materials are therefore larger than Mg$_{2}$SiO$_{4}$[s] and MgSiO$_{3}$[s], since there are less gas phase Mg atoms to condense when grains are transported to the equator.
Mg$_{2}$SiO$_{4}$[s] and MgSiO$_{3}$[s] constitute less of the the grain volume as the supersaturation of SiO[s] and SiO$_{2}$[s] increases in these regions while Mg$_{2}$SiO$_{4}$[s] and MgSiO$_{3}$[s] decreases.
That is, Mg/Si material can become thermo-chemically unstable at equatorial regions while Si/O material remains stable.
SiO[s] and SiO$_{2}$[s] are more efficient growth materials than Mg$_{2}$SiO$_{4}$[s] and MgSiO$_{3}$[s] at the equatorial thermo-chemical conditions and become the most dominant material compositions at the equator.
Over time, this leads to an upper atmosphere equatorial band of SiO[s] and SiO$_{2}$[s].
A longitude, latitude and height inhomogeneous element abundance structure develops as the evaporation and growth windows are different for each species as they travel around the atmosphere.

In summary, the atmosphere is depleted of elements due the cloud formation processes.
Local regions of cloud material evaporation on the dayside replenish the gas phase of elements.
Replenishment of the nightside of elements is governed through the equatorial jet, which transports uncondensed material from the dayside to the nightside.
The gas phase elements are then depleted by growth of material at the cooler regions. 
A consequence of the dayside thermal instability and hydrodynamic transport of material is that the $\phi$ = 90\degr\ terminator region is replenished in metallic elements while the $\phi$ = 270\degr\ terminator is severely depleted.
The atmosphere never returns to solar element abundances for those elements involved in cloud formation.

\subsection{Summary of dynamic cloud formation results}
\label{ref:ressummary}

The previous sections presented each of our cloud formation results individually. 
In this section, we examine the physics of our results as a whole and explain the cloud formation process on a global scale.

The nucleation of seed particles early in the simulation begins the cloud formation process.
After a few minutes/hours of simulation, the rate of seed particle formation becomes negligible throughout the entire atmosphere due to the depletion of Ti elements from the gas phase.
Meanwhile, cloud particles that grow $\gtrsim$ 1 $\mu$m begin to settle rapidly from the upper atmosphere to deeper layers ($\sim$ 1 bar), to their pressure supported levels.
Settling of cloud particles results in a globally uniform cloud layer of maximum number density at $\sim$ 1 bar.
These settling grains deplete elements from the upper atmosphere.
This leaves sub-micron cloud particles in the upper atmosphere from p$_{\rm gas}$ $<$ 1 bar.
Cloud particles are transported by meridional gas motions towards the equatorial jet where the largest number density n$_{\rm d}$ of cloud particles can be found.
Cloud particles then follow a circulation cycle as they are transported across the globe, dependent on the local temperature and element abundance conditions.
This large scale cloud formation circulation cycle can be summarised as follows:

\begin{itemize}
\item As cloud particles are transported towards the night-day terminator ($\phi$ = 270\degr), the increase in gas temperature leads material to become thermally unstable.
 The most volatile material evaporates first, while the more stable material evaporates closer to the sub-stellar point ($\phi$ = 0\degr).
 TiO$_{2}$ seed particles ($\sim$ 0.001 $\mu$m) remain thermally stable. 
 The local gas phase is replenished in elements from the evaporating material.
\item The equatorial jet transports the element replenished gas phase and the thermally stable seed particles ($\sim$ 0.001 $\mu$m) past the sub-stellar point ($\phi$ = 0\degr) and to the $\phi$ = 90\degr\ day-night terminator.
 This replenishes the nightside regions near the day-night terminator of uncondensed elements. 
\item From longitudes $\phi$ = 90-180\degr; due to the colder nightside temperatures and the replenishment of gas phase elements and transport of seed particles from the equatorial jet, silicate material is thermally stable and condenses onto the seed particles.
 This depletes the gas phase of elements on the nightside.
 Due to this cloud formation, particles are $\sim$ 0.01-0.1 $\mu$m in these regions.
\item Cloud particles remain thermally stable as they are transported through longitudes $\phi$ = 180-270\degr. 
 The gas phase is severely depleted of elements but the cooler temperatures keep the cloud material thermally stable.
 The cloud particles are then transported to the night-day terminator ($\phi$ = 270\degr) and the cloud formation cycle repeats.
 \end{itemize}

A characteristic of this element cycle is that the amount of metallic elements returned to the gas phase by dayside grain evaporation is not enough to grow the grains to large sizes ($>$ 1 $\mu$m) when they are thermally stable on the nightside.
Over time, an upper atmosphere equatorial band of SiO$_{2}$[s] and SiO[s] rich grains develops due to greater number of uncondensed Si atoms available at the equator, while Mg atoms condense at higher latitudes. 

\section{Cloud/gas opacity and radiative effects of clouds}
\label{sec:CloudOpacity}

\begin{figure*}
\begin{center}
\includegraphics[width=0.49\textwidth]{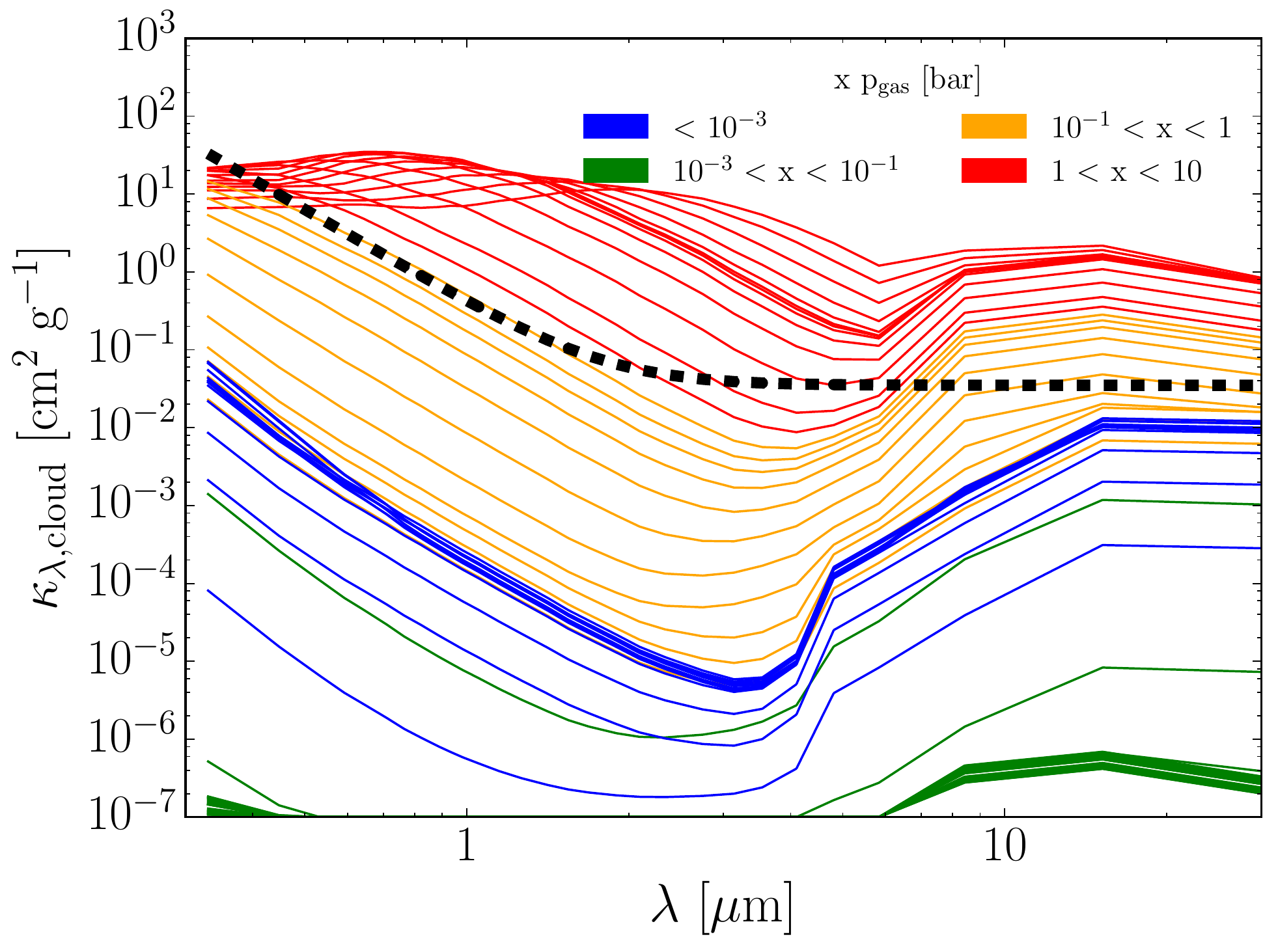}
\includegraphics[width=0.49\textwidth]{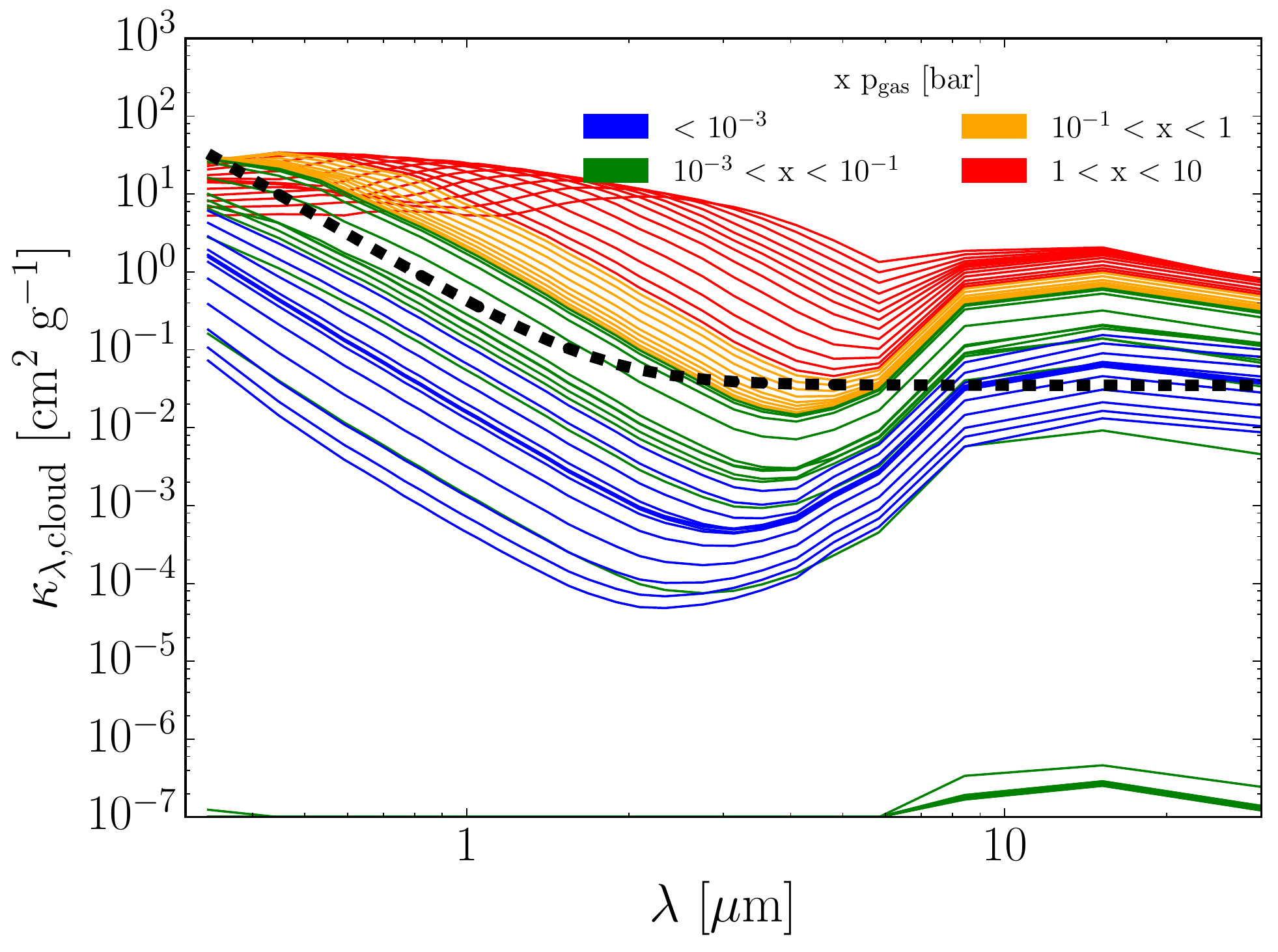}
\includegraphics[width=0.49\textwidth]{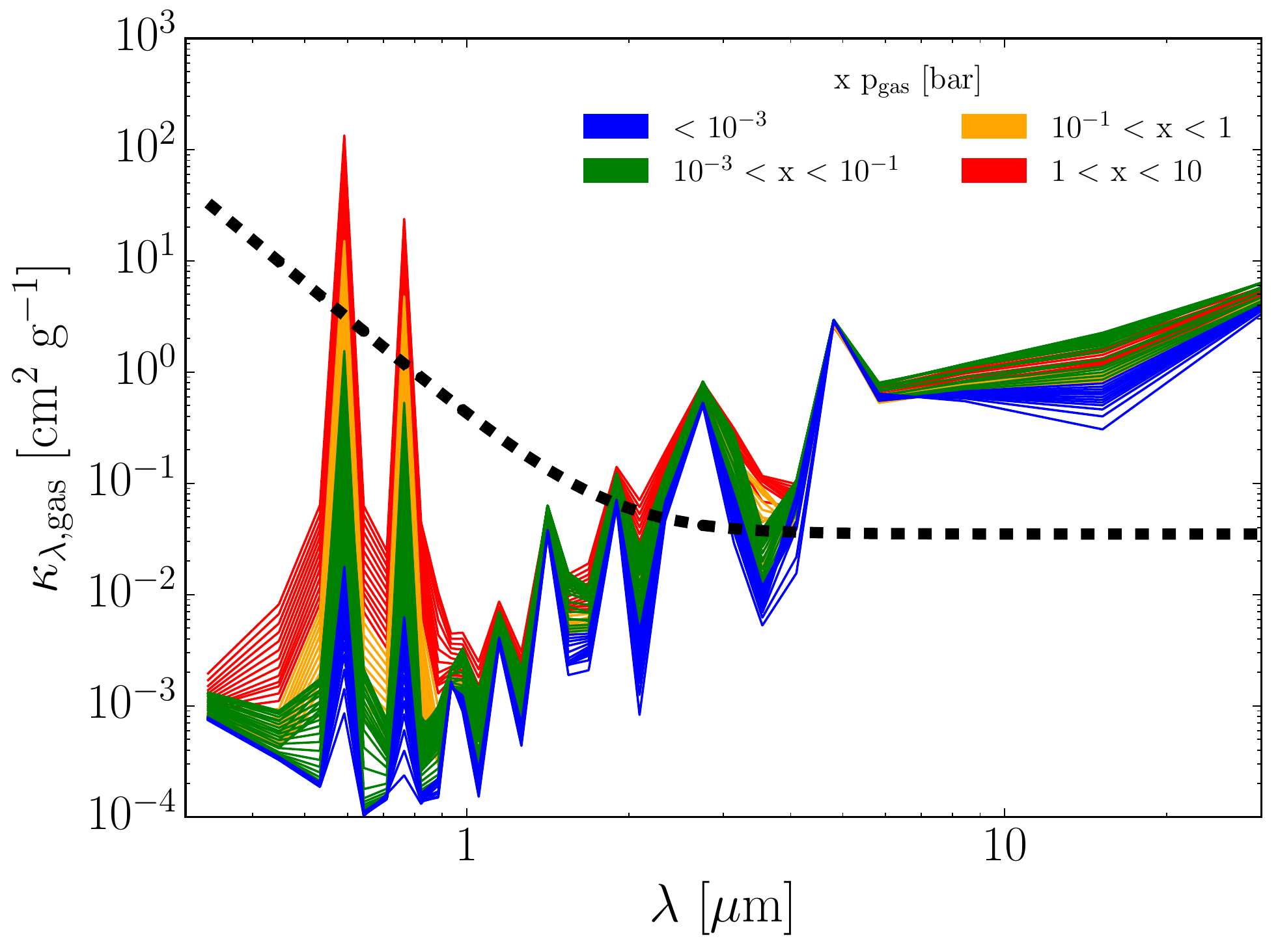}
\includegraphics[width=0.49\textwidth]{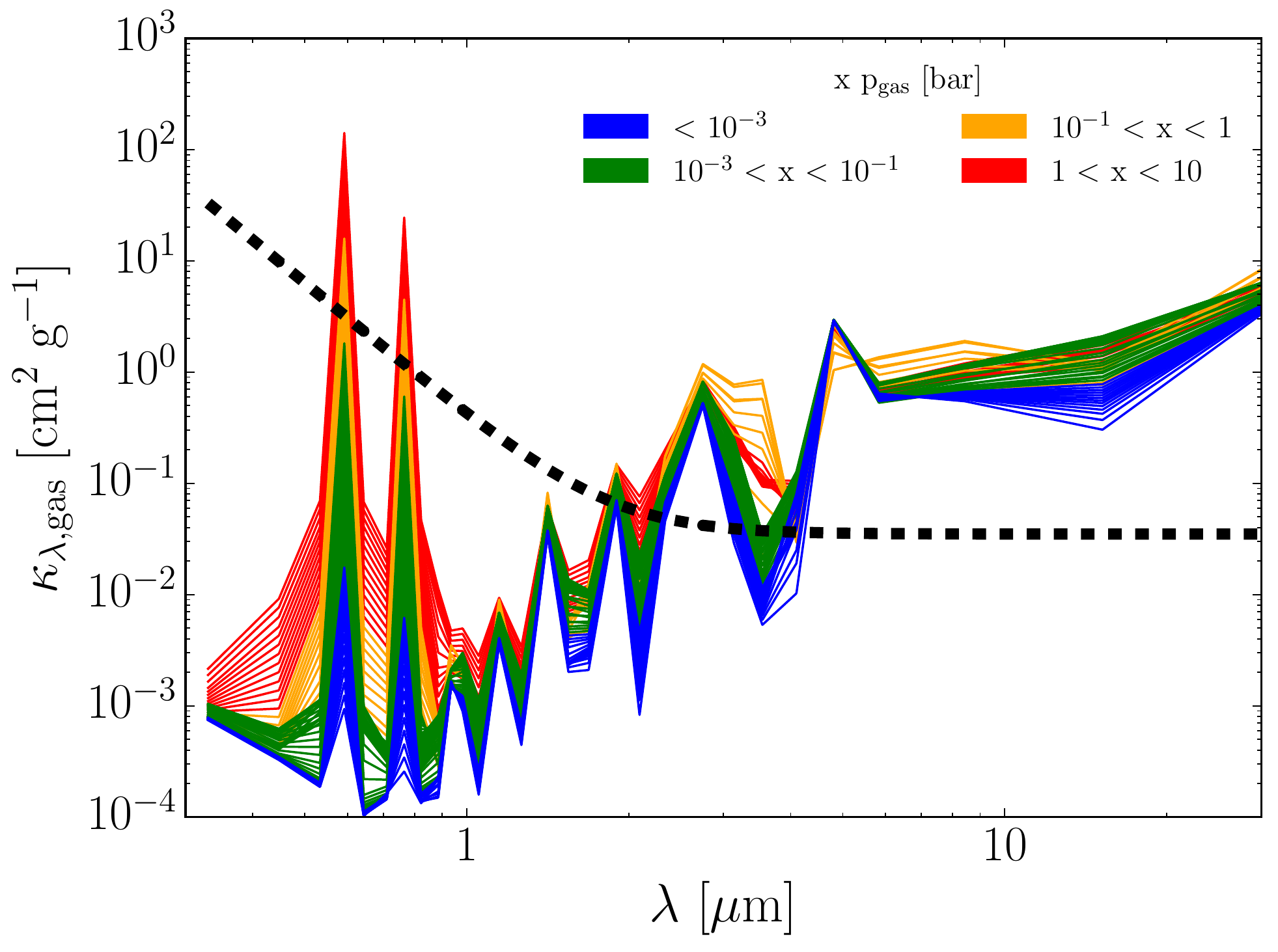}
\includegraphics[width=0.49\textwidth]{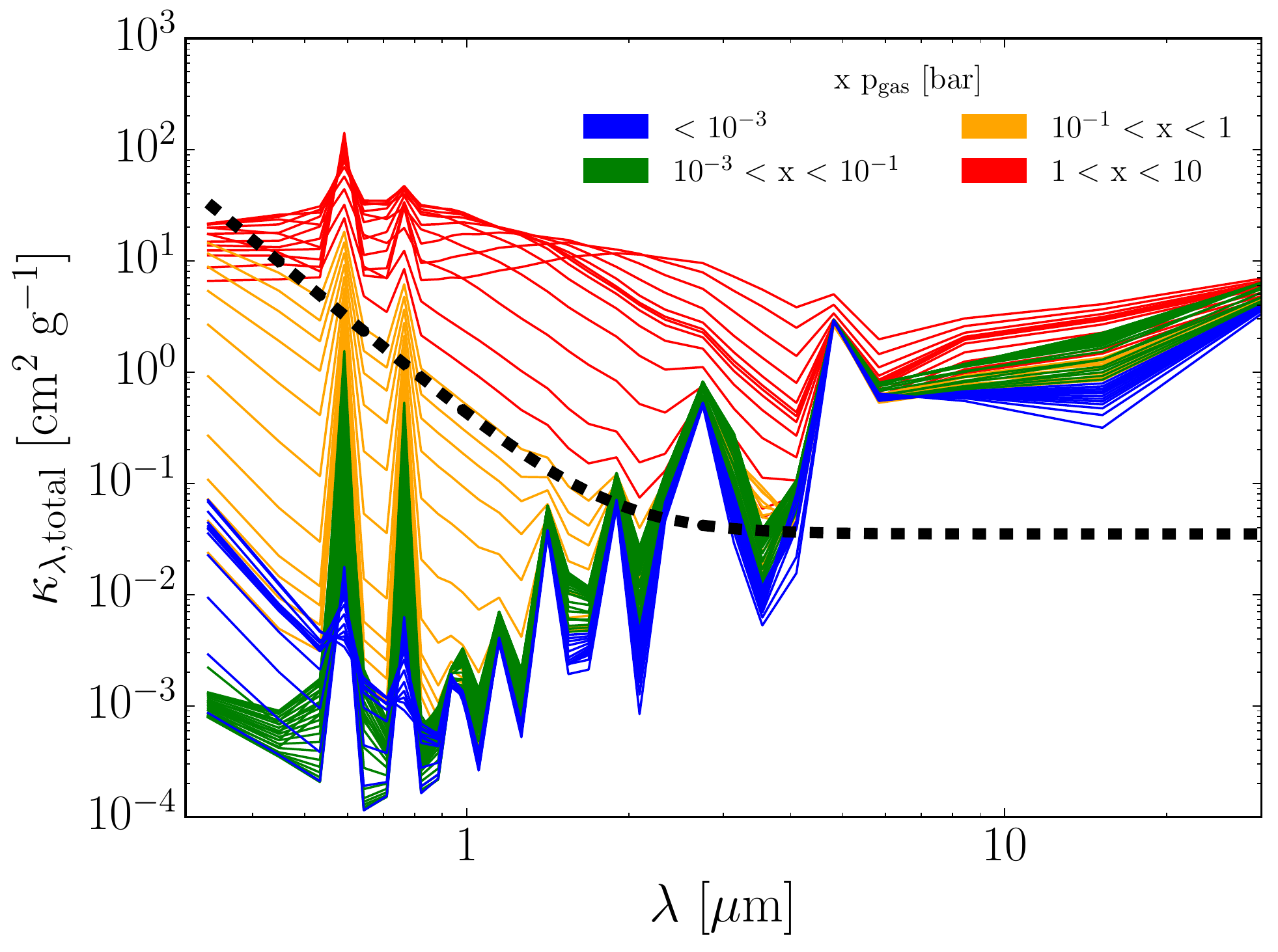}
\includegraphics[width=0.49\textwidth]{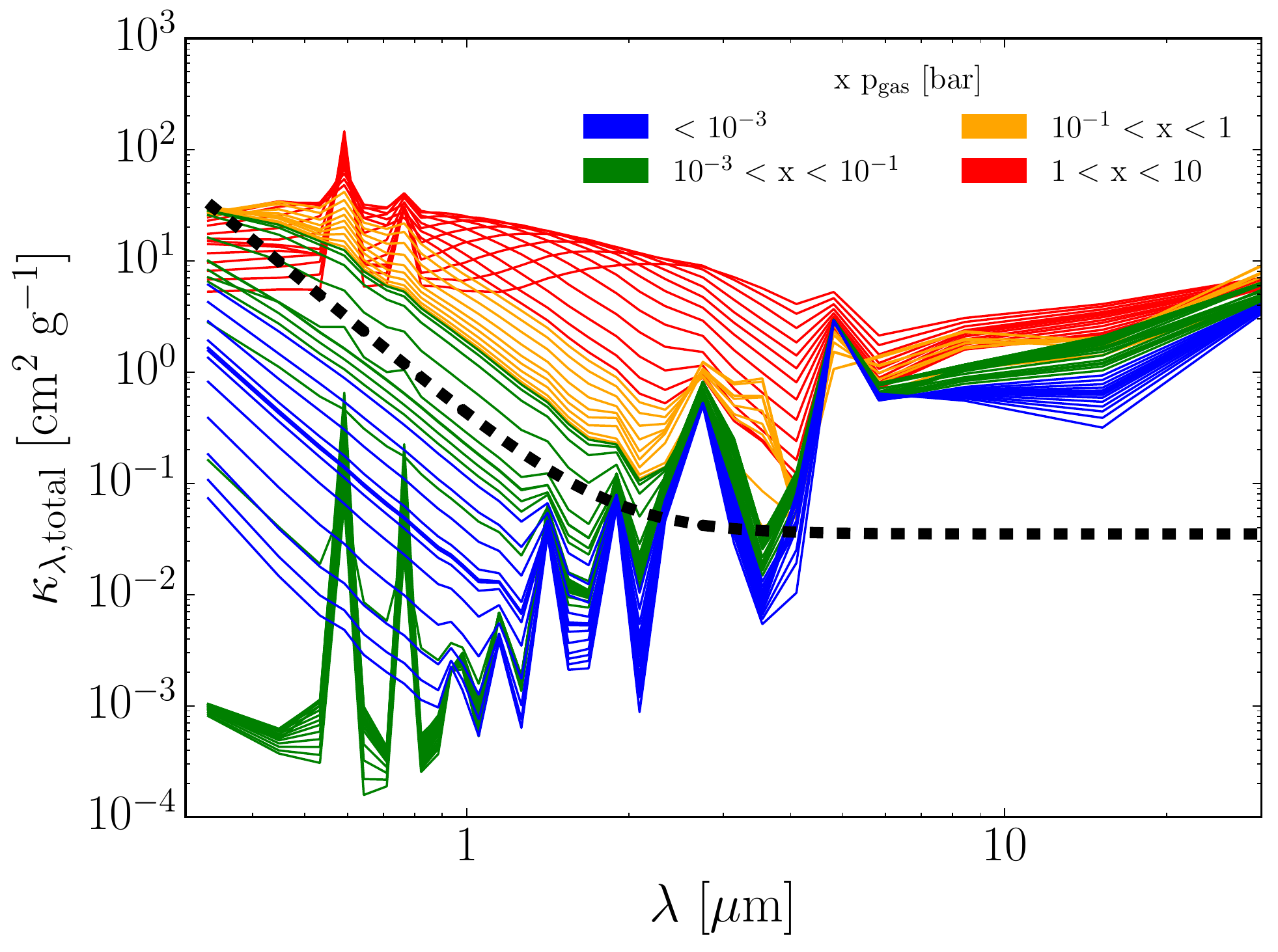}
\caption{ Cloud (top), gas (middle) and total (bottom) opacities at the sub-stellar point (left) and $\phi$ = 0\degr, $\theta$ = 45\degr\ (right) at the centre of each of the wavelength bands used in the RHD radiative transfer scheme.
The thick dashed line shows the parameterised cloud opacity used in \citet{Dobbs-Dixon2013}, Eq. \eqref{eq:paraopac}. The addition of the the size and composition dependent cloud opacity results in a more inhomogeneous opacity structure.
}
\label{fig:ext1}
\end{center}
\end{figure*}

\begin{figure*}
\begin{center}
\includegraphics[width=0.49\textwidth]{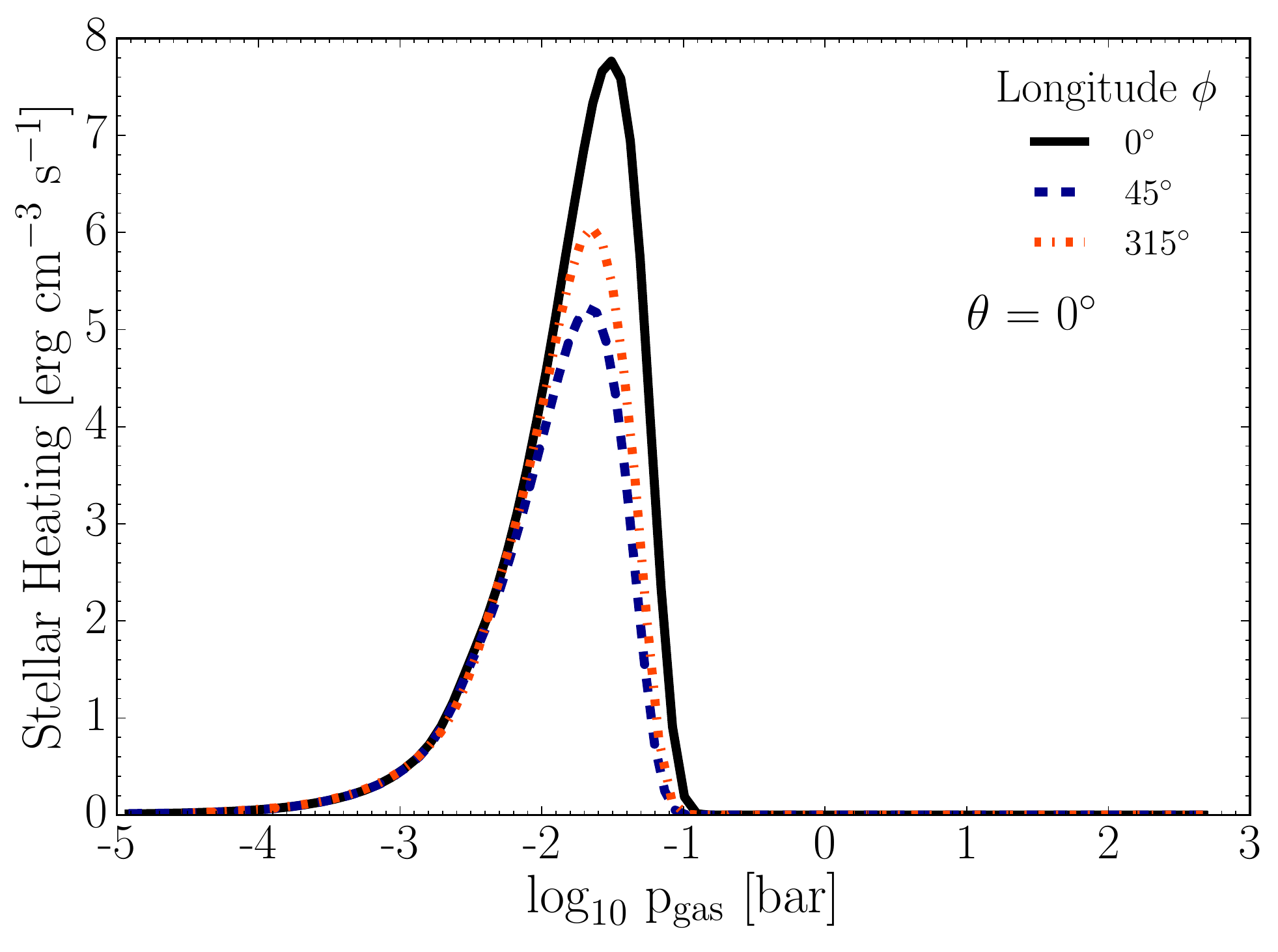}
\includegraphics[width=0.49\textwidth]{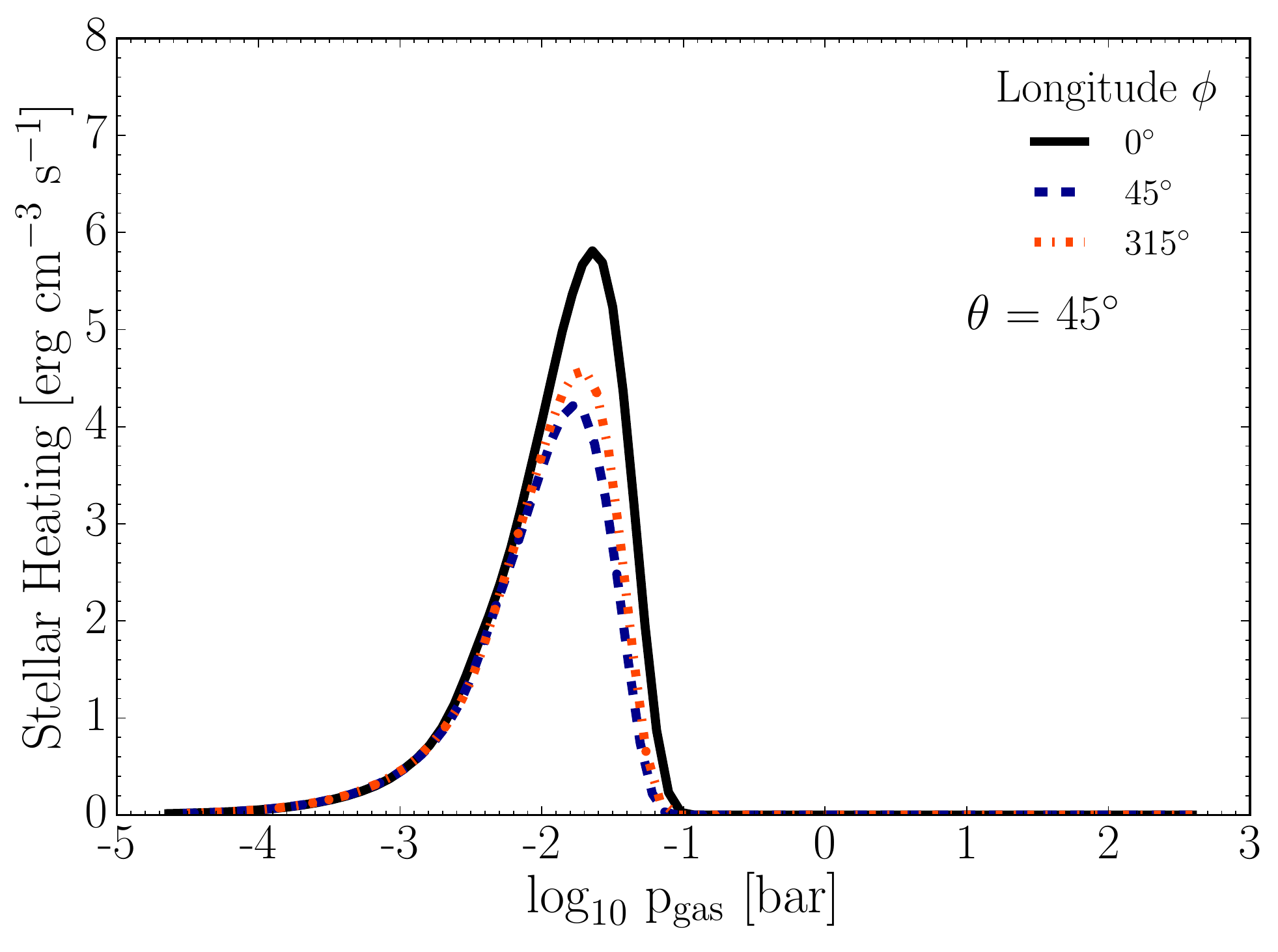}
\caption{Stellar heating rate S$_{\star}$ [erg cm$^{-3}$ s$^{-1}$] (Eq. \ref{eq:stellarheat}) at the equator and $\theta$ =  45\degr\ latitude, $\phi$ = 0\degr, 45\degr, 315\degr\ longitude as a function of pressure.
The heating rate steadily rises up to a maximum at $\sim$10 mbar corresponding to the highest temperature regions at the upper atmosphere.
The heating rate drops off rapidly at 100 mbar.
}
\label{fig:insol1}
\end{center}
\end{figure*}

From Sect. \ref{sec:tprof} the temperature structure of the planet is affected by the presence of cloud particles.
The cloud particles add an inhomogeneous opacity distribution to the atmosphere, altering the local radiation fields.
A specific feature due to the presence of cloud particles is the equatorial regions temperature bump at $\sim$1 bar of $>$100 K.
Figure \ref{fig:ext1} shows the cloud, gas and total opacity at the wavelength midpoints of the wavelength opacity bands in \citet{Showman2009}, applied in this paper, at the sub-stellar point and $\phi$ = 0\degr, $\theta$ = 45\degr.
Figure \ref{fig:ext1} also shows the parameterised cloud opacity used in \citet{Dobbs-Dixon2013} (black dashed line),

\begin{equation}
\label{eq:paraopac}
\kappa_{\rm cloud}(\lambda) = \kappa_{\rm grey} + \kappa_{\rm RS}\left(\frac{\lambda}{0.9 \mu m}\right)^{-4},
\end{equation}

where $\kappa_{\rm grey}$ = 0.035 cm$^{2}$ g$^{-1}$ and  $\kappa_{\rm RS}$ = 0.6 cm$^{2}$ g$^{-1}$.
The size and composition dependent cloud opacity that results from the present simulations is typically lower than the \citet{Dobbs-Dixon2013} opacity in the upper dayside atmosphere (p$_{\rm gas}$ $<$ 1 bar) but larger and greyer, i.e. less strongly wavelength dependent, at all wavelengths from p$_{\rm gas}$ $>$ 1.

Fig. \ref{fig:insol1} shows the stellar heating rate S$_{\star}$ (Eq. \ref{eq:stellarheat}) at the sub-stellar point, $\theta$ = 45\degr\  at $\phi$ = 45\degr, 315\degr\ longitudes.
The peak of energy deposition by stellar photons occurs at $\sim$10 mbar, where the highest dayside temperatures occur.
The stellar energy deposition drops off rapidly at 100 mbar where some of the coolest dayside temperatures can be found.
Compared to a self-consistent gas phase opacity simulation \citep[Fig. 1, ][]{Tsai2014}, the peak of stellar energy deposition is at the same pressure level $\sim$10 mbar.
However, the peak of the heating at the sub-stellar point for the gas phase opacity simulation is $\sim$ 3 erg cm$^{-3}$ s$^{-1}$ less and there is a more gradual drop off in heating to $\sim$ 1 bar.
The \citet{Tsai2014} simulation was found to be too cold when compared to the observations \citep{Dobbs-Dixon2013}.
Our microphysical cloud structure maintains similar stellar heating regions seen in \citet{Dobbs-Dixon2013}, which suggests that a cloud opacity (parameterised or microphysical) pushes the stellar energy deposition further upward on the dayside atmosphere. 
The upper atmosphere gas temperature is typically cooler on the dayside and nightside with the microphysical cloud model compared to the parameterised cloud in \citet{Dobbs-Dixon2013}. 
This indicates that the lower cloud opacity in the microphysical model allows the gas to cool more efficiently in these regions than in the \citet{Dobbs-Dixon2013} simulation.

\citet{Amundsen2014} suggest that the Planck averaged gas opacities used in the current study can lead to greater uncertainties in the stellar heating rate compared to other methods. 
The addition of cloud opacity may reduce this error by muting or washing out the rich molecular lines when the cloud opacity approaches the gas opacity, which can be seen in Fig. \ref{fig:ext1}.
However, in regions of low cloud opacity (e.g. seed particle regions) the results of \citet{Amundsen2014} suggest that the maximum of the stellar heating rate may occur in deeper atmospheric layers.
We hypothesise that this would lead to a smaller or larger vertical extension of the seed particle region on the dayside, depending if this change decreased or increased the temperature at the seed particle region boundaries.
Energy deposited at greater depth would be advected more efficiently in the vertical and horizontal directions, which may impact the overall trends of the cloud particles.

Since the stellar energy deposition is negligible at p$_{\rm gas}$ > 100 mbar, the T$_{\rm gas}$ $>$ 100 K temperature bump seen at the equator (Fig. \ref{fig:Tstrucmean}) cannot be due to stellar heating.
A backwarming effect due to the presence of the opaque cloud base at $\sim$ 1 bar occurs. 
This backwarming was not seen in \citet{Dobbs-Dixon2013}, suggesting that the increased cloud opacity in these regions  (Fig. \ref{fig:ext1}) is responsible for this feature.
The gas irradiated by the host star at $\sim$10 mbar radiates with a Planck function B($\lambda$, T) peak at $\lambda$ $\sim$ 1-2 $\mu$m.
This emitted radiation is then absorbed deeper in the atmosphere where the cloud opacity at $\sim$1 bar is largest at these infrared wavelengths, heating the local gas phase.
The backwarming effect is not seen at mid-latitudes due to the larger grain sizes of the cloud at p$_{\rm gas}$ $<$ 1 bar, producing larger infrared opacity (by up to 3 magnitudes) compared to the equator.
The remitted infrared at mid-latitude regions is then absorbed closer to the temperature peak at 10 mbar, which results in a slightly flatter temperature inversion. 
Therefore, the 100-200 K temperature bump is not seen for mid-high latitude regions.
Similar backwarming effects due to cloud particle and/or gaseous opacity have been investigated in previous studies \cite[e.g.][]{Helling2000,Tsuji2002,Burrows2006,Witte2009,Heng2012,Heng2013}.

\section{Discussion}
\label{sec:Discussion}

\begin{figure*}
\begin{center}
\includegraphics[width=0.49\textwidth]{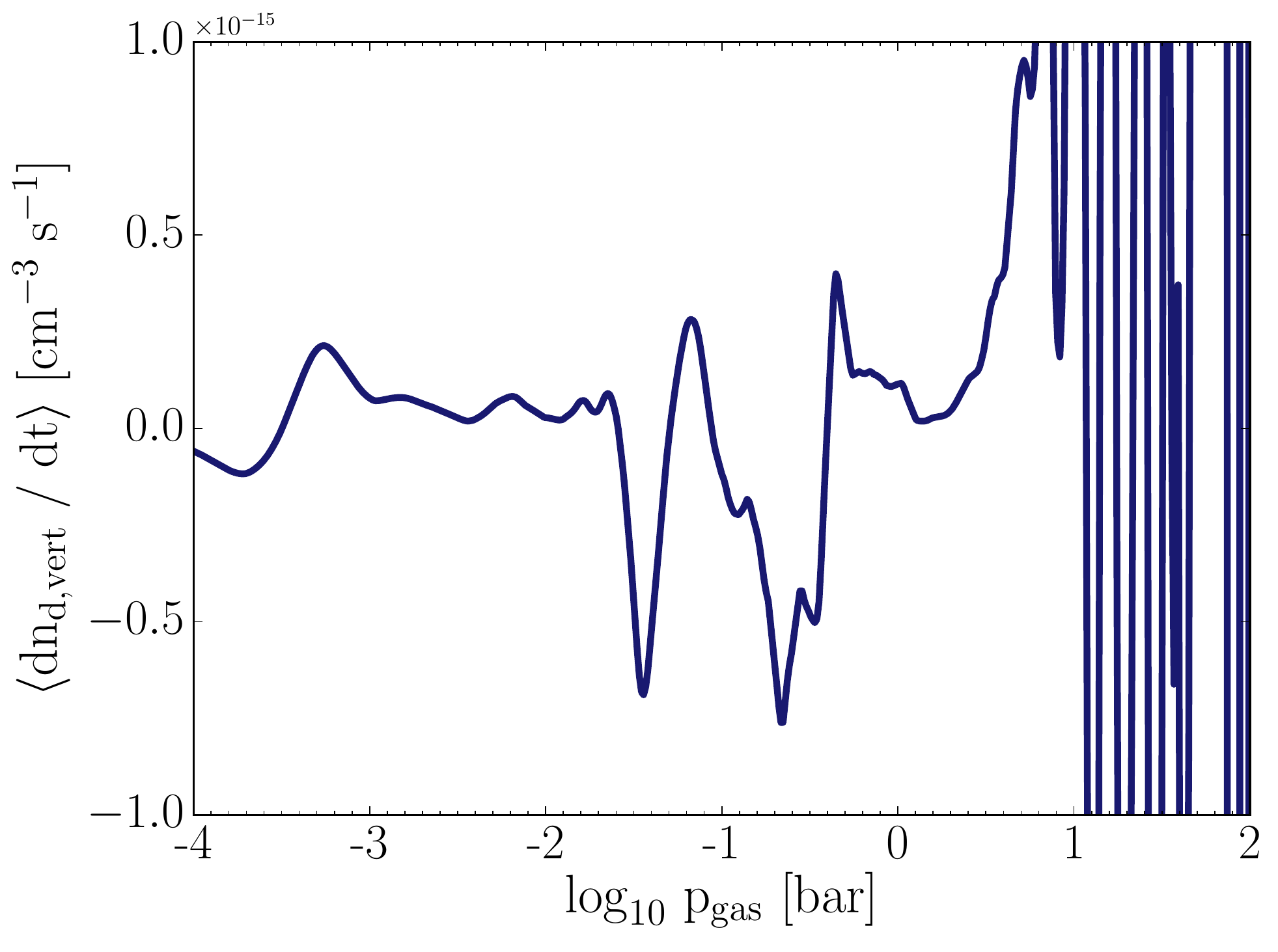}
\includegraphics[width=0.49\textwidth]{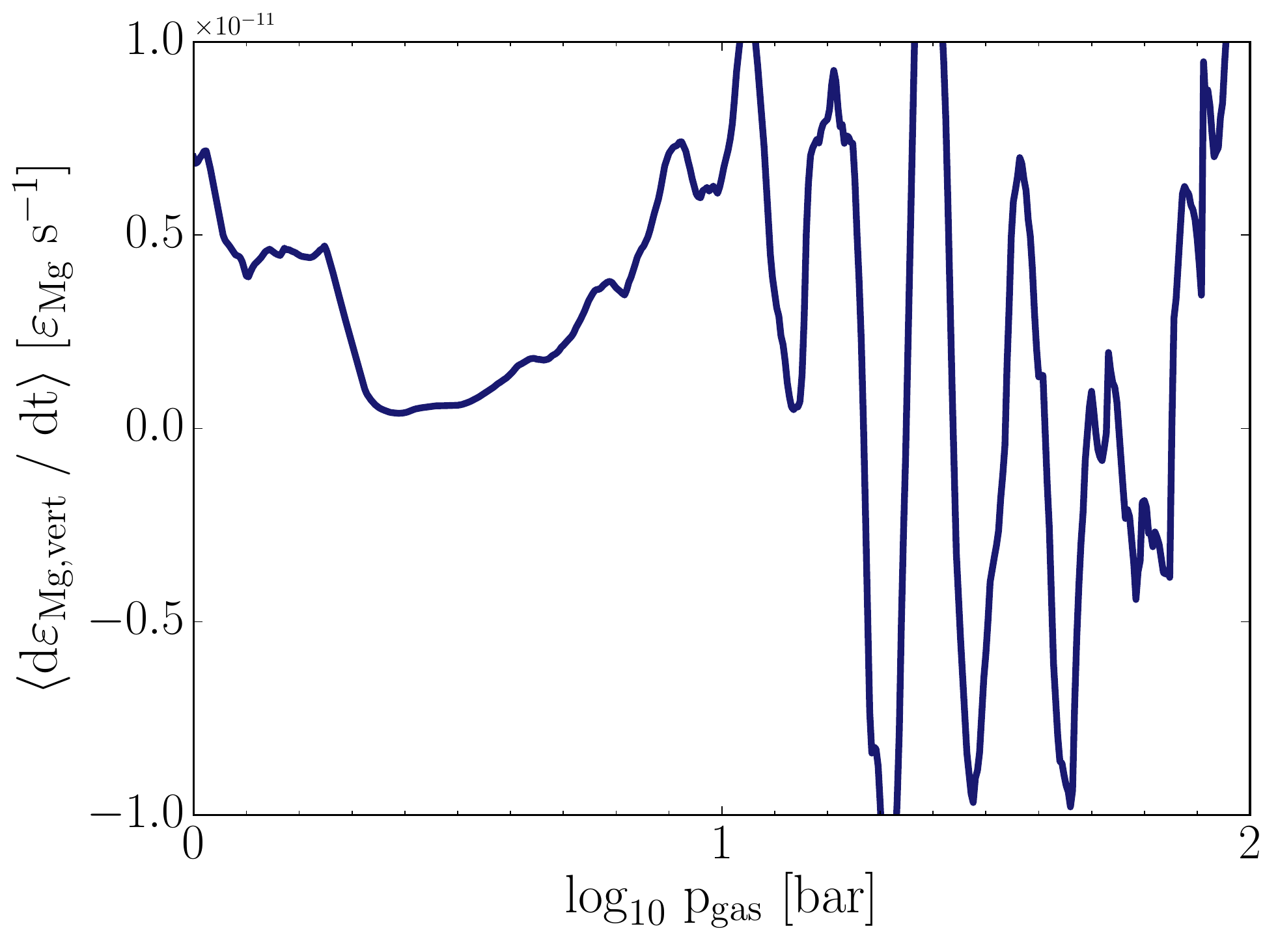}
\caption{ Snapshot horizontally and meridionally averaged iso-bars of the time dependent changes in n$_{\rm d}$ (left) and $\varepsilon_{\rm Mg}$ (right) due to the vertical advection.
$\varepsilon_{\rm Mg}$ is given in the range 1 - 100 bar to avoid the skewing of the global averages by dayside particle evaporation in the upper atmosphere.
Settling of grains by the drift velocity is taken into account for the n$_{\rm d}$ plot.}
\label{fig:vflux}
\end{center}
\end{figure*}

Our results suggest that the atmosphere of HD 189733b contains a silicate mineral cloud component in line with interpretation of transit spectra and albedo observations \citep[e.g.][]{Lecavelier2008, Sing2011b, Pont2013, Evans2013, Wakeford2015, Sing2016}.
\citet{Lecavelier2008} suggest that a sub-micron ($\sim$ 0.01$\,\ldots\,$0.1 $\mu$m) grain size of silicate composition such as MgSiO$_{3}$[s] located at local gas pressures of $\sim$ 10$^{-6}$$\,\ldots\,$ 10$^{-3}$ bar can fit the optical wavelength Rayleigh slope. 
The results of our model suggest that transit spectra observations would sample a variety of grain sizes at different pressure levels within the size and pressure ranges suggested by \citet{Lecavelier2008} and \citet{Wakeford2015}.
Additionally, the composition of the grains changes with longitude and latitude, meaning these observations would also sample differences in cloud particle composition.

We neglect possible condensation of more volatile material such as ZnS[s], KCl[s] and Na$_{2}$S[s].
These materials have been used for modelling cooler objects such as GJ 1214b \citep{Charnay2015b} and T Brown Dwarfs \citep{Morley2012}, where the more stable species considered in this study are found deeper (below $\tau$ $\sim$1) in the atmosphere. 
However, if the atmosphere can efficiently mix solid/gas material upwards, we can expect the more stable condensates to also be present at high altitudes for these cooler objects.
\citet{Sing2015a} suggest that Na$_{2}$S[s] and KCl[s] condensation could be responsible for the sub-solar Na to K ratio observed on hot Jupiter WASP-31b (T$_{\rm eq}$ = 1575 K).
SiO[s], the most volatile species in the current set-up, is only abundant at specific regions on the nightside which are thermo-chemically stable for it.
The atmosphere is generally highly depleted ($>$ 10 orders of magnitude less than solar) of the Ti, Si, Mg elements which take part in the cloud formation, however, longitude, latitude and depth differences in atomic abundance are present.
From our model, the dayside-nightside terminator region is replenished of elements by the equatorial jet after material has evaporated at the hottest dayside regions.
We suggest from the presented results, that Na$_{2}$S[s] and/or KCl[s] condensation could occur on the nightside and deplete Na/K on the nightside-dayside terminator boundary for both HD 189733b and WASP-31b.
Even if Na$_{2}$S[s] and/or KCl[s] are a minor component of the total dust volume, the condensation of the materials can cause a large decrease in elemental abundance, as discussed in \citet{Helling2016}.
The grains would evaporate their Na/K content once they travel to the dayside which would replenish the Na/K atomic abundance for the dayside-nightside terminator.
Additionally, the different thermo-chemical kinetics of Na$_{2}$S[s] and KCl[s] could lead to latitudinal variance, similar to our equatorial band of Si/O and mid-latitude Mg/Si material dominated regions.

As noted in \citet{Mayne2014}, for most GCM/RHD modelling of hot Jupiters, the deep atmosphere (p $\gtrsim$ 1 bar) takes longer to reach a steady state due to the slow (t $>$ 1200 days) momentum exchange between the lower and upper layers.
There is evidence from Fig. \ref{fig:convergence} that the velocity structure is still evolving slowly at these deep regions from the effect of the cloud opacity.
Due to the added cost of the microphysical cloud model, the effect of the upper and lower atmospheric cloud opacity on the dynamics of deeper regions may take many more months or years of simulation, beyond the scope of this early investigation. 

The mixing of replenished gas material upward from the deeper depths ($\sim$ 100 bar) where cloud particles evaporate their volatile contents is also expected to occur on the momentum exchange timescales suggested by \citet{Mayne2014}; 
as it is this timescale where the information of the gaseous elemental content is exchanged between the deep and upper atmosphere.
The replenishment rate over 1 scale height can be approximated by the mixing timescale $\tau_{\rm mix}$ at these depths. 
In \citet{Lee2015b} we estimated that the mixing timescale would be on the order of $\tau_{\rm mix}$ $\sim$ 10$^{8}$ s at $\sim$ 100 bar.
The replenishment rate for Mg abundances at these depths can be estimated from  $\varepsilon^{0}_{\rm Mg}$  by $\tau_{\rm mix}$,  $\varepsilon^{0}_{\rm Mg}$ / $\tau_{\rm mix}$ $\approx$ 4 $\cdot$ 10$^{-13}$ $\varepsilon_{\rm Mg}$ s$^{-1}$; $\varepsilon^{0}_{\rm Mg}$ = 10$^{-4.4}$ and $\tau_{\rm mix}$ = 10$^{8}$ s. 
This value would be many times smaller as the mixing material travels several atmospheric scale heights before reaching the upper atmosphere.
\citet{Agundez2014} suggest that the mixing timescale may be on the order $\tau_{\rm mix}$ $\sim$ 10$^{9}$ s at these depths for HD 189733b, calculated from the GCM mixing tracing method of \citet{Parmentier2013}.
To illustrate this point, Fig. \ref{fig:vflux} shows snapshot horizontal and meridional mean dn$_{\rm d}$ / dt and d$\varepsilon_{\rm Mg}$ / dt due to vertical advection at gas pressure iso-bars.
The small magnitudes of these changes compared to the absolute values suggest that vertical advection may not significantly alter upper atmosphere cloud particle results during the epoch of the simulation discussed here.
Longer integration times ($>$1000 days) will be required to better understand the effect of deeper mixing of gas phase elements and the settling of smaller particles.
However, the results of \citet{Parmentier2013} suggest, for HD 209458b, that sub-micron sized cloud particles may remain present in the upper atmospheric ($>$ 1 bar) layers over longer timescales.
Overall, this is in contrast to more convective atmospheres, e.g. Brown Dwarfs, where $\tau_{\rm mix}$ is estimated to be $\sim$ 300 s at $\sim$ 10 bar \citep{Woitke2004}, which increases the resupply rate of elements to the upper atmosphere.

Through our modelling we have shown that mineral cloud particles can survive for many advective timescales in the atmosphere of HD 189733b.
Cloud particles are continually transported between warmer and cooler regions by either horizontal motions from warm dayside to cool nightside, or from vertical flows from the cooler upper atmosphere to the warmer inner atmosphere and back again.
\citet{Helling2009b} show that SiO$_{2}$[s], MgSiO$_{3}$[s] and Mg$_{2}$SiO$_{4}$[s] can crystallise efficiently in supersaturated atmospheric conditions similar to that of hot Jupiters, especially if grains are transported from hotter to cooler regions and vice-versa.
In addition, since the cloud particles can survive for long periods in the atmosphere, the internal structure of the grains have time to rearrange into a crystalline structure.
This would change the optical properties of the grains, where they would appear more crystalline compared to amorphous grains.
This may increase the wavelength dependent back-scattering of photons.

Our presented cloud modelling also lends weight to a patchy cloud scenario to explain the Westward offsets in optical phase-curves of some Kepler hot Jupiter exoplanets \citep{Heng2013, Hu2015,Esteves2015, Shporer2015}.
The evaporation window of the reflecting silicate materials in this study occurs $\phi$ =  45-90\degr\ longitude Westward from the sub-stellar point, leading to smaller grain sizes, a changed material composition and a large drop in cloud opacity across large areas of the mid and Eastern dayside.
Due to the different atmospheric properties of the Kepler planets, should this evaporation window occur closer to the sub-stellar point, the difference in reflectivity of the East and West cloud properties would be become more noticeable.
Our current study therefore provides a microphysical basis to the analysis in \citet{Oreshenko2016} and \citet{Parmentier2016}.
By examining the kinetics of the cloud formation with radiative feedback effects for the Kepler planets, a detailed picture of the composition and size of the optical backscattering material can be calculated.
The variability in the reflectivity of Kepler-7b has been shown to be small over the Kepler observing period \citep{Demory2013}, suggesting that a long-term, stable cloud particle hydrodynamic circulation and thermo-chemical cycle is a likely possibility for these atmospheres.

\subsection{Comparison to 1D results}
In this section we compare our 3D coupled model to our previous 1D post-processing, non-global approach in \citet{Lee2015b}.
In the current study we have investigated 5 cloud species (TiO$_{2}$[s], SiO[s], SiO$_{2}$[s], MgSiO$_{3}$[s], MgSiO$_{4}$[s]), neglecting 7 materials (Fe[s], FeO[s], FeS[s], Fe$_{2}$O$_{3}$[s], CaTiO$_{3}$[s], MgO[s], Al$_{2}$O$_{3}$[s]) which were included in previous modelling efforts \citep{Helling2006,Helling2008,Lee2015b,Helling2016}.
The previous 1D study \citet{Lee2015b} showed that some of the neglected species can have significant volume fractions in certain parts of the atmosphere.
Al$_{2}$O$_{3}$[s]  and CaTiO$_{3}$[s] were found to be thermally stable at regions of the hot dayside, this is likely to be the same in the 3D RHD case. 
This suggests that the grain sizes could be underestimated for these regions in the RHD model. 
However, Al$_{2}$O$_{3}$[s] and CaTiO$_{3}$[s] are typically not efficient growth species and are unlikely to grow the grain to significantly larger ($>$ 0.01 $\mu$m) sizes.
The addition of Fe[s] will effect the thermal stability of the cloud particles, grain sizes and drift velocity and therefore the local degree of element depletion, especially in the deeper regions where other species are less thermally stable compared to Fe[s]. 
A richer chemical composition can be expected when additional high-temperature condensates are included.

\citet{Lee2015b} generally reproduces the regions of efficient nucleation, growth and evaporation compared to the 3D RHD model, indicating that the chemical processes are accurately captured by the 1D models.
The influence of dynamics on the specific cloud properties is large however, which leads to differences in predicted grain sizes between the two approaches.
The time dependent settling of $\sim$ 1 $\mu$m grains to their pressure supported regions near 1 bar results in a higher cloud opacity deeper in the atmosphere for the RHD model compared to the 1D models.

In more general terms, our 1D cloud simulations, {\sc Drift}, are a valuable analysis tool because they are fast at providing a stationary solution to cloud properties with a substantial degree of chemical details. 
The 3D RHD simulations with our cloud formation module, are time-consuming but allows us to resolve the time and spacial evolution of a cloud-forming hot Jupiter atmosphere.
At present, the time evolution of the current study has focused on the first 60 Earth days and the spacial resolution is limited to that of the RHD/GCM cells.
Long-term studies which address the limitations of the current implementation are under development.

\section{Summary and conclusions}
\label{sec:Conclusion}

We have developed a 3D kinetic cloud formation module for exoplanet RHD/GCM simulations.
Through our coupled RHD and cloud model we have shown that HD 189733b has extremely favourable thermodynamic and hydrodynamic conditions for efficient cloud formation and growth.
We lend weight to previous interpretations of observations of a thick mineral cloud component containing sub-micron sized particles in the upper atmosphere.
The interplay between the hydrodynamical motions and the cloud formation produces an inhomogeneous opacity structure which has effects on the global atmospheric conditions.
A summary of key results include:

\begin{itemize}
\item Grain sizes are sub-micron at atmospheric pressures and terminator regions probed by transmission spectroscopy.
\item Silicate materials are thermally stable in regions probed by transmission spectroscopy.
\item Grain sizes, number density and opacity of cloud particles are non uniform across the globe with significant differences in longitude, latitude and depth.
\item Due to the global elemental depletion from cloud formation, the equatorial regions are dominated by SiO[s] and SiO$_{2}$[s] while mid-upper latitudes mostly contain MgSiO$_{3}$[s] and Mg$_{2}$SiO$_{4}$[s].
\item Hydrodynamic motions primarily govern the global distribution of cloud particles, transporting cloud particles towards nightside equatorial regions.
\item The existence of the clouds as well as the particle sizes and material mixes are the result of the kinetic cloud formation processes.
\item The atmosphere is severely depleted ($\ge$10 orders of magnitude) of elements Ti, Si, Mg used in the cloud formation theory while O is only depleted by $\sim$30 $\%$.
\item Mid-high latitude nightside regions are not efficiently replenished in elements and contain the most reduced gas phase elemental abundances.
\item Thermally unstable materials on the dayside replenish elements to the gas phase; these uncondensed elements are then transported to the nightside via the equatorial jet.
\item Maximum thermal stability of the cloud particles are found at the coolest parts of the nightside inside the large scale vortex regions.
\end{itemize}

We improve on previous 1D approaches by developing an atmospheric RHD model consistently coupled to cloud formation model of with self consistent opacity feedback, element mixing and cloud particle transport and settling. 
We emphasise that we do not rely on any mixing parameterisation of gaseous /solid material (e.g. by use of K$_{\rm zz}$), assumptions about the grain sizes or particle size distributions. 
Through this early study, we have demonstrated that our kinetic cloud formation model is well suited to be applied to 3D hydrodynamic studies of exoplanet atmospheres.
However, long-term studies are required to address the limitations of the current implementation of the cloud module.
Potential future applications of our model are 3D Brown Dwarf atmosphere simulations such as those in \citet{Showman2013a, Zhang2014}.
The model can be extended for warm-Neptune studies with the addition of more volatile cloud species such as ZnS[s], KCl[s] and Na$_{2}$S[s] to the chemical scheme.

\begin{acknowledgements}
We thank the anonymous referee for insightful comments and suggestions which improved the manuscript substantially.
GL and ChH highlight the financial support of the European community under the FP7 ERC starting grant 257431.
Our local HPC computational support at Abu Dhabi and St Andrews is highly acknowledged.
Most plots were produced using the community open-source Python packages Matplotlib, SciPy and AstroPy.
We thank J. Blecic and current/former members of the LEAP team for insightful discussions and suggestions.
\end{acknowledgements}

\bibliographystyle{aa}
\bibliography{bib2}{}

\end{document}